\documentclass[11pt]{article}
\usepackage{epsfig}
\usepackage{amsfonts}
\usepackage{amsmath}
\usepackage{amssymb}
\usepackage{bbm,bm}
\usepackage{graphicx}
\usepackage{slashed}
\allowdisplaybreaks[1]

\usepackage{tikz}
\newdimen\nodeDist
\usepackage{datetime}
\usepackage{xcolor}
\usepackage{hyperref}
\usepackage[numbers,sort&compress]{natbib}
\input pix.sty

\usepackage{myhighlight}

\usepackage{multirow}

\usepackage{bm}
\newcommand{\repr}[3]{(\bm{#1},\bm{#2})_{#3}}

\newcommand{\myaut}{\textsc{AutoTherm}}
\newcommand{\myfr}{\textsc{FeynRules}}
\newcommand{\myfa}{\textsc{FeynArts}}
\newcommand{\myfc}{\textsc{FormCalc}}
\newcommand{\mymath}{\textsc{Mathematica}}
\newcommand{\anpart}{field-theoretical}
\newcommand{\numpart}{convolution}
\newcommand{\myautdir}{autotherm-v1.0}

\newcommand{\masym}{m_{\infty}}
\newcommand{\qperp}{q_{\perp}}
\newcommand{\polvec}[2]{\epsilon^\lambda_{#1}(\mathbf{#2})}
\newcommand{\projt}[1]{\mathbb{P}^{\rm T}_{#1}}
\def\exppar{y_\vartheta}
\def\extpart{\boldsymbol{\pi}}
\def\bathpart{\boldsymbol{\sigma}}
\def\twotwo{2\leftrightarrow 2}

\usepackage{caption}
\usepackage{subcaption}

\usepackage{slashed}

\usepackage{mwe}

\usepackage{array}

\newcolumntype{C}[1]{>{\centering\let\newline\\\arraybackslash\hspace{0pt}}m{#1}}

\newcommand{\mD}{m_\rmii{D}}

\newcommand{\qm}{q_-}
\newcommand{\qp}{q_+}
\newcommand{\qmp}{q_\mp}
\newcommand{\qpm}{q_\pm}

\def\lsi{\raise0.3ex\hbox{$<$\kern-0.75em\raise-1.1ex\hbox{$\sim$}}}
\def\gsi{\raise0.3ex\hbox{$>$\kern-0.75em\raise-1.1ex\hbox{$\sim$}}}

\newcommand{\rmii}[1]{{\mbox{\tiny\rm{#1}}}}

\newcommand{\Tint}[1]{{\hbox{$\sum$}\!\!\!\!\!\!\!\int\,}_{\!\!\!\!\raise-0.9ex\hbox{$\scriptstyle{#1}$}}}
\newcommand{\Tinti}[1]{{{\Sigma}\!\!\!\!\raise0.3ex\hbox{$\int$}_\rmii{${#1}$}}}

\newcommand{\bi}{\begin{itemize}}
\newcommand{\ei}{\end{itemize}}
\newcommand{\hide}[1]{ }

\newcommand{\opdim}{{d}} 

\def\twotwo{2\leftrightarrow 2}

\makeatletter
\renewcommand\section{\@startsection{section}{1}{\z@}%
  {-5.5ex \@plus -1ex \@minus -.2ex}
  {2.3ex \@plus.2ex}%
  {\normalfont\large\bfseries}}
\renewcommand\subsection{\@startsection{subsection}{2}{\z@}%
  {-3.25ex\@plus -1ex \@minus -.2ex}%
  {1.5ex \@plus .2ex}%
  {\normalfont\normalsize\bfseries}}
\renewcommand\thesection{\@arabic\c@section}
\renewcommand\thesubsection{\thesection.\@arabic\c@subsection}
\renewcommand{\@seccntformat}[1]{%
  \csname the#1\endcsname.\hspace{1.0em}}
\makeatother
\begin{document}

\flushbottom

\begin{titlepage}

\begin{flushright}
September~2026
\end{flushright}
\begin{centering}

\includegraphics[width=2.5cm]{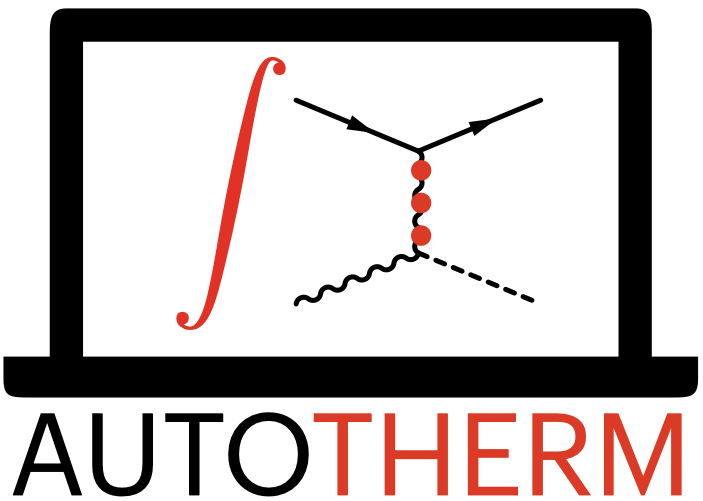}

\vfill

{\Large{\bf
 Automated Thermal Field Theory rates for cosmology
}}

\vspace{0.8cm}
Killian~Bouzoud$^{\rm a}$,
Jacopo~Ghiglieri$^{\rm a}$ 
and Greg~Jackson$^{\rm a}$ 

\vspace{0.8cm}

$^{\rm a}$%
{\em
SUBATECH, Nantes Universit\'e, IMT Atlantique, IN2P3/CNRS,\\
4 rue Alfred Kastler, La Chantrerie BP 20722, 44307 Nantes, France\\
}

\vspace*{0.8cm}

\mbox{\bf Abstract}

\end{centering}

\vspace*{0.3cm}

\noindent
\myaut{} is a modular code automating the determination of 
thermal production and interaction rates for cosmology from 
Thermal Field Theory. 
This first release computes fully automatically 
the leading-order contribution from ultrarelativistic 
$\twotwo$ processes to the production rate of 
an off-equilibrium $\vartheta$ particle coupled to a thermal bath. 
It takes as input a \myfr{} model file describing 
the Lagrangian of the bath-$\vartheta$ system. 
The \myaut{} program consists of a \textsc{Wolfram} package, 
which utilises \myfa{}/\myfc{}, together with a few Python modules 
for symbolic and numerical routines. 
Their combined usage leads from the model file to the rate, 
but each component can also be used in a standalone manner. 

The main strength of this release is the automated, 
model-independent handling of Hard Thermal Loop resummation, 
which is necessary whenever a $\twotwo$ process is mediated by a 
$t$-channel massless mediator. 
Developments in Thermal Field Theory reduce the apparently large 
model dependence to the determination of the thermal mass 
of the mediator, which we automate successfully. 
They further allow us to provide three 
leading-order-equivalent implementations of HTL resummation. 
Their spread provides a first estimate of the theory uncertainty 
from higher-order thermal corrections. 

We showcase the strength of our \myaut{} framework by 
successfully reproducing or correcting a host of results in 
the literature, such as the thermal production rate of 
ultrarelativistic right-handed neutrinos, of gravitons 
and gravitinos, of axions and of dark photons. 

\vfill


\vfill
\end{titlepage}

\tableofcontents
\clearpage

%
\section*{Program summary}
\noindent
{\em Program title}: \myaut
\\[1mm]%
{\em Version}: 1.0
\\[1mm]%
{\em Developer's repository link:} \href{https://autotherm.in2p3.fr}{https://autotherm.in2p3.fr} 
\\[1mm]%
{\em Licensing provisions:} GNU General Public License 3 (GPLv3)
\\[1mm]%
{\em Programming languages:} Python, Wolfram Language, Cython, C
\\[1mm]%
{\em Packages/external routines:} \myfr{}, \myfa{}, \myfc{}, NumPy, SciPy, SymPy, 
Polylogarithm, Cubature
\\[1mm]%
{\em Operating systems:} Linux, macOS
\\[1mm]%

\clearpage

%
\section{Introduction}
\label{sec_intro}

Our current understanding of the thermal history of the early universe, 
from reheating down to 
neutrino decoupling and later phenomena, features many open questions. 
For starters, 
the magnitude of the reheating temperature $T_\mathrm{RH}$ is very 
poorly constrained, from $\mathcal{O}(10^{15})$ GeV
down to a few MeV~\cite{Hannestad:2004px,Planck:2018jri,BICEP:2021xfz}. 
This in turn reverberates on an uncertainty
on the nature of the earliest particle degrees of freedom (d.o.f.s) 
in thermal equilibrium, ranging from 
speculative particle physics at very high energies at the high end 
to well-known hadrons, leptons and photons at the opposite.
Other currently unsettled issues include the nature and production 
channels of particle-physics candidates for 
Dark Matter (DM) and the mechanism addressing 
the Baryon Asymmetry of the Universe (BAU),
both of which require new physics beyond the 
Standard Model (SM) of particle physics.

However high is $T_\mathrm{RH}$ and 
whichever particle physics exists there, 
the presence of a thermal-equilibrium radiation epoch is undisputed:
most models address DM and/or the BAU by relying on it. 
For instance, the presence of extra thermal d.o.f.s
at some early enough stage and their later \emph{freeze out} is the
centerpiece of the Weakly Interacting Massive Particle
(WIMP) DM scenario \cite{Jungman:1995df}.
It is also  possible to construct economical
scenarios where the Beyond the Standard Model (BSM) d.o.f.s 
never thermalize in the history of the universe, 
giving rise to \emph{freeze-in} thermal 
production~\cite{Hall:2009bx,Bernal:2017kxu}. 
We refer to~\cite{Asadi:2022njl,Antel:2023hkf,Cirelli:2024ssz} 
for recent overviews.

Key ingredients in the theoretical description of these 
and many other examples are 
\emph{thermal production and interaction rates}, which encode how fast 
any degree of freedom can interact with others and/or itself so as to 
reach, maintain or leave thermal equilibrium. 
Precise knowledge of these rates is then 
of the utmost importance for furthering our understanding 
of the nature of the fundamental interactions in 
the  earliest epochs of our universe. 

This is our main motivation for the development and 
release of \myaut{}: 
this first public version provides a consistent, 
automated pipeline for the perturbative 
determination of these rates at leading-order (LO) 
within a modern framework 
based on the state of the art in Thermal Field Theory (TFT). 
In its current form, it can determine rates from $\twotwo$ processes 
where all particles are ultrarelativistic (UR) and their interactions 
are described by an arbitrary model of particle physics. 
Ideal applications are those models where production, 
equilibration or freeze out happen in the UR regime. Examples include 
dark radiation, some models of dark matter, 
Akhmedov--Rubakov--Smirnov (ARS) leptogenesis~\cite{Akhmedov:1998qx}.
We refer to \cite{Arza:2026rsl} for a recent white paper on 
light particles and their cosmological consequences.

As we shall discuss at length, \myaut{} is  particularly suited 
for models with $t$-channel massless mediators, 
see~\cite{Arina:2025zpi} for a review. 
These models would naively give rise to infrared (IR) divergent rates. 
\myaut{} can automatically handle 
the necessary Hard Thermal Loop (HTL) 
resummations~\cite{Braaten:1989mz,Frenkel:1989br,Taylor:1990ia,Braaten:1991gm}
that are needed to incorporate the emergence of 
\emph{collective medium effects} in the IR. 
This is carried out using a host 
of TFT developments that took place over the past two decades 
in   hot QCD and early-universe cosmology. 
As we shall show, the works 
of~\cite{Aurenche:2002pd,CaronHuot:2008ni,Besak:2012qm,Ghiglieri:2013gia}
led to the introduction of \emph{light-cone techniques} rooted 
in the causal properties of thermal amplitudes. 
These have been used to derive closed-form analytical expressions 
for the HTL-resummed $t$-channel-mediator contribution to the rates.

These light-cone techniques will then be the linchpin of 
our automation methodology. 
As we shall show, 
\myaut{} provides three LO-equivalent implementations of 
HTL resummation; 
their spread can then be used as a proxy for the theory uncertainty 
of the rate from unknown, potentially large, 
higher-order corrections from TFT, see 
e.g.~\cite{CaronHuot:2008ni,Bouzoud:2026rur}
for examples in hot QCD and cosmology respectively. 

We shall illustrate how this is achieved by
exploiting the leading-order factorization of TFT production rates 
and of the thermal masses needed for $t$-channel mediators into 
a convolution of
in-vacuum  matrix elements and thermal distributions.
By suitably building upon 
existing in-vacuum automation tools \myfr{}~\cite{Alloul:2013bka},
\myfa{}~\cite{Hahn:2000kx} and \myfc{}~\cite{Hahn:2016ebn}
we then automate the generation and evaluation of those in-vacuum
matrix elements; these are then handed over to dedicated 
Python, Cython and C routines for the numerical convolution. From a 
technical standpoint, the current release of \myaut{} has the 
form of a \mymath{} package and of Python modules. Their combined 
use allows the determination of the thermal rates from 
a \emph{model file} describing the Lagrangian of the model; 
they can also be used as standalone 
modules for other uses, such as determination of the thermal masses. 

As we shall show, with this first release we are able to rapidly
reproduce a host of results from the literature, such as 
the thermal production rate of 
hot axions~\cite{Graf:2010tv,Bouzoud:2024bom},
gravitons~\cite{Ghiglieri:2020mhm,Ringwald:2020ist}, 
dark photons~\cite{Salvio:2022hfa}, 
UR right-handed neutrinos~\cite{Besak:2012qm} and 
Majorana DM candidates~\cite{Biondini:2020ric}.
In all these cases we bundle detailed examples with our release, 
together with their model files, as well as that for the
SM in the electroweak symmetric phase and for the 
Minimal Supersymmetric Standard Model (MSSM) 
in its supersymmetric phase.

The latter is used in our companion paper~\cite{gravitinopaper},
dedicated to the reanalysis of the 
gravitino production rate. This is a particularly good example 
of the tedious, error-prone and technically intricate 
nature of the calculations that \myaut{} tackles, greatly reducing the 
opportunities for  computational or methodological
blunders. 

This paper is organized as follows. 
In Sec.~\ref{sec:theory} we lay out the 
main definitions for  production and interaction rates in TFT. 
The later Sec.~\ref{sec:naivelo} connects 
their perturbative LO evaluation with 
$\twotwo$ processes and shows the appearance of IR divergences 
with $t$-channel massless mediators. Sec.~\ref{sec:soft} describes how 
this divergence can be addressed by HTL resummation and how 
the latter can 
be automated  in a model-independent way. Secs.~\ref{sec:struct}
and \ref{sec:tuto} are dedicated to the description of 
the structure and usage of \myaut{}. 
Sec.~\ref{sec:validation} describes the results we can 
reproduce with our current release; future extensions, 
as well as a summary, 
are outlined in our concluding Sec.~\ref{sec:concl}. 
Further technical detail, as 
well as more detailed usage instructions for the code, 
are found in the appendices. 

%
\section{Particle production in a thermal bath}
\label{sec:theory}

Let us consider a thermalized plasma coupled with 
a particle species $\vartheta$. 
If the bath-$\vartheta$ interactions are slow compared with 
the characteristic interaction timescale in the thermal bath, 
$\vartheta$ is not in equilibrium. 
Let us define $d_\vartheta$ as the degeneracy of $\vartheta$, 
e.g. $d_\gamma=2$ for the two polarization states of the photon. 
We can then define the one-particle phase-space density 
\emph{for each degenerate degree of freedom} as
\begin{equation}
  f_\vartheta(t,{\bf k})
  \; \equiv \;
  \frac{1}{d_\vartheta} 
  \frac{
    (2\pi)^3 {\rm d} N_\vartheta
  }{ 
    {\rm d}^3\mathbf{k}\,{\rm d}^3\mathbf{x}
  }\,,
  \label{eq:def_f}
\end{equation}
where we  omit the ${\bf x}$ argument for $f_\vartheta$ 
assuming homogeneity. 
$N_\vartheta$ is the number of $\vartheta$ particles 
of all degeneracies. 
In cases where $\vartheta$ is not self-conjugate, 
the phase-space density $f_{\bar\vartheta}(t,{\bf k})$ 
of antiparticles is defined analogously. 
The $\vartheta$ number density reads
\begin{equation}
  n_\vartheta(t)
  \; \equiv \;
  \frac{ 
    {\rm d} N_\vartheta
  }{
    {\rm d}^3\mathbf{x}
  }
  \; = \;
  d_\vartheta \int\frac{ {\rm d}^3\mathbf{k} }{ (2\pi)^3 }\,
  f_\vartheta(t,{\bf k})\,,
  \label{eq:def_n}
\end{equation}
and similarly for $n_{\bar\vartheta}(t)$ in the 
non self-conjugate case. 
In what follows we will only
consider isotropic and homogeneous media, 
so that $f_\vartheta$ is a function of 
$t$ and $k \equiv \vert \mathbf{k}\vert$ only. 
We take the opportunity to introduce some of our conventions: 
we call $\eta_{\mu\nu}={\rm diag}(1,-1,-1,-1)$ 
the flat Minkowski metric. 
Four-vectors are denoted with calligraphic uppercase letters, i.e. 
$\mathcal{K}^\mu=(k^0,\mathbf{k})$.

Our aim is then to describe the time evolution of $f_\vartheta(t, k)$. 
To this end, let us write the Lagrangian of this system as follows
\begin{equation}
  \mathcal{L}_{\rm tot}
  \; = \; 
  \mathcal{L}_{\rm med}
  \; + \; \mathcal{L}_{\vartheta}
  \; + \; \mathcal{L}_{\rm int},
  \label{eq:Ltot}
\end{equation}
where $\mathcal{L}_{\rm med}$ describes the degrees of freedom 
of the thermal bath, 
$\mathcal{L}_{\vartheta}$ is the kinetic term for the particle 
$\vartheta$ and $\mathcal{L}_{\rm int}$ encodes 
the plasma-$\vartheta$ interactions.

For example, when studying photon production in heavy-ion collisions 
$\mathcal{L}_{\rm med} = \mathcal{L}_{\rm QCD}$ describes 
the quark-gluon plasma (QGP), 
$\vartheta$ is a photon with $\mathcal{L}_{\vartheta}$ 
the Maxwell term and $\mathcal{L}_{\rm int}$ 
the minimal coupling to (light) quarks.
In cosmology, we could have 
$\mathcal{L}_{\rm med} = \mathcal{L}_{\rm SM}$, 
describing a Standard Model (SM) thermal bath, 
with $\vartheta$ being a relic BSM particle, 
as well as cases where $\mathcal{L}_{\rm med}$ contains 
extra equilibrated BSM degrees of freedom. 

From an Effective Field Theory (EFT) perspective,
the interactions in $\mathcal{L}_{\rm int}$ can be written as
\begin{equation}
  \mathcal{L}_{\rm int}
  \; = \;
  \frac{1}{\Lambda^{\opdim-4}} \,
  \sum_i c_i^{(\opdim)}\hat{\mathcal{O}}_i^{(\opdim)},
  \label{eq:lint}
\end{equation}
where ${\hat {\cal O}}_i^{(\opdim)}$ are the allowed operators of 
mass dimension $\opdim$.
The scale $\Lambda$ assures that the Wilson coefficients 
$c_i^{(\opdim)}$ are dimensionless. 
For $\opdim \geq 5\,$, $\Lambda$ is the high-energy scale of the 
ultraviolet (UV) complete theory.

In Eq.~\eqref{eq:lint} we have assumed that 
$\mathcal{L}_\mathrm{int}$ is homogeneous in $\Lambda$ or 
equivalently considered a single $\opdim$. 
As we shall show, the current release of \myaut{} can work 
with any $\opdim\ge 4$; IR effects are correctly accounted for 
in the most common $\opdim=4$ and $\opdim=5$ cases; 

As we have hinted, our assumption of $\vartheta$ having 
slower interactions than 
the thermalized degrees of freedom is predicated on 
the size of its coupling with the latter. 
Let us introduce the appropriate dimensionless quantity
\begin{equation}
  \exppar
  \; \equiv \; 
  \sum_i c_i^{(\opdim)} \biggl(\frac{E_i}{\Lambda}\biggr)^{\opdim -4}
  \; \stackrel{E_i \sim T}{\sim} \;
  \sum_i c_i^{(\opdim)}\biggl(\frac{T}{\Lambda}\biggr)^{\opdim -4}
  \,,
  \label{eq:exppar}
\end{equation}
where $E_i\ll \Lambda$ is the typical energy scale of 
the considered process mediated 
by the operator $\hat{\mathcal{O}}_i^{(\opdim)}$. 
When considering thermal production in the ultrarelativistic 
regime, we will then have $E_i\sim T$. 
Hence, our requirement of slower interactions 
is necessarily a requirement on the smallness of $\exppar\,$; 
as the corresponding constraint is in general model dependent,
we do not comment on it further here. 

Let us consider operators that are \emph{linear} in $\vartheta$.
That is
\begin{equation}
  c_i^{(\opdim)} \hat{\mathcal{O}}_i^{(\opdim)}
  \; = \;
  \Biggl\{\begin{array}{l l}
    \hat{\vartheta}_{\alpha\cdots\beta}^{}
    \hat{J}_{\alpha\cdots\beta}^{i} & 
    \text{ if } \vartheta=\bar\vartheta \\[1mm]
    \hat{\bar\vartheta}_{\alpha\cdots\beta}^{}
    \hat{J}_{\alpha\cdots\beta}^{i} +\text{h.c.} &
    \text{ if } \vartheta\ne\bar\vartheta
  \end{array}
  \, ,
\end{equation}
where $\alpha\cdots\beta$ represent the Lorentz indices and/or 
any other indices (spin, color, etc.) of $\vartheta$. 
Addition of the Hermitean conjugate (h.c.) is needed only 
when $\vartheta$ is not self-conjugate, 
$\vartheta\ne\bar\vartheta$. 
The $\hat{J}$ operators are then entirely built out of bath fields 
in $\mathcal{L}_\mathrm{med}$. 
This allows us to rewrite $\mathcal{L}_{\rm int}$ as
\begin{equation}
  \mathcal{L}_{\rm int}
  \; \equiv \;
  \frac{1}{ \Lambda^{\opdim-4} }
  \times\Biggl\{\begin{array}{l l}
    \hat{\vartheta}_{\alpha\cdots\beta}
    \hat{J}_{\alpha\cdots\beta} & 
    \text{ if }\vartheta=\bar\vartheta  \\[1mm]
    \hat{\bar\vartheta}_{\alpha\cdots\beta}
    \hat{J}_{\alpha\cdots\beta} +\text{h.c.} & 
    \text{ if }\vartheta\ne\bar\vartheta  \\
  \end{array} \,.
  \label{eq:lintJ}
\end{equation}
Operators that are quadratic (or more) in $\vartheta$ are relevant 
for double (or triple, etc.) production of $\vartheta$. 
In its current form, \myaut{} can handle in 
a fully automated fashion
\emph{single production} of $\vartheta$ from interaction terms 
of the form of Eq.~\eqref{eq:lintJ}. 
This is the class of processes most sensitive to IR medium effects; 
IR-finite double production is partially supported in
a semi-automated fashion, as shown in 
our example of Sec.~\ref{sub:d6val} handling double 
dark-photon production as in Ref.~\cite{Salvio:2022hfa}. 

With  $\mathcal{L}_{\rm int}$ in the form of Eq.~\eqref{eq:lintJ}, 
the methods of~\cite{Bodeker:2015exa} 
--- see~\cite{Laine:2016hma} for a textbook derivation --- 
lead to an evolution equation for $f_\vartheta$. 
Let us introduce 
\begin{equation}
  \Pi_\lambda^{<}(\mathcal{K})
  \; \equiv \;
  \frac{\tau}{\Lambda^{2(\opdim-4)}}\,
  \bar v_{\alpha';\lambda}(\mathcal{K})
  v_{\alpha;\lambda}(\mathcal{K}) 
  \int_{\mathcal{X}} e^{ i\mathcal{K}\cdot\mathcal{X} }
  \Bigl\langle 
    \hat{J}_{\alpha'\cdots\beta}^{\dagger}(0)
    \hat{J}_{\alpha\cdots\beta}^{}(\mathcal{X})
  \Bigr\rangle
  \;, 
  \label{eq:defpi}
\end{equation}
where $\langle\ldots\rangle$ denotes a thermal expectation value. 
$\Pi_\lambda^{<}(\mathcal{K})$ is then
the lesser thermal Wightman function of the $\hat{J}$ current for 
an on-shell external $\vartheta$ state with 
$\mathcal{K}^\mu=(E_k,\mathbf{k})$, 
with $E_k=\sqrt{k^2+M_\vartheta^2}$. 
For a self-conjugate $\vartheta$ then $\hat{J}^\dagger=\hat{J}$. 
The statistics of $\vartheta$ is defined by $\tau$, 
i.e. $\tau=1$ for a boson, $\tau=-1$ for a fermion. 
All non-spin indices 
are contracted between $\hat{J}^\dagger(0)$ and $\hat{J}(\mathcal{X})$, 
while the  $v_{\alpha;\lambda}(\mathcal{K})$
and $\bar v_{\alpha';\lambda}(\mathcal{K})$
account for the possible spin of $\vartheta$: 
for a fermion they are the spinors $u_\lambda(\mathcal{K})$ and 
$\bar u_\lambda(\mathcal{K})$, 
where $\lambda$ labels one of the two spin states 
and the index $\alpha$ is then a Dirac index. 
Similarly, for vector bosons $v$  and $\bar v$ will be 
a standard polarization vector and its complex conjugate 
and for a spin-two massless field (the graviton) 
a polarization tensor  and its complex conjugate; 
see App.~\ref{sub:graviton} for our handling of their properties. 

Equation~\eqref{eq:defpi} can then be understood as 
singling out a single spin/polarization 
state while tracing over any other possible degeneracy; 
in addition, the introduction of the $v$ structures gives 
$\Pi_\lambda$ dimensions of mass squared in 
all cases, given that the dimensionality of the 
$u_\lambda(\mathcal{K})$ ($[u_\lambda(\mathcal{K})]=1/2$) 
compensates the smaller dimensionality of $\hat{J}$ when 
$\vartheta$ is a spin-1/2 fermion.

One then has~\cite{Bodeker:2015exa,Laine:2016hma}
\begin{equation}
  \label{eq:boltzmann}
  \frac{\mathrm{d}}{\mathrm{d}t}{f}_\vartheta 
  \; = \; 
  \Gamma_\vartheta
  \bigl[ 
    f_\mathrm{eq} - f_\vartheta 
  \bigr]
  \, + \,
  \mathcal{O}\bigl(\exppar^4\bigr)
  \;.
\end{equation}
Here $f_\mathrm{eq}=(\exp(E_k/T)- \tau)^{-1}$ is 
the equilibrium distribution for $\vartheta\,$. 
In an expanding FLRW universe 
$\frac{\mathrm{d}}{\mathrm{d}t}=\partial_t-H k\partial_k$, 
with $H$ the Hubble rate. 
We have typographically suppressed the 
$k$ dependence of $f_\vartheta$ and $\Gamma_\vartheta$ as well. 
The latter is the production or equilibration rate; it is given as 
\begin{equation}
  \Gamma_\vartheta
  \; = \;
  \frac{
    1
  }{2 d_\vartheta E_k } \, \sum_\lambda
  \frac{\Pi^{<}_\lambda(\mathcal{K})}{\tau\, f_\mathrm{eq}}
  \;, \label{rategen}
\end{equation}
where we are now tracing over spin, when present. 
In most applications of our code spin is actually degenerate;
although computing $\Pi^{<}_\lambda$ for a single value 
of $\lambda$ would suffice, it is in general 
more practical to trace, 
as it is something that the zero-temperature automation tools 
we rely on readily implement.

Antiparticles, when distinct from particles, evolve according to
\begin{equation}
  \label{eq:boltzmannbar}
  \frac{\mathrm{d}}{\mathrm{d}t} {f}_{\bar\vartheta}
  \; = \;
  \Gamma_{\bar\vartheta}
  \bigl[
    f_\mathrm{eq}-f_{\bar\vartheta}
  \bigr]
  \, + \, \mathcal{O}\bigl(\exppar^4\bigr).
\end{equation}
As we consider a CP-symmetric plasma without chemical potentials 
in this first release, one can then show~\cite{Laine:2016hma} 
that 
$\Gamma_{\bar\vartheta}\stackrel{\mathrm{CP}}{=}\Gamma_{\vartheta}$. 

Hence, Eqs.~\eqref{eq:boltzmann} and \eqref{eq:boltzmannbar} state 
that $\Gamma_\vartheta$
describes the time evolution of $\vartheta$ at first order in 
the $\exppar$ coupling between $\vartheta$ and the medium; 
the expression in Eq.~\eqref{rategen} is, on the other hand, 
valid to all orders in the couplings in 
$\mathcal{L}_\mathrm{med}$ between 
the equilibrated particles~\cite{Bodeker:2015exa}. 
As we shall show, a perturbative expansion in those couplings 
will reduce, in the appropriate 
setting, to a well-known Boltzmann picture. 
In other words, our 
definitions in Eqs.~\eqref{eq:boltzmann}, \eqref{eq:boltzmannbar} 
and \eqref{rategen} require 
the existence of particle states for $\vartheta$ and 
thus of a meaningful single-particle density $f_\vartheta$; 
no such requirement is necessary for the equilibrated d.o.f.s 
in $\hat{J}$, which may very well be described 
by strongly-coupled fields far away from any (quasi)-particle 
picture. 

One possible such example is provided by the QGP, where the 
gauge coupling $g_3\sim 2$ is not small, 
making a weak-coupling expansion around quark and gluon 
particle states potentially unreliable close to the QCD crossover. 
A non-perturbative lattice-based approach is in principle possible. 
However, only the Euclidean $\Pi_E$ correlator can 
be determined and the analytical continuation to 
Minkowskian-time $\Pi^<$ represents a great source of uncertainty, 
see e.g.~\cite{Meyer:2011gj}. 
In the SM, lattice approaches are further hampered 
by its chiral interactions. 
Calculations in $\mathcal{N}=4$ supersymmetric Yang-Mills 
through the AdS/CFT 
correspondence~\cite{Maldacena:1997re,Witten:1998qj,Gubser:1998bc} 
--- see e.g.~\cite{Castells-Tiestos:2022qgu} 
for gravitational-wave production --- 
can give qualitative strong-coupling insight.

Furthermore, it has been shown in Ref.~\cite{Bodeker:2015exa} 
that $\Gamma_\vartheta$, as determined from 
Thermal Field Theory in Eq.~\eqref{rategen}, 
is valid both as a \emph{production rate}, 
i.e. when $f_\mathrm{eq}\gg f_{\vartheta}$ in 
Eq.~\eqref{eq:boltzmann}, 
so that $f_{\vartheta}$ can be neglected on its right-hand side, 
and as an \emph{equilibration rate}, 
applicable when $f_\mathrm{eq}\sim f_{\vartheta}$. 
In the context of cosmological applications, 
the former case corresponds to freeze-in production, 
while the latter may apply to a freeze out. 
In the $f_\mathrm{eq}\gg f_{\vartheta}$ literature, 
the production rate is oftentimes defined as 
$
  \Gamma_{\vartheta}^\text{prod} 
  \; \equiv \; 
  \Gamma_{\vartheta} \, f_\mathrm{eq}
$. 
Here we do not make this distinction and call 
$\Gamma_{\vartheta}$ interchangeably a production, 
interaction or equilibration rate.

%
\section{Perturbative determination of the \texorpdfstring{$\twotwo$}{2<->2} rate 
at leading order}
\label{sec:naivelo}

In this section we shall show how the 
perturbative evaluation of Eq.~\eqref{rategen} is related to 
a naive convolution of tree-level matrix elements squared and 
distribution functions. We shall show when this naive approach 
fails, requiring more sophisticated resummations rooted 
in Thermal Field Theory.
We shall also discuss the theoretical principles underlying
the automated generation, evaluation and decomposition of 
these matrix elements squared. In Sec.~\ref{sub:naivedef}
we examine the perturbative expansion of Eq.~\eqref{rategen},
to then concentrate in Sec.~\ref{sub:naive} on the $\twotwo$
contribution. Finally, in Sec.~\ref{sub:naivediv} we shall show
how the naive convolution results in IR divergences when
considering massless $t$-channel mediators.

Before we delve into the details, we remark that this first 
release implements an ultrarelativistic $\vartheta$ interacting 
with a bath of equally ultrarelativistic particles. 
In what follows, we thus take $\mathcal{K}^\mu=(k,\mathbf{k})$ 
and $E_k=k$ in Eq.~\eqref{rategen}. 
We shall furthermore always consider $k\gtrsim T$ for 
the momentum of $\vartheta$; this is the momentum range that dominates 
equilibrium thermodynamical functions 
such as number and energy densities. 
The rates we shall obtain are not consistent for $k\ll T$, 
where calculations become significantly more intricate 
and model-dependent: see for instance Ref.~\cite{Bouzoud:2026rur}
for a detailed analysis of $k\lesssim g_3 T$ axion production 
using perturbative and non-perturbative methods.

%
\subsection{Definition of leading order}
\label{sub:naivedef}

%
\begin{figure}[t]

\begin{center}
    \includegraphics[width=0.7\linewidth]{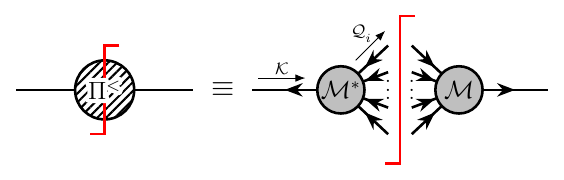}
\end{center}

\vspace{-5mm}

\caption[a]{\small
  Graphical representation of cutting rules for $\Pi^<_\lambda$. 
  The gray blobs on the right represent the fully retarded matrix 
  elements. Cut lines are replaced by the Wightman ${}^<$ propagators. 
  Arrows denote the flow of causality, 
  as in~\cite{Caron-Huot:2007zhp,Caron-Huot:2007cma} 
  and in Fig.~\ref{fig:1to1causal}.
}
\label{fig:cutsgeneric}
\end{figure}
%

A practical way to compute perturbatively the Wightman function 
$\Pi^{<}_\lambda$ and by extension the $\vartheta$ production 
rate $\Gamma_\vartheta$ is
through finite-temperature cutting rules~\cite{Weldon:1983jn} 
in the  formulation of~\cite{Caron-Huot:2007zhp,Caron-Huot:2007cma} 
--- see the review in \cite{Ghiglieri:2020dpq} --- 
yielding directly 
the desired Wightman functions in terms 
of fully retarded amplitudes and Wightman (cut) $<$ propagators.
An example is shown in Fig.~\ref{fig:cutsgeneric}. 
By cutting diagrams for the two-point function $\Pi^{<}_\lambda$, 
each cut line is put on shell and multiplied by 
the equilibrium statistical distribution for
the corresponding particle. 
$\Gamma_\vartheta$ can then be expressed as a sum of 
the phase-space convolution of fully retarded matrix elements squared
for on-shell $m\to n$ processes with 
the appropriate equilibrium statistical functions for the 
on-shell external particles. 

%
\begin{figure}[t]

\begin{center}
    \includegraphics[width=0.7\linewidth]{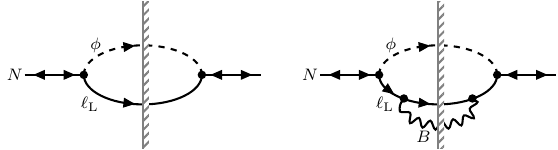}
\end{center}

\vspace{-3mm}

\caption[a]{\small
  Left: a one-loop diagram for $\Pi^<$ in the $\opdim=4$ 
  right-handed neutrino model described in Sec.~\ref{sub:d4val}. 
  The symbol $N$ denotes this right-handed Majorana state, 
  $\phi$ and $\ell_\mathrm{L}$ the SM Higgs and lepton doublets. 
  The gray vertical line represents the only possible cut. 
  Right: a two-loop diagram in the same model, where $B$ is 
  the SM U(1) gauge boson. 
  If the one cut of the one-loop diagram vanishes kinematically, 
  then so do the two other cuts of this two-loop diagram 
  that do not go through the $B$ line.
}
\label{fig:cuts}
\end{figure}
%

What are the most relevant processes for $\vartheta$ production? 
Let us assume that the $\hat{J}$
operators in Eq.~\eqref{eq:defpi} contain quadratic terms in 
the bath fields, e.g. $\hat{J}=h\psi\varphi$, 
with $h$ a dimensionless coupling, $\psi$  and $\varphi$
an equilibrated fermion and scalar respectively. 
Tree-level $1\leftrightarrow 2$ processes, 
also known as \emph{Born processes}, would then naively  be present. 
These correspond to cuts of the one-loop $\Pi^{<}_\lambda$; 
an example is shown on the left in Fig.~\ref{fig:cuts}. 
They are however kinematically forbidden when  all particles 
are exactly massless, 
moving the first order for production processes 
to $\twotwo$.\footnote{\label{foot_j}%
  This is also trivially true in models where the $\hat{J}$
  operators do not contain any quadratic term in the bath fields. In 
  those models no non-trivial thermal effects such as the emergence 
  of collectivity is expected at leading order; 
  provided they have a $\twotwo$ tree-level channel for 
  $\vartheta$ production, they are then easily handled by \myaut{}. 
} 
These correspond to cuts 
of two-loop diagrams for $\Pi^{<}_\lambda$. An example is shown 
in Fig.~\ref{fig:cuts} on the right; it corresponds to the 
square of the tree-level $\twotwo$ diagram  shown 
on the left in Fig.~\ref{fig:d4diags}. 
The $1\leftrightarrow 3$ crossings of these processes 
are vanishing at vanishing masses. 

These tree-level $\twotwo$ processes scale necessarily like
$g^2 \exppar^2$, where $g$ denotes collectively the couplings between
the equilibrated particles. 
In most use cases this can be the parametrically 
largest subset thereof; for instance, when considering the 
SM in the symmetric phase, 
it is common to consider $g^2=(g_1^2,g_2^2,g_3^2,\vert h_t\vert^2)$, 
with $g_i$ the three 
gauge couplings and $h_t$ the top Yukawa,\footnote{%
  Normally the  Higgs quartic coupling $\lambda$ is also taken 
  to be of parametric size $g^2$, 
  e.g. in determining the thermal mass of the Higgs doublet 
  from its effective potential. 
  In the specific case of $\twotwo$ processes 
  with one external $\vartheta$, 
  quartic couplings between four other particles 
  play no role at leading order.
}
even though all other Yukawa couplings 
mediate equilibrated interactions in most of 
the symmetric phase~\cite{Bodeker:2019ajh}. 
As we shall discuss in App.~\ref{app_sm},  
$g_i$, $h_t$ and $\lambda$ are the only couplings included 
in our implementation of the symmetric-phase SM. 

In a medium, all equilibrated particles develop 
a \emph{thermal mass} of order $gT$. 
What is the impact of these masses, 
as well of potential  mass terms in $\mathcal{L}_\mathrm{med}$ 
of comparable size? 
In the case of $\twotwo$ processes, 
the $k\gtrsim T$ rate is dominated by 
external states with energies and momenta of 
the order of the temperature, so that this correction 
to the dispersion relation can be safely neglected. 
This is far from true for 
intermediate states; 
Sec.~\ref{sec:soft} will be dedicated precisely to the proper 
handling of thermal effects therein.

In the case of $1\leftrightarrow 2$ processes, nonzero masses 
can in principle open up a Born channel, 
e.g. a $1\to 2$ process with $\vartheta$ 
in the final state becomes allowed if the initial-state particle 
has a nonzero but small mass $m_1$. 
The matrix element squared will scale like 
$\exppar^2 m_1^{2(1+\opdim-4)}/T^{2(\opdim-4)}$ 
on dimensional and Lorentz-invariance grounds.
The potential factors of $T$ in the denominator 
(i.e. when $\opdim > 4$) 
stem from the regime whereby $E\sim m_1$ 
in the definition~\eqref{eq:exppar}.
The most pertinent case is $\opdim=4$, where 
this expression then scales like 
$\exppar^2 m_1^2\sim \exppar^2 (gT)^2$. 
In this case the thermal convolution of this
Born process is of the same size of that of the 
$2\leftrightarrow 2$ ones, while it remains 
perturbatively smaller if $\opdim>4$.

It is important to note here that  collinear enhancements 
may cause all $1+n\leftrightarrow 2+n$ 
processes, with $n\ge 1$, to exist even when 
the $n=0$ Born case is disallowed and to 
contribute at the same order as $\twotwo$ processes. 
This was first noticed in \cite{Aurenche:1998nw} 
in the context of photon production in the QGP. 
There it was shown that $2\leftrightarrow 3$ processes contribute 
at the same order as $\twotwo$ processes.
Those $1+n\leftrightarrow 2+n$ processes can then be resummed into an 
``effective'' $1\leftrightarrow 2$ process through a procedure called 
Landau--Pomeranchuk--Migdal (LPM) 
resummation~\cite{Landau:1953gr,Landau:1953um,Migdal:1956tc}. 
We refer to~\cite{Anisimov:2010gy} for its introduction 
in cosmological settings in the context of right-handed 
neutrino production. See also~\cite{Laine:2022pgk} 
for a review of rate determinations for these particles. 

This  collinear enhancement is a  feature of $\opdim = 4$ models only. 
The collinear nature of these processes implies small-angle 
$1\leftrightarrow 2$ splitting, 
which is suppressed in higher-dimensional $\opdim>4$ couplings. 
See Refs.~\cite{Salvio:2013iaa,Ghiglieri:2020mhm} for a more detailed
discussion of suppressed small-angle emission in 
dimension-five couplings of gravitons, 
gravitinos and axions to fermions, scalars and gauge bosons. 

In this first version of \myaut{}, 
we focus on $\twotwo$ processes that account for the entire 
leading order contribution to the $\vartheta$ production rate 
for irrelevant operators and part of it for marginal ones; 
we defer the implementation of LPM resummation to a future release.

Anticipating that \myaut{} will be used primarily in cosmology,
we shall from now on assume that $\mathcal{L}_\mathrm{med}$ 
describing the equilibrated 
degrees of freedom in Eq.~\eqref{eq:Ltot} is:
\begin{itemize}
    \item 
    Renormalizeable, i.e. it only contains relevant or 
    marginal operators. We remind that this need not apply to 
    $\mathcal{L}_\mathrm{int}$.
    \item
    With marginal couplings only, 
    as is the case for the SM and for many extensions thereof. 
    The addition of relevant cubic couplings in addition 
    to marginal ones is at the moment unsupported by \myaut{}. 
\end{itemize}

%
\subsection{The \texorpdfstring{$\twotwo$}{2<->2} rate at naive leading order}
\label{sub:naive}

%
\begin{figure}[t]

\begin{center}
    \includegraphics[width=0.7\linewidth]{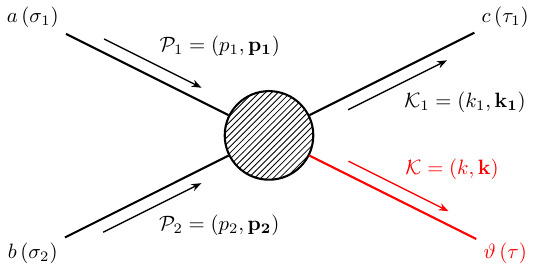}
\end{center}

\vspace{-3mm}

\caption[a]{\small
  Our notation  for the $\twotwo$ interaction. 
  See the main text for the definition of 
  $\sigma_1,\sigma_2,\tau_1,\tau$.
}
\label{fig:kinematics}
\end{figure}
%

Let us start by defining the kinematics of $\twotwo$ processes of 
the form $ab\leftrightarrow c\vartheta\,$.
The four-momenta of $a,b,c,\vartheta$ are respectively 
$\mathcal{P}_1,\mathcal{P}_2,\mathcal{K}_1,\mathcal{K}$. 
As we argued, we can take 
$\mathcal{P}_1^2=\mathcal{P}_2^2=\mathcal{K}_1^2=\mathcal{K}^2=0$ 
even when 
external states would receive slight dressed masses 
(smaller than the temperature). 
We shall work in the rest frame of the medium. 
See Fig.~\ref{fig:kinematics} for a pictorial summary 
of the conventions. 

The quantum statistics of the equilibrated bath particles $a,b,c$ 
are encoded by the indices $\sigma_1,\sigma_2,\tau_1$ respectively. 
If we define the distribution function $f_\sigma(E)$ as
\begin{equation}
  f_\sigma(E)
  \; \equiv \;
  \frac{1}{e^{E/T}-\sigma} \, ,
  \qquad 
  n_\sigma(E)
  \; \equiv \; 
  \sigma f_\sigma(E)
  \; = \;
  \frac{\sigma}{e^{E/T}-\sigma}, 
\end{equation}
then bosons have $\sigma=+1$ and fermions have $\sigma=-1$ 
(i.e. they represent the eigenvalue under particle exchange). 
For future convenience we also introduced $n_\sigma(E)$. 

As we mentioned, the $\twotwo$ 
production rate at naive leading order emerges from cutting rules as 
a well-known convolution of statistical functions and on-shell 
\emph{fully retarded} matrix elements squared. 
The latter agree --- at tree level --- with the standard 
\emph{time-ordered} matrix elements squared obtained from 
in-vacuum perturbation theory. 
This is a central tenet for our adoption of techniques and tools 
developed in that setting. We thus have 
\begin{equation}
  \Gamma_\vartheta^\text{naive}
  \; \equiv \;
  \frac{1}{2}\frac{1}{2k}\frac{1}{d_\vartheta}
  \int{\rm d}\Omega_{2\to2}
  \sum_{a,b,c}\tau_1\,\mathcal{N}_{\tau_1;\sigma_1,\sigma_2}
  \left\vert
    \mathcal{M}_{ab\to c\vartheta}
  \right\vert^2\,,
  \label{eq:prodrate}
\end{equation}
where the phase-space element ${\rm d}\Omega_{2\to2}$ is given 
in Eq.~\eqref{eq:phaselem} and we defined\footnote{%
  \label{foot:boltz}
  Readers familiar with the standard Boltzmann equation would expect
  \begin{align}
    \frac{\mathrm{d}}{\mathrm{d}t}{f}_\vartheta=
    \frac{1}{4kd_\vartheta}\int{\rm d}\Omega_{2\to2}
    \sum_{a,b,c}
    \left\vert\mathcal{M}_{ab\to c\vartheta}\right\vert^2\,
    \Bigl[&f_{\sigma_1}(p_1)f_{\sigma_2}(p_2){\bar n}_{\tau_1}(k_1)(1+\tau f_\vartheta(t,k))\nonumber\\
   & -f_{\tau_1}(k_1)f_{\vartheta}(t,k){\bar n}_{\sigma_1}(p_1){\bar n}_{\sigma_2}(p_2)\Bigr],
    \label{eq:prodratetrad}
  \end{align}
  for an off-equilibrium $f_\vartheta$ and equilibrated 
  distributions for the other external states
  (where $\tau \equiv \tau_1 \sigma_1 \sigma_2$). 
  Energy conservation and detailed balance, 
  reflected by the explicit form of the Bose--Einstein and 
  Fermi--Dirac functions, ensure that 
  plugging Eq.~\eqref{eq:prodrate} in Eq.~\eqref{eq:boltzmann} 
  reproduces Eq.~\eqref{eq:prodratetrad}. 
  This is an example of how the approach in Eq.~\eqref{eq:prodrate}
  reduces to the Boltzmann picture when 
  the same underlying assumptions are taken. 
}
\begin{equation}
  \mathcal{N}^{ }_{\tau_1;\sigma_1,\sigma_2}
  \; \equiv \;
  \frac{
    n_{\sigma_1}(p_1)n_{\sigma_2}(p_2){\bar n}_{\tau_1}(k_1)
  }{
    n_{\sigma_1\sigma_2\tau_1} (k)
  }\,,
  \quad \text{with} \quad 
  {\bar n}_\sigma(p)\equiv 1+n_\sigma(p).
\end{equation}
${\bar n}_{\tau_1}(k_1)$ encodes the final-state effect of 
quantum statistics; 
when $c$ is a boson we have $\bar{n}= 1+f_{+}$ (Bose enhancement) 
and when $c$ is a fermion, we have 
$\bar{n}= 1-f_{-}$ (Pauli blocking). 
Note that angular-momentum conservation 
requires an even number of fermions in this $\twotwo$ process. 
This in turn 
implies that the statistics $\tau$ of $\vartheta$ obeys 
$\tau=\sigma_1\sigma_2\tau_1\,$.

To briefly comment on the factors in Eq.~\eqref{eq:prodrate}:  
$1/2$  is the symmetry factor for identical 
initial-state particles. 
For different initial-state particles it is compensated by 
the double counting in the sum over $a$ and $b$. 
$1/2k$ comes from the Lorentz-invariant phase space measure. 
Finally, $\tau_1$ compensates the overall sign coming from 
$\mathcal{N}^{ }_{\tau_1;\sigma_1,\sigma_2}$; 
this is because $n_{-}=-f_{-}$. 

%
\begin{figure}[t]

\begin{center}
    \includegraphics[width=0.7\linewidth]{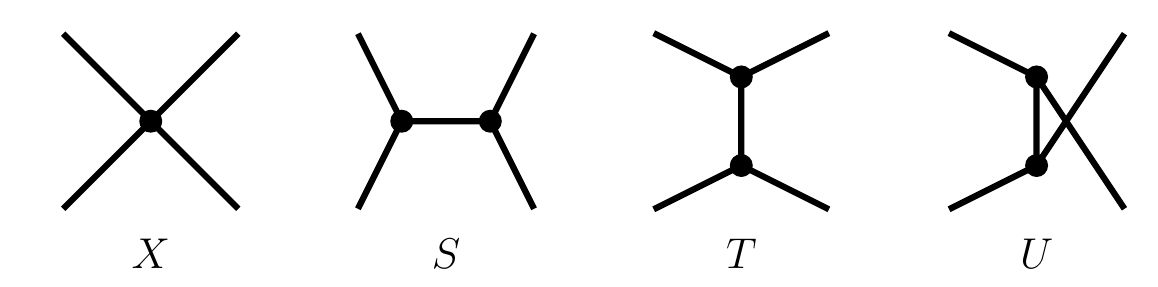}
\end{center}

\vspace{-4mm}

\caption[a]{\small
  All possible tree-level $2\leftrightarrow 2$ topologies. 
  Solid lines represent any propagator and black dots 
  represent vertices. 
}
\label{fig:top2to2}
\end{figure}
%

The  matrix elements squared 
$\left\vert\mathcal{M}_{ab\to c\vartheta}\right\vert^2$ 
in Eq.~\eqref{eq:prodrate} are summed over the 
degeneracies of \emph{both} initial and final states, 
in keeping with the conventions in 
Eqs.~\eqref{eq:def_f} and \eqref{rategen}. 
\myaut{} currently deals with tree-level matrix elements only, 
corresponding to the topologies in Fig.~\ref{fig:top2to2}.
Loop processes dominated by a loop scale $\Lambda\gg T$
can be recast by the user in effective local operators with 
appropriate matching coefficients in $\mathcal{L}_\mathrm{int}$.

The  $\left\vert\mathcal{M}_{ab\to c\vartheta}\right\vert^2$ 
are functions of the usual Mandelstam invariants
\begin{align}
  s=(\mathcal{P}_1+\mathcal{P}_2)^2,\quad
  t=(\mathcal{P}_1-\mathcal{K}_1)^2,\quad
  u=(\mathcal{P}_2-\mathcal{K}_1)^2,
\end{align}
and in the UR limit  $s+t+u=0$ holds. 
Matrix elements squared or terms therein 
with positive powers of $u$ in their denominators 
can be reshuffled into $t$ denominators 
by exploiting the fact that upon exchanging 
$\sigma_1$ and $\sigma_2$ in Eq.~\eqref{eq:prodrate}, 
we have $t\leftrightarrow u$. 
This is carried out by \myaut{} after having computed the 
$\left\vert\mathcal{M}_{ab\to c\vartheta}\right\vert^2$ 
through the routines described in 
Secs.~\ref{sec:struct}--\ref{sec:tuto}.\footnote{%
  \label{foot_mess}
  UR matrix elements squared can be written in different, 
  equivalent ways by exploiting 
  $s+t+u=0$, so that denominators in the form of products of 
  positive powers of two Mandelstam invariants, e.g. $u^n t^m$, 
  might appear when naively parsing terms in a matrix element squared. 
  However, for $\opdim\ge 4$ the numerator must necessarily be a 
  positive power of the remaining Mandelstam variable, 
  in this case $s^{n+m+\opdim-4}$. By rewriting  
  the numerator as $s^{\opdim-4}(-t-u)^{n+m}$ this 
  apparently problematic term can then 
  be recast in the form of $t^a$ and $u^b$ denominators.
}

After this step, for a given dimensionality $\opdim$ of 
the interaction operator and under our assumptions of 
a set of marginal couplings between bath particles and of massless 
mediators, the matrix elements squared 
with massless external states will then be a polynomial 
in $s^mt^n$, with $m+n=\opdim-4$. 
For each term  in these polynomials it is possible, as shown in 
\cite{Besak:2012qm,Ghiglieri:2016xye,Ghiglieri:2020mhm} 
and summarized in  App.~\ref{sec:phaseint}, 
to perform all but two of the integrations 
in Eq.~\eqref{eq:prodrate} analytically. 
\myaut{} leverages the fact that the number of possible terms 
in these polynomials is small for $\opdim=4$ and $\opdim=5$, 
so that a complete basis can be constructed in either case. 
The matrix element squared are then projected onto 
the corresponding bases by \myaut{}.

In more detail, for a $\opdim=4$ interaction, 
we follow the work of \cite{Besak:2012qm} and use
\begin{equation}
  \left\vert
    \mathcal{M}^{(4)}_{ab\to c\vartheta}
  \right\vert^2
  \; \equiv \;  
  \underbrace{a_1\frac{s}{t}+a_2}_{t\;{\rm param.}}
  \, + \, 
  \underbrace{b_1\frac{u}{s}}_{s\;{\rm param.}}.
  \label{eq:basis4}
\end{equation}
``$t$ param.'' and ``$s$ param.'' stand for 
two different parametrizations of the integration 
variables in $\int{\rm d}\Omega_{2\to2}$ 
that are better suited for integrands where respectively 
$t$ or $s$ sit at the denominator. 
They are described in  App.~\ref{sec:phaseint}.

For a $\opdim=5$ interaction, we follow the work of 
Ref.~\cite{Ghiglieri:2020mhm} and use 
\begin{equation}
  \left\vert\mathcal{M}^{(5)}_{ab\to c\vartheta}\right\vert^2
  \; \equiv \;
  \underbrace{a_1\frac{s^2+u^2}{t}+a_2\, t}_{t\;{\rm param.}}
  \, +\, 
  \underbrace{b_1\frac{t^2}{s}+b_2\, s}_{s\;{\rm param.}}.
  \label{eq:basis5}
\end{equation}
For interactions with $\opdim >5$, or should \myaut{} fail 
to properly decompose $\left\vert\mathcal{M}\right\vert^2$, 
a \textit{fallback} integration routine in 
App.~\ref{sec_fallback} will be used.

%
\begin{figure}[t]

\begin{center}
    \includegraphics[width=0.65\linewidth,trim=0 0 0 3.1cm, clip]{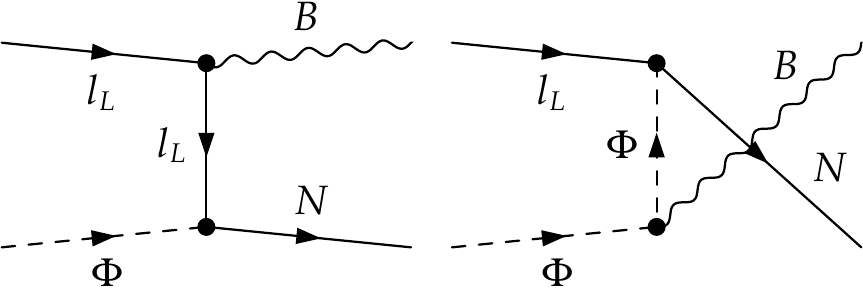}
\end{center}

\vspace{-4mm}

\caption[a]{\small
  Tree-level diagrams for an example $\ell_\mathrm{L}\phi\to B N$ 
  process for the $\opdim=4$ right-handed neutrino 
  model described in Sec.~\ref{sub:d4val} and already used for 
  Fig.~\ref{fig:cuts}. 
  Unlike that figure, the present one has been generated 
  by \myaut{} through  \myfa{}: 
  this explains the slight differences in notation and 
  graphical conventions.
}
\label{fig:d4diags}
\end{figure}
%

The attentive reader might feel our bases to be incomplete: 
simple tree-level processes mediated
by gauge bosons naturally feature \emph{squares}
of Mandelstam variables at the denominator, e.g. the 
$(s^2+u^2)/t^2+(s^2+t^2)/u^2$ structure in M{\o}ller 
scattering or its $u\leftrightarrow s$ Bhabha crossing. 
Any process of this
kind would however either require 
to have an incoming $\vartheta$ 
along with an outgoing one 
({\em elastic scattering}, for $t^2$ or $u^2$ 
denominators), or, 
to produce $\vartheta\bar\vartheta$ in the final state
({\em pair production}, for an $s^2$ denominator). 
These cases are outside the scope of 
our code~\footnote{\label{foot_broken}%
  Spontaneously broken gauge theories
  allow $\vartheta$ to be a single component of a gauge multiplet, 
  e.g. an active neutrino in the SM. 
  In that case a $1/t^2$-proportional 
  $ab\leftrightarrow c \vartheta$ process would be present. 
  In the SM at temperatures such that $T\sim v(T)$, with $v(T)$ 
  the Higgs expectation value,
  the gauge-boson mass $\propto g_i v$ originating from 
  symmetry breaking is not sufficient for a proper LO description 
  of the would-be  $1/t^2$ power divergence. 
  One would indeed need to resum the  $\propto g_i^2 v^2$ mass  squared
  together with gauge-boson Hard Thermal Loop self-energy 
  $\propto g_i^2 T^2$, 
  see~\cite{Ghiglieri:2016xye,Jackson:2019tnr} for 
  an explicit derivation and calculation in the SM and 
  \cite{vandeVis:2025plm} for a more generic discussion 
  of this type of processes. 
  As \myaut{} treats gauge bosons as massless before HTL resummation, 
  these processes are currently unsupported.
} 
Fermionic mediators, as in tree-level Compton scattering,
do give rise to reciprocals of $s$, $t$ and $u$ and are compatible 
with a single external $\vartheta$. Conversely, 
scalar-mediated processes, though compatible with a single external 
$\vartheta$ as well, cannot give rise to 
inverse powers of the Mandelstam variables. A concrete example of 
fermion and scalar mediators is provided in Fig.~\ref{fig:d4diags}
for a model that shall be described in 
more detail in Sec.~\ref{sub:d4val}.

In the case of $\opdim>4$ couplings, 
squares of the Mandelstam invariants are no longer expected 
at the denominators. 
For $\opdim=4$ we have for instance a $(s^2+u^2)/t^2$ M{\o}ller-like 
gauge-mediated 
$f_1 f_2\leftrightarrow f_1 f_2$
scattering between two different fermion species, 
which comes from the $T$ topology in Fig.~\ref{fig:top2to2}. 
If now
one of the two $f_i f_i V$ gauge vertices in that topology 
--- $V$ being the intermediate gauge boson --- 
is replaced by an $f_i \vartheta V$ vertex coming from a 
dimension-five operator, it will necessarily add one extra power 
of $t$ at the numerator of the matrix element squared. 
The extra power of a Mandelstam variable is expected 
on dimensional and Lorentz-invariance
grounds, as argued before. The fact 
that it must be $t$ follows from the fact that, 
in the notation of Fig.~\ref{fig:kinematics}, 
the $f_i \vartheta V$ vertex will have extra dependence 
on $\mathcal{P}_2^{\mu_1}$ and $\mathcal{K}^{\mu_2}$ with respect to 
the $f_i f_i V$ vertex. These two 
vectors will be contracted in an operator- and model-dependent 
Lorentz structure; 
however, once the diagram is squared and spins/polarizations 
traced over,
Lorentz invariance dictates that the only available structures are 
the scalar products $\mathcal{P}_2^2=0$, 
$\mathcal{K}^2=0$ and $\mathcal{P}_2\cdot{K}=-t/2$. 
The argument holds also when multiple diagrams contribute to 
the process, as in general only a single 
topology is responsible for the $1/t^2$, $1/u^2$ or $1/s^2$ behavior.

For what concerns $\opdim=3$ couplings in $\mathcal{L}_\mathrm{int}$, 
they could only arise 
in the form of cubic scalar couplings, e.g. of the form 
$\lambda_3\phi^2\vartheta$, with $\phi$
an equilibrated real scalar and $\lambda_3$ 
a coupling of mass dimension one. 
In this case  $\lambda_3^2g^2/t$ matrix elements squared 
can arise from, e.g., a $g$-proportional
Yukawa coupling of $\phi$ to some equilibrated fermions. 
The resulting divergence would need to be 
addressed by methods different from those we handle, 
as the scalar HTL self-energy 
from marginal interactions is purely real.

Finally, for $\opdim>5$ couplings,  
the arguments we presented imply that 
no divergence can arise in $ab\leftrightarrow c\vartheta$ processes. 
They can thus be handled by \myaut{} through the fallback method.

Regardless of which integration method is chosen, 
once the $\left\vert\mathcal{M}\right\vert^2$ have been obtained, 
we can write
\begin{equation}
  \Gamma_\vartheta^\text{naive}
  \; = \; 
  \frac{1}{4 k d_\vartheta}\sum_{a,b,c}
  \left[
    \gamma^{t(\opdim)}_{c;a,b}(a_i)
    \, +\, 
    \gamma^{s(\opdim)}_{c;a,b}(b_i)
  \right]
  \; ,
  \label{eq:paramrateproc}
\end{equation}
where $\gamma^t$ and $\gamma^s$ are obtained by plugging the $t$- or 
$s$-parametrization component of $\tau_1$ times the matrix element 
squared in Eq.~\eqref{eq:prodrate} 
and multiplying by $1/(4k d_\vartheta)\,$. 
For example, with $d=4$ from Eq.~\eqref{eq:basis4} we have
\begin{equation}
  \gamma^{t(4)}_{c;a,b}(a_i)
  \; = \; 
  \tau_1
  \int{\rm d}\Omega_{2\to2} \,
  \mathcal{N}_{\tau_1;\sigma_1,\sigma_2}\,
  \biggl[
    a_1\frac{s}{t}
    \, + \,
    a_2
  \biggr]\,.
  \label{eq:g4texample}
\end{equation}
In the interest of a lightweight notation, 
we are not showing explicitly that the $a_i$ coefficients 
and the $\tau_1;\sigma_1,\sigma_2$ statistics depend on 
the chosen $c;a,b$ process, e.g. $a_i(c;a,b)$, $\tau_1(c)$.  
As $\mathcal{N}$
is a function of the statistics only of the $a,b,c$ particles, 
it is convenient to factor out this dependence. 
We can thus rewrite Eq.~\eqref{eq:paramrateproc} as 
\begin{equation}
  \Gamma_\vartheta^\text{naive}
  \; = \;
  \frac{1}{4 k d_\vartheta}
  \sum_{\tau_1,\sigma_1,\sigma_2} 
  \left[
    \gamma^{t(\opdim)}_{\tau_1;\sigma_1,\sigma_2}(\tilde{a}_i)
    \, + \,
    \gamma^{s(\opdim)}_{\tau_1;\sigma_1,\sigma_2}(\tilde{b}_i)
  \right]
  \;,
  \label{eq:paramrate}
\end{equation}
where $\tilde{a}_i$ and $\tilde{b}_i$ are shorthand for 
$\tilde{a}_i(\tau_1;\sigma_1,\sigma_2)$ 
and $\tilde{b}_i(\tau_1;\sigma_1,\sigma_2)$. 
They originate from summing over processes with the same statistics, 
i.e. 
\begin{equation}
  \tilde{a}_i(\tau_1;\sigma_1,\sigma_2)
  \; = \;
  \sum_{a,b,c} a_i(c;a,b)\;
    \delta_{\tau_1,\tau_1(c)}\,
    \delta_{\sigma_1,\sigma_1(a)}\,
    \delta_{\sigma_2,\sigma_2(b)}\,,
\end{equation}
and analogously for $\tilde{b}_i$. 
In what follows we will often omit the 
$\tilde{a}_i$ and $\tilde{b}_i$ dependence in 
$\gamma^{t(\opdim)}_{\tau_1;\sigma_1,\sigma_2}$ and 
$\gamma^{s(\opdim)}_{\tau_1;\sigma_1,\sigma_2}$. 
In practice, 
\myaut{} will first generate tree level diagrams and compute 
matrix elements squared for all $ab\to c\vartheta$ processes; 
it is then convenient to define similarly
\begin{equation}
  \left\vert 
    \mathcal{M}
    (\tau_1;\sigma_1,\sigma_2)
  \right\vert^2
  \; = \;
  \sum_{a,b,c} 
  \left\vert
    \mathcal{M}_{ab\to c\vartheta}
  \right\vert^2\;
    \delta_{\tau_1,\tau_1(c)}\,
    \delta_{\sigma_1,\sigma_1(a)}\,
    \delta_{\sigma_2,\sigma_2(b)}\,.
  \label{eq:totalmat}
\end{equation}

%
\subsection{Emergence of IR divergences}
\label{sub:naivediv}

Using the tools in  App.~\ref{sec:phaseint}, 
the various $\gamma^t$ and $\gamma^s$ 
appear to be ready for numerical integration. 
However, the presence of $\propto1/t$ terms is a problem. 
Note that the $\propto 1/s$ case is safe: 
if the center-of-mass (c.o.m.) energy of the process goes to zero, 
then particle production cannot take place and 
the contribution to the rate vanishes. 

%
\begin{figure}[t]

\begin{center}
    \begin{subfigure}{0.49\textwidth}
        \centering
        \includegraphics[width=\textwidth]{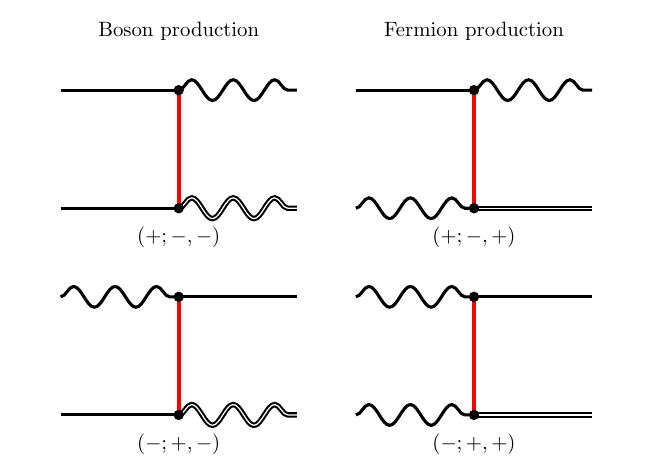}
        \caption{\small For a $\opdim=4$ interaction.}
        \label{fig:allprocd4}
    \end{subfigure}
    \hfill
    \begin{subfigure}{0.49\textwidth}
        \centering
        \includegraphics[width=\textwidth]{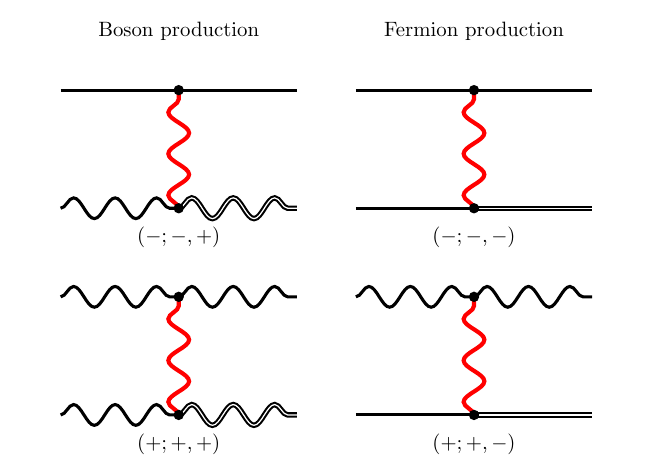}
        \caption{\small For a $\opdim=5$ interaction.}
        \label{fig:allprocd5}
    \end{subfigure}
\end{center}

\vspace{-4mm}

\caption[a]{\small
  All possible $(\tau_1;\sigma_1,\sigma_2)$ $t$-channel diagrams.
  Wavy lines represent bosons, solid lines represent fermions.
  Thick red lines correspond to mediators and double lines to 
  $\vartheta$.
}
\label{fig:allproc}
\end{figure}
%

On the other hand, the 
$\int{\rm d}\Omega_{2\to2} \,\mathcal{N}_{\tau_1;\sigma_1,\sigma_2}$ 
measure is non-vanishing around $t=0$, 
so the $\propto1/t$ (or equivalently $\tilde{a}_1\neq 0$) case 
will \emph{naively} lead to a divergent rate. 
By the arguments made previously, 
we expect to find such naively divergent processes 
in the forms shown in Fig.~\ref{fig:allproc}, corresponding 
to fermion exchange for $\opdim=4$ and gauge-boson exchange 
for $\opdim=5$. 
Physically, the $t\to 0$ limit  at constant $s$  corresponds to 
the exchange of a massless 
mediator with very small momentum at fixed c.o.m. energy. 

To illustrate how this divergence emerges, let us start with
the $\opdim=5$ case; as per
Fig.~\ref{fig:allprocd5}, particles
$a$ and $c$, which couple to the $t$-channel gauge boson, 
have the same $\tau_1$ statistics, 
whereas $\sigma_2$ must agree 
with that of $\vartheta$, $\sigma_2=\tau$. 
Hence, the most IR sensitive contribution comes from
\begin{align}
  \Gamma_\vartheta^{(5)\text{ IR}}
  \; = \; 
  & \frac{1}{4 k d_\vartheta} \sum_{\tau_1=\pm 1}
  \tau_1 \tilde{a}_1(\tau_1;\tau_1,\tau)
  \int{\rm d}\Omega_{2\to2}\,
  \mathcal{N}_{\tau_1;\tau_1,\tau}\,
  \frac{2s^2}{t}
  \nonumber \\
  \; \equiv \;
  & \frac{1}{4 k d_\vartheta} 
  \sum_{\tau_1=\pm 1} 
  \gamma^{t(5)\;\mathrm{IR}}_{\tau_1;\tau_1,\tau}
  \;,
  \label{eq:paramrateIR}
\end{align}
where we have used the fact that the matrix element $(s^2+u^2)/t$
approximates in the soft $\vert t\vert\ll s$ limit to $2s^2/t$. 
In the second line 
we have introduced implicitly 
$\gamma^{t(5)\;\mathrm{IR}}_{\tau_1;\tau_1,\tau}$,
which can be computed following App.~\ref{sec_dim5}. 
Starting from Eq.~\eqref{eq:gammatsbase} and taking the  
$\vert q_0\vert <q\ll T$ limit where 
the exchanged gauge boson has small energy $q_0$ 
and three-momentum $q$, we find
\begin{equation}
  \gamma^{t(5)\;\mathrm{IR}}_{\tau_1;\tau_1,\tau}
  \; \approx \;
  \frac{
    \vert 2^{\tau_1}{-}1\vert \tilde{a}_1 k T^2
  }{
    16\pi
  } 
  \int_{-\infty}^k {\rm d}q_0\,q_0
  \int_{|q_0|}^{2k-q_0}{\rm d}q\,
  [ 1 + n_{\tau_1^2}(q_0) + n_{\tau}(k{-}q_0) ]\,
  \frac{\mathcal{Q}^2}{q^4}
  \; ,
  \label{eq:d5pppstart}
\end{equation}
where $\mathcal{Q}^\mu=(q_0,\mathbf{q})\,$. 
While the derivation in App.~\ref{sec_dim5} is necessary for 
a proper accounting of all factors, 
we can understand $\mathcal{Q}^2/q^4\propto s^2/t$ intuitively: 
$t=\mathcal{Q}^2$ with this choice of coordinates, 
while 
$
  s = 2(\mathcal{P}_1-\mathcal{Q})\cdot\mathcal{K}
    \approx 2\mathcal{P}_1\cdot \mathcal{K}
$.
By writing this as 
$
  2\mathcal{P}_1\cdot \mathcal{K}
  =
  2p_1 k(1-\cos\theta_{\mathbf{p}_1\mathbf{k}})
  =
  -2p_1 k \mathcal{Q}^2/q^2(1-\cos\phi)
$  
one gets to the desired factor. We exploited the fact that $\phi$ 
is the azimuthal angle between $\mathbf{p}_1$ and $\mathbf{k}$ and 
that the on-shell requirements for the external 
states at soft $\mathcal{Q}$ imply 
$
  \cos\theta_{\mathbf{p}_1\mathbf{q}}
  \approx \cos\theta_{\mathbf{k}\mathbf{q}}
  \approx q_0/q \,
$.
The $\approx$ symbol implies that subleading terms in 
the soft limit have been dropped with 
respect to the definition in Eq.~\eqref{eq:paramrateIR}. 

The expression in square brackets encodes 
the residual dependence on statistical 
factors after all but two integrals have been carried out; 
it follows from the identity in Eq.~\eqref{calnt}.
Since $\tau_1^2=1$, it contains $n_{+}(q_0)$, 
the Bose--Einstein distribution for the gauge boson. 
As  $n_{+}(q_0\ll T)\approx T/q_0\gg 1$, the 
expression in square brackets approximates to $T/q_0$ 
and is $\sigma_2$-independent. We denote this limit as 
\begin{equation}
  \gamma^{t(5)\;\mathrm{div}}_{\tau_1;\tau_1,\tau}
  \; = \;
  \frac{
    \vert 2^{\tau_1} - 1\vert 
    \tilde{a}_1(\tau_1;\tau_1,\tau) k T^3
  }{
    16\pi
  }
  \int_{-\infty}^k {\rm d}q_0
  \int_{|q_0|}^{2k-q_0}{\rm d}q\,
  \frac{\mathcal{Q}^2}{q^4}
  \;,
  \label{eq:d5pppdiv}
\end{equation}
where ``div'' labels this logarithmically IR-divergent part.

In the dimension-four case, the divergence shows up in the 
small-$t$ limit of $\gamma^{t(4)}_{\tau_1;-\tau_1,-\tau}$, 
where the statistics now reflect the fermion exchange illustrated 
in Fig.~\ref{fig:allproc}. 
In this case the most IR-sensitive contribution reads 
\begin{align}
  \Gamma_\vartheta^{(4)\text{ IR}}
  \; = \;
  & \frac{1}{4 k d_\vartheta}
  \sum_{\tau_1=\pm 1}
  \tau_1 \tilde{a}_1(\tau_1;-\tau_1,-\tau)
  \int{\rm d}\Omega_{2\to2}\,
  \mathcal{N}_{\tau_1;-\tau_1,-\tau}\,\frac{s}{t}
  \nonumber \\
  \; \equiv \; 
  &\frac{1}{4 k d_\vartheta}
  \times 2
  \times \gamma^{t(4)\;\mathrm{IR}}_{+;-,-\tau}
  \; ,
  \label{eq:paramrateIR4}
\end{align}
where in the second line 
we have introduced implicitly 
$\gamma^{t(4)\;\mathrm{IR}}_{+;-,-\tau}\,$. 
The factor of two there is a consequence of crossing symmetry 
between the $(+;-,-\tau)$ and $(-;+,-\tau)$ amplitudes squared, 
corresponding to the two diagrams in each column of 
Fig.~\ref{fig:allprocd4}.
\myaut{} will thus raise an error if  
$
  \tilde{a}_{1}(+;-,-\tau) 
  \ne 
  \tilde{a}_{1}(-;+,-\tau) \,
$.
We also note that, while 
$
  \mathcal{N}_{+;-,-\tau} 
  \ne 
  - \mathcal{N}_{-;+,-\tau} \,
$,
the two sides become equal up to higher order terms 
when expanding for $q_0\ll T$.

$\gamma^{t(4)\;\mathrm{IR}}_{+;-,-\tau}$ can be computed 
following App.~\ref{sec_dim4}. 
Taking the limit of Eq.~\eqref{eq:gammatsbase} where 
all components of the momentum 
$\mathcal{Q}^\mu=(q_0,\mathbf{q})$
of the exchanged fermion are small, we find
\begin{align}
  \gamma^{t(4)\;\mathrm{IR}}_{+;-,-\tau}
  \; \approx \; &
  - \frac{ \tilde{a}_1 T^2 }{ 128\pi }
    \int_{-\infty}^k {\rm d}q_0
    \int_{|q_0|}^{2k-q_0}{\rm d}q 
  \bigl[ 1 + n_{-}(q_0) + n_{-\tau}(k-q_0) \bigr]
  \frac{1}{q^2}
  \; ,
  \label{eq:d4pppstart}
\end{align}
Following the arguments previously employed, 
the $s/t$ matrix element is dominated by a term 
proportional to $p_1k/q^2$ for soft $\mathcal{Q}\,$.
In Eq.~\eqref{eq:d4pppstart} the residual statistical dependence 
contains $n_{-}(q_0)\,$, in accordance with fermion exchange.
In the soft limit $n_{-}(q_0\ll T)\approx-1/2\,$, so that 
\begin{align}
   \gamma^{t(4)\;\mathrm{div}}_{+;-,-\tau}
   \; = \; &
   - \frac{ \tilde{a}_1(+;-,-\tau) T^2 }{ 128\pi }
     \biggl[
       \frac12 + n_{-\tau}(k)
     \biggr]
  \int_{-\infty}^k {\rm d}q_0
  \int_{|q_0|}^{2k-q_0}\frac{{\rm d}q }{q^2} 
  \,.
  \label{eq:d4pppdiv}
\end{align}
This expression, just like that for 
$\gamma^{t(5)}_{\tau_1;\tau_1,\tau}$, 
is logarithmically IR divergent for $q\to 0$. 
Physically, both cases correspond to exchanging a particle with 
very large wavelength. 
In such a case, the mediator would be unable to resolve 
the particles in the thermal bath individually. 
Rather, it is sensitive to the collective dynamics of the medium. 
In the next Section, we shall explain how the collective dynamics 
that have been neglected so far cure the  would-be divergence 
in Eqs.~\eqref{eq:d5pppdiv} and \eqref{eq:d4pppdiv}. 
Before we delve into that, we note 
that in cases where all $\tilde{a}_1$ vanish the naive rate is 
the final result; 
details on the numerical integration are provided in 
App.~\ref{sec:phaseint}.

%
\section{Model-independent automation of HTL resummation}
\label{sec:soft}

Let us now illustrate how collective effects \emph{need} to
be included for a consistent LO determination, 
before then explaining our implementations. 
Readers familiar with Hard Thermal Loop resummation can skip 
directly to Sec.~\ref{sub:matching}.

%
\subsection{The emergence of collective effects}

Let us look more carefully at what happens when the 
components of the four-momentum $\mathcal{Q}\,$, 
associated with the mediator, 
are smaller than the temperature. 
The corresponding wavelength is then larger than $1/T$, which is also 
the typical spatial separation between modes with 
$\mathcal{P}\sim T$, which 
are those that dominate the thermodynamics of the medium. 
Hence, this soft mediator is no 
longer able to resolve individual constituents; 
it rather becomes sensitive to their collective behavior. 

More quantitatively, it is well known that the first IR scale to appear 
is the \textit{soft scale} $\mathcal{Q}\sim gT$, where 
$g$ is a gauge coupling for a gauge-boson mediator and can be a set of 
gauge and Yukawa couplings for a fermionic mediator. 
Let us consider for illustration the case of a gauge boson. 
Its free, soft retarded propagator scales like 
$1/\mathcal{Q}^2\sim\mathcal{O}((gT)^{-2})$.
If we now look at higher-order corrections to this propagator, 
the loop expansion breaks down:
the retarded self-energy is dominated by the gauge-invariant 
\emph{Hard Thermal Loop} term of order $g^2 T^2$ 
from  \textit{hard} loop momenta $\mathcal{P}\sim T$. 
If we want to obtain the correct contribution at order $(gT)^{-2}$, 
resumming the Hard Thermal Loop in the propagator becomes necessary. 
A similar argument holds for fermions, where the 
free retarded propagator is $\mathcal{O}((gT)^{-1})$ and the 
retarded self-energy can be shown to be $\mathcal{O}(gT)$. 

The Hard Thermal Loop (HTL) theory~\cite{Braaten:1989mz,Braaten:1991gm} 
provides a consistent, gauge-invariant description of 
the physics of $n \ge 2$-point functions with 
soft external legs by integrating out hard modes. 
It is thus the EFT which allows one to perform 
the aforementioned resummation. 
A key feature of this EFT is that these 
$n$-point functions stem from a small set of non-local 
effective operators~\cite{Braaten:1991gm}. 
In a nutshell, there is one for each bath particle; it factors 
into a rather universal part, depending only on the spin 
of the bath particle and, for $n>2$-point functions, 
on its gauge interactions, and a 
Wilson coefficient known as the \emph{thermal mass}. 
These coefficients are of order $gT$ and depend on 
the interactions of that particle with the thermal bath. 
Multiple conventions exist in the literature; 
for gauge bosons we shall use --- and compute --- the 
\textit{Debye mass} $\mD$, while for fermions 
we will express the HTL in terms of the asymptotic mass $\masym$. 
We will discuss how to automate their evaluation in any 
$\mathcal{L}_\mathrm{med}$ 
with marginal couplings only in Sec.~\ref{sec:masses}. 
We give the expressions of the $n=2$-point functions, 
i.e. the HTL propagators,\footnote{%
  Under our $k\gtrsim T$ assumption, $\vartheta$ is hard.
  Therefore, if the bath particles are also hard, 
  all vertices in the diagrams
  in Fig.~\ref{fig:top2to2} will have at least one hard leg.
  If the bath particles are soft instead, 
  it can be shown \cite{Ghiglieri:2013gia,Bouzoud:2026rur} 
  that they give rise to a subleading contribution to the rate.
  Hence, at LO we do not need $n>2$-point functions, 
  i.e. HTL-resummed vertices.
} for fermions and gauge bosons in App.~\ref{sec:htlprops}.

The collective effects described by these propagators 
and vertices are well known.
For time-like $\mathcal{Q}^2>0$, 
the dispersion relation of the particle 
is modified: it acquires a momentum-dependent mass of order $gT$.
We are more interested however in what happens when 
$\mathcal{Q}$ is space-like since we are looking at 
$t$-channel diagrams where the exchanged particle is such that 
$\mathcal{Q}^2=t<0$.
In this region, the self-energy develops an imaginary part: 
this is the  well-known 
phenomenon of \textit{Landau damping}, 
sometimes known as dynamical screening~\cite{Baym:1990uj}.  
HTL resummation correctly describes soft-mediator exchanges, 
rendering them finite. 

This procedure has been applied, with slight technical differences, 
to a host of production 
rates such as axions~\cite{Braaten:1991dd,Graf:2010tv}, 
gravitinos~\cite{Bolz:2000fu,Pradler:2006qh}, 
photons~\cite{Baier:1991em,Kapusta:1991qp} and 
right-handed neutrinos~\cite{Besak:2012qm}. 
In what follows we describe how it can be 
automated for any generic $\opdim=4,5$ 
interaction and we illustrate the three different 
schemes we implement. 

%
\subsection{Model-independent HTL resummation}
\label{sub:matching}

%
\begin{figure}[t]

\begin{center}
    \begin{subfigure}{0.49\textwidth}
        \centering
        \includegraphics[width=0.7\textwidth]{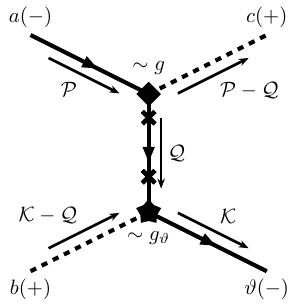}
        \caption{\small Soft fermion exchange, $\opdim=4$ interactions.}
        \label{fig:amydiagd4}
    \end{subfigure}
    \hfill
    \begin{subfigure}{0.49\textwidth}
        \centering
        \includegraphics[width=0.7\textwidth]{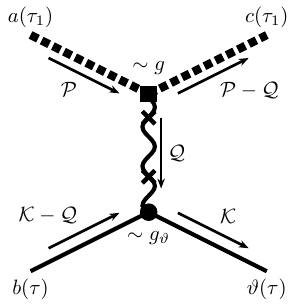}
        \caption{\small Soft gauge boson exchange, $\opdim=5$ interactions.}
        \label{fig:amydiagd5}
    \end{subfigure}
\end{center}

\vspace{-4mm}

\caption[a]{\small
  Example diagrams for the soft mediator exchange. 
  In both cases the crosses in the propagator of 
  the mediator denote HTL resummation. 
  On the left, dashed lines represent bosons,
  arrowed solid lines represent fermions. On the right,
  wavy lines represent gauge bosons, while solid and dotted lines 
  represent arbitrary, potentially different 
  statistics $\tau$ and $\tau_1$ respectively.
  We distinguish between $g$, the couplings of the mediator 
  to the bath particles and $g_\vartheta$, 
  the (model-dependent) coupling of the mediator to $\vartheta$.
  In the $\opdim=4$ case the $a\leftrightarrow c$ crossing 
  is not shown; neither are the two diagrams 
  for a bosonic $\vartheta$.
}
\label{fig:amydiags}
\end{figure}
%

In Fig.~\ref{fig:amydiags} we sketch the class of IR-sensitive 
diagrams we need to deal with in 
the $\opdim=4$ and $\opdim=5$ case respectively. 
In Eqs.~\eqref{eq:d4pppdiv} and \eqref{eq:d5pppdiv}
we have seen how the naive, 
unresummed calculation results in a divergence. 
By replacing the
bare propagator of the mediator with an HTL-resummed one for 
$\mathcal{Q}\lesssim gT$ we then find a finite rate. 
While the specifics of what precisely is resummed and of how 
the resummation is or is not switched off for 
$\mathcal{Q}\gtrsim T$ differ between the schemes, 
the main properties do not. 
Importantly,
in all cases we shall make use of light-cone techniques and 
the associated closed-form expressions.

Let us then inspect these diagrams and see the potential sources 
of model dependence. 
Hard Thermal Loop 
propagators are themselves model-dependent, 
but as we argued that dependence is entirely encoded in 
the thermal masses,
which we determine automatically as explained in Sec.~\ref{sec:masses}. 

We then examine the incoming and outgoing thermal scatterer on 
the upper half of each diagram 
in Fig.~\ref{fig:amydiags} 
and the vertex connecting
them to the mediator, which is denoted by a $g$ coupling  
(diamond in Fig.~\ref{fig:amydiagd4} for $\opdim=4$, 
square in Fig.~\ref{fig:amydiagd5} for $\opdim=5$). 
This is a coupling coming from $\mathcal{L}_\mathrm{med}$ and 
contains several sources of model dependence. 
A first one is the spin of the thermal scatterer, 
as a fermion-fermion-scalar coupling differs from 
a fermion-fermion-gauge-boson one (for $\opdim=4$). 
Similarly, for $\opdim=5$, the minimal coupling of gauge bosons 
to themselves, to scalars and to fermions has 
different Lorentz structures. 
However, it is well known that in the limit of 
soft $\mathcal{Q}$ and hard $\mathcal{P}$
this spin dependence vanishes and 
all Lorentz structures 
``eikonalize''~\cite{Arnold:2002zm,Arnold:2003zc}, 
reducing to the simplest one, 
fermion-fermion-scalar (for $\opdim=4$) and 
scalar-scalar-gauge (for $\opdim=5$). 
This is indeed what 
underpins the universality of Hard Thermal Loops. 
The most striking example is given by gauge bosons: 
their one-loop thermal self-energy has different Lorentz structures 
for the contributions from scalar, fermion and gauge loops. 
The HTL limit $\mathcal{Q} \ll \mathcal{P}$ is however 
the same for the three cases.

The outstanding model dependence in this part of the diagrams 
then comes from the statistics, gauge transformations
and, when applicable, Yukawa couplings of the scatterers. 
Upon summing over all possible scatterers, we can absorb
this dependence in a model-dependent factor $C^{(\opdim)}$.

In the lower half of each diagram 
in Fig.~\ref{fig:amydiags} 
we depict the vertex where the resummed mediator attaches to 
$\vartheta$ and to another incoming equilibrated particle 
(star in Fig.~\ref{fig:amydiagd4} for $\opdim=4$, 
circle in Fig.~\ref{fig:amydiagd5} for $\opdim=5$). 
We associate an 
arbitrary $g_\vartheta$ coupling constant with this interaction.  
The vertex emerges from $\mathcal{L}_\mathrm{int}$
and is thus very much model dependent in terms of nature 
and spin/statistics 
of the  outgoing  and incoming states. 
In what follows we will show 
how in the soft $\mathcal{Q}$ limit the entire model dependence 
can be absorbed by $g_\vartheta$.

In this soft limit we will then be able to write the 
resummed matrix elements squared,
summed over all equilibrated $a,b,c$ degrees of freedom,
as a function of two unknown model-dependent parameters $g_\vartheta$ 
and $C^{(\opdim)}$. 
By requiring that the $\mathcal{Q} \gg g T$ limit of 
this resummed calculation
agrees with the $\mathcal{Q} \ll  T$ limit of the naive calculation, 
we can then determine these model-dependent parameters.

Let us start from the $\opdim=5$ case. 
As in our analysis in and around Eq.~\eqref{eq:paramrateIR},
the relevant statistics are $\tau_1;\tau_1,\tau$. 
We can then write the contribution to the rate coming 
from Fig.~\ref{fig:amydiagd5} as 
\begin{equation}
  \Gamma^{(5)\text{ HTL}}_\vartheta
  \; = \; 
  \frac{1}{4d_\vartheta k} \sum_{\tau_1=\pm 1}
  \tau_1\,C^{(5)}_{\tau_1}
  \int{\rm d}\Omega_{2\to2}\,
  \mathcal{N}_{\tau_1;\tau_1,\tau}\,
  \left\vert
    \mathcal{M}^{(5)}_\mathrm{HTL}
  \right\vert^2
  \;,
  \label{eq:htlcontrstart}
\end{equation}
where the factors $C^{(5)}_{\tau_1}$ keep track of the 
model-dependent multiplicity resulting from 
the sum over thermal scatterers of statistics $\tau_1$
in Fig.~\ref{fig:amydiagd5}. 

For what concerns the matrix element squared, 
we follow \cite{Arnold:2002zm,Arnold:2003zc}, who 
showed that for a gauge-mediated exchange between two scalars 
$\phi_1,\phi_2\,$, HTL resummation amounts to replacing 
$
  \left\vert\mathcal{M}\right\vert^2
  \propto 
  g^4 (s-u)^2/t^2
$ with 
\begin{align}
  \label{amyscalar}
  \left\vert
    \mathcal{M}^{R\,\mathrm{HTL}}_{\phi_1\phi_2\to\phi_1\phi_2}
  \right\vert^2
  & \; \propto \;
  g^4 \left\vert 
    (2\mathcal{K}-\mathcal{Q})^\mu
    (2\mathcal{P}-\mathcal{Q})^\nu 
    G^{R\,\mathrm{HTL}}_{\mu\nu}(\mathcal{Q})
  \right\vert^2 \\[2mm]
  & \hspace{-4mm}
  \; \stackrel{\mathcal{K},\mathcal{P}\gg\mathcal{Q}}{\approx} \;
  16g^4 \left\vert 
    \mathcal{K}^\mu \mathcal{P}^\nu 
    G^{R\,\mathrm{HTL}}_{\mu\nu}(\mathcal{Q})
  \right\vert^2
  \;, 
  \label{amyscalarsoft}
\end{align}
with $g$ the gauge coupling. 
$G^{R\,\mathrm{HTL}}_{\mu\nu}(\mathcal{Q})$ 
is the retarded HTL propagator: as the matrix elements emerging 
from cutting rules are fully retarded, 
causal propagators only can appear therein 
at tree level. We are labeling the momenta as in 
Fig.~\ref{fig:amydiagd5}. 
As shown in~\cite{Arnold:2002zm,Arnold:2003zc}, 
replacing the scalars with fermion or vector scatterers 
gives rise to a matrix element squared which, 
in the $\mathcal{P},\mathcal{K}\gg \mathcal{Q}$ limit, 
is proportional to Eq.~\eqref{amyscalarsoft}, 
the corresponding limit of Eq.~\eqref{amyscalar}. 

We then take the following ansatz for 
$\big\vert\mathcal{M}^{(5)}_\mathrm{HTL}\big\vert^2$, namely 
\begin{equation}
  \label{msqdim5}
  \left\vert\mathcal{M}^{(5)}_\mathrm{HTL}\right\vert^2
  \; \equiv \;
  16 g^2 g_\vartheta^2 (q^2-q_0^2)
  \left\vert 
    \mathcal{K}^\mu 
    \mathcal{P}^\nu 
    G^{R\,\mathrm{HTL}}_{\mu\nu}(\mathcal{Q})
  \right\vert^2
  \; .
\end{equation}
This corresponds to taking Eq.~\eqref{amyscalarsoft} 
and replacing two powers of the gauge coupling $g$ with two of 
the dimensionful $g_\vartheta\,$. 
For the arguments presented after Eq.~\eqref{eq:basis5}, 
this dimensionality must be compensated 
by a $t$-proportional factor, which we simply choose to be 
$q^2-q_0^2=-t>0\,$. 

In order to carry out the matching with the naive calculation, 
we take the $\mathcal{Q}\gg gT$ 
limit of Eq.~\eqref{msqdim5}, 
which at first order is tantamount to replacing the HTL-resummed 
propagator with its free counterpart, yielding 
\begin{align}
  \left\vert \mathcal{M}^{(5)}_\text{HTL UV} \right\vert^2
  \; = \; &
  16 g^2 g_\vartheta^2(q^2-q_0^2)
  \left\vert 
    \mathcal{K}^\mu\mathcal{P}^\nu 
    G^{R\,{\rm free}}_{\mu\nu}(\mathcal{Q})
  \right\vert^2
  \nonumber \\[1mm]
  \; = \; &
  -4 g^2 g_\vartheta^2 
  \frac{(2\mathcal{K}\cdot\mathcal{P})^2}{\mathcal{Q}^2}
  \; = \;
  -4 g^2 g_\vartheta^2 \frac{u^2}{t}
  \; \approx \;
  -2 g^2 g_\vartheta^2 \frac{2s^2}{t}
  \;.
  \label{msqdim5UV}
\end{align}
As argued around Eq.~\eqref{eq:paramrateIR}, 
the $\tilde{a}_1$-proportional $(s^2+u^2)/t$ structure 
reduces to $2s^2/t$ in the soft limit. 
Hence, the UV limit of Eq.~\eqref{eq:htlcontrstart} reduces to
\begin{equation}
  \Gamma^{(5)\text{ HTL}}_{\vartheta\text{ UV}}
  \; = \; 
  -\frac{2 g^2 g_\vartheta^2}{4d_\vartheta k}
  \sum_{\tau_1=\pm 1}
  \tau_1\,C^{(5)}_{\tau_1}
  \int{\rm d}\Omega_{2\to2}
     \, \mathcal{N}_{\tau_1;\tau_1,\tau}\,
  \frac{2s^2}{t}
  \;.
  \label{eq:htlcontrstartUV}
\end{equation}

By equating this with Eq.~\eqref{eq:paramrateIR} we find
\begin{align}
 g_\vartheta^2 C^{(5)}_{\tau_1}
 \; = \;
 -\frac{\tilde{a}_1(\tau_1;\tau_1,\tau)}{2 g^2 } 
 \; ,
  \label{eq:d5coeffs}
\end{align}
which fixes the $g_\vartheta^2 C^{(5)}_{\tau_1}$ model dependence.
Let us note that, in models where $\vartheta$ couples to 
multiple gauge bosons, 
this procedure has to be carried out for each gauge group, 
thus introducing multiple 
$g_\vartheta^2 C^{(5)}_{\tau_1}$. 
In order not to overburden this illustration, we do not 
show this explicitly here; \myaut{} takes care of this automatically. 

Let us now turn to $\opdim=4$ and  HTL resummation for 
fermion exchange, 
as per Fig. \ref{fig:amydiagd4}. 
The resulting contribution to the rate, summed 
over all processes, yields 
\begin{equation}
  \Gamma^{(4)\text{ HTL}}_\vartheta
  \; = \;
  \frac{2C^{(4)}}{4d_\vartheta k}
  \int{\rm d}\Omega_{2\to2} \,
  \mathcal{N}_{+;-,-\tau}\,
  \left\vert\mathcal{M}^{(4)}_\mathrm{HTL}\right\vert^2
  \; ,
  \label{eq:htlcontrstart4}
\end{equation}
where $C^{(4)}$ accounts for the 
model dependence multiplicity in  the sum over processes in 
the upper half of Fig.~\ref{fig:amydiagd4}. 
As in Eq.~\eqref{eq:paramrateIR4}, 
we have exploited $a\leftrightarrow c$ crossing 
symmetry in the diagrams in each column of 
Fig.~\ref{fig:allprocd4}  and in 
the soft limit of 
$\tau_1\,\mathcal{N}_{\tau_1;-\tau_1,-\tau}$ to consider 
twice the $+;-,-\tau$ statistics. 

In this case the matrix element squared 
$\big\vert\mathcal{M}^{(4)}_\mathrm{HTL}\big\vert^2$ 
can be taken directly from \cite{Arnold:2002zm,Arnold:2003zc}, 
which showed how to dress $s/t$ matrix elements squared with 
fermionic HTLs. We then have 
\begin{equation}
  \label{msqdim4}
  \left\vert\mathcal{M}^{(4)}_\text{HTL}\right\vert^2
  \; = \;
  g^2 g_\vartheta^2
  \mathrm{Tr}\,\bigl[\,
    \slashed{\mathcal{K}}\,
    S_R^{\mathrm{HTL}}(\mathcal{Q})\,
    \slashed{\mathcal{P}}\,
    \bigl(S_R^{\mathrm{HTL}}(\mathcal{Q})\bigr)^*
  \,\bigr]
  \; . 
\end{equation}
Here $S_R^{\mathrm{HTL}}(\mathcal{Q})$ is 
the retarded fermion HTL propagator, 
given explicitly in Eq.~\eqref{htlS}. 
Here $\mathcal{P}$ and $\mathcal{K}$ are to be intended as 
the momenta of the fermionic external states. 
In this case the model-dependent coupling parameter $g_\vartheta$ 
is dimensionless, whereas $g$ is not necessarily a single coupling 
but could be a polynomial 
in multiple gauge and/or Yukawa couplings. 

As in the $\opdim=5$ case, 
we take the UV limit of Eq.~\eqref{msqdim4}, 
which amounts to undoing HTL resummation. We have
\begin{equation}
  \label{msqdim4UV}
  \left\vert\mathcal{M}^{(4)}_\text{HTL UV}\right\vert^2
  = \;
  4g^2 g_\vartheta^2 
  \frac{
  2\mathcal{K}\cdot\mathcal{Q}\,
   \mathcal{P}\cdot\mathcal{Q} - 
   \mathcal{K}\cdot\mathcal{P}\, \mathcal{Q}^2
  }{
    \mathcal{Q}^4
  }
  \approx 
  -4g^2 g_\vartheta^2 
  \frac{\mathcal{K}\cdot\mathcal{P}}{\mathcal{Q}^2}
  \approx
  -2g^2 g_\vartheta^2 \frac{s}{t}
  \;,
\end{equation}
where we used the fact that, for soft $\mathcal{Q}$, 
the on-shell requirements $(\mathcal{K}-\mathcal{Q})^2=0$ 
and $(\mathcal{P}-\mathcal{Q})^2=0$ 
for the corresponding external states 
suppress $\mathcal{K}\cdot\mathcal{Q}\,\mathcal{P}\cdot\mathcal{Q}$. 

We then have, for the UV limit of Eq.~\eqref{eq:htlcontrstart4} 
\begin{equation}
  \Gamma^{(4)\text{ HTL}}_{\vartheta\text{ UV}}
  \; = \; 
  - \frac{4g^2g_\vartheta^2C^{(4)}}{4d_\vartheta k}
  \int{\rm d}\Omega_{2\to2}\,
  \mathcal{N}_{+;-,-\tau}\,
  \frac{s}{t}
  \;,
  \label{eq:htlcontrstart4UV}
\end{equation}
By equating this with Eq.~\eqref{eq:paramrateIR4} 
we fix the $g^2 g_\vartheta^2 C^{(4)}$ model dependence as 
\begin{align}
  g_\vartheta^2 C^{(4)}
  \; = \; 
  - \frac{\tilde{a}_1(+;-,-\tau)}{2 g^2} \,.
  \label{eq:d4coeffs}
\end{align}

%
\subsection{Schemes for automated HTL resummation}
\label{sub:schemes}

In the previous section, 
we have matched our model-dependent naively divergent processes 
to universal fermion or gauge boson 
HTL-resummed matrix elements squared, thereby avoiding 
the otherwise tedious task of directly automating the HTL EFT. 
We can 
now use these universal HTL-resummed matrix elements squared  
to correctly describe soft mediator exchanges. 
To this end, we 
now introduce the three schemes implemented in \myaut{}. 
More extensive details are found in the appendices.

%
\subsubsection{Subtracted scheme}
\label{sec:subtr}

In a nutshell, the \emph{subtracted} scheme consists in removing the 
divergent part of $\gamma^{t(\opdim)}_{\tau_1;\sigma_1,\sigma_2}$,
identified  in 
Eq.~\eqref{eq:d5pppdiv} for $\opdim=5$ and 
Eq.~\eqref{eq:d4pppdiv} for $\opdim=4$.
We can then add back the soft  exchange and compute it using the
explicit form of the HTL-resummed matrix elements 
$\big\vert\mathcal{M}^{(\opdim)}_\text{HTL}\big\vert^2$
in Eq.~\eqref{eq:htlcontrstart} for $\opdim=5$ and 
Eq.~\eqref{eq:htlcontrstart4} for $\opdim=4$.

In the former case ($d=5$), the evaluation of 
Eq.~\eqref{eq:htlcontrstart} has been carried out in detail
using light-cone techniques~\cite{Aurenche:2002pd,CaronHuot:2008ni}
in~\cite{Ghiglieri:2020mhm,Bouzoud:2024bom}. 
It is summarized in  App.~\ref{sec:dim5app}.
The end result there is Eq.~\eqref{eq:htlcontrd5}. 
By using the determination 
of the model-dependent coefficients in Eq.~\eqref{eq:d5coeffs} 
we can rewrite it as
\begin{equation}
  \Gamma^{(5)\text{ HTL}}_\vartheta
  \; = \; 
  -\frac{ T^3 }{96\pi d_\vartheta} 
  \biggl[
    \tilde{a}_1(+;+,\tau)
    +\frac12 \tilde{a}_1(-;-,\tau)
  \biggr]
  \ln\left( 1 + \frac{4k^2}{\mD^2} \right)
  \; ,
  \label{eq:htlcontrd5a}
\end{equation}
where the logarithm signals the two-scale nature of this contribution, 
namely the 
``hard'' scale $k$ (which is $\sim T$ or larger) 
and the ``soft'' scale $\mD\,$.
An important cross-check is performed by 
\myaut{} while computing the $\opdim=5$ HTL contribution.
Namely, since the coefficient of the logarithm in 
Eq.~\eqref{eq:htlcontrd5a} can be determined entirely from 
the soft $\mathcal{Q}$ regime only using the HTL theory, 
it is expected on dimensional reasons that it is 
proportional to $T\mD^2$, with the single power of $T$ arising from 
the soft limit of $n_+(q_0)$ and the $\mD^2$ from the 
HTL effective theory, as discussed after 
Eq.~\eqref{eq:d5pppstart}. 
Therefore, \myaut{} will check, for each gauge group, 
if the corresponding $\tilde{a}_1(+;+,\tau)+\tilde{a}_1(-;-,\tau)/2$ 
contribution is proportional to its $\mD^2/T^2$
(obtained through the means described in \ref{sec:masses}) and 
raise a warning if that is not the case.

In the $\opdim=4$ case, the evaluation of 
Eq.~\eqref{eq:htlcontrstart4} has been carried out 
in  App.~\ref{sec:dim4app}
using the light-cone techniques of 
Refs.~\cite{Besak:2012qm,Ghiglieri:2013gia,Ghiglieri:2016xye},
yielding Eq.~\eqref{eq:htlcontrd4}. 
Plugging in it the model-dependent coefficients as matched in 
Eq.~\eqref{eq:d4coeffs} we have
\begin{equation}
  \label{eq:htlcontrd4a}
  \Gamma^{(4)\text{ HTL}}_\vartheta
  \; = \;
  -\frac{ \tilde{a}_1(+;-,-\tau)  T^2 }{ 256\pi d_\vartheta k } 
  \left[
    \frac{1}{2}+n_{-\tau}(k)
  \right]
  \ln\left(1+\frac{4k^2}{\masym^2}\right)
  \; .
\end{equation}
Once again, \myaut{} will check that the HTL contribution is 
indeed proportional 
to the thermal mass squared, in this case $\masym^2$.

For both the $\opdim=4,5$ cases we can then  
amend the naive rate in Eq.~\eqref{eq:paramrate}
by carrying out the aforementioned subtraction, i.e. 
\begin{equation}
  \Gamma_\vartheta^{\text{subtr}}
  \; = \;
  \frac{1}{4 k d_\vartheta}
  \left\{
  \sum_{\tau_1,\sigma_1,\sigma_2}
  \left[
      \gamma^{t(\opdim)}_{\tau_1;\sigma_1,\sigma_2}
    - \gamma^{t(\opdim)\;\mathrm{div}}_{\tau_1;\sigma_1,\sigma_2}
    + \gamma^{s(\opdim)}_{\tau_1;\sigma_1,\sigma_2}
  \right]
  \right\}
  \, + \,
  \Gamma^{(\opdim)\text{ HTL}}_\vartheta
  \; ,
  \label{eq:paramratesubtr}
\end{equation}
where the $\opdim=5$ subtraction term 
$\gamma^{t(5)\;\mathrm{div}}_{\tau_1;\tau_1,\tau}$ is given in 
Eq.~\eqref{eq:d5pppdiv}. 
It is understood above that 
$\gamma^{t(5)\;\mathrm{div}}_{\tau_1;\sigma_1,\sigma_2}$ 
vanishes for $\tau_1\ne \sigma_1$, as these processes, if present, 
are necessarily finite (and the would-be HTL version is absent). 
For $\opdim=4$, 
$
  \gamma^{t(4)\;\mathrm{div}}_{+;-,-\tau}
  =
  \gamma^{t(4)\;\mathrm{div}}_{-;+,-\tau}
$
is given in Eq.~\eqref{eq:d4pppdiv}, 
with all 
$
  \gamma^{t(4)\;\mathrm{div}}_{\tau_1;\tau_1,\sigma_2}
  =
  0 \,
$.
Numerically, the remaining $q_0,q$ integrals in 
$\gamma^{t(\opdim)}_{\tau_1;\sigma_1,\sigma_2}$ 
and 
$\gamma^{t(\opdim)\;\mathrm{div}}_{\tau_1;\sigma_1,\sigma_2}$ 
must be dealt with by performing
the subtraction under the integral sign, so as to remove the 
problematic divergent part.

%
\subsubsection{Strict LO  scheme}
\label{sec:strict}

The strict LO  scheme, introduced in \cite{Braaten:1991dd}, 
uses an intermediate scale 
$q^{\star}$  such that $gT\ll q^{\star}\ll T$. 
For $q>q^{\star}$, one computes the rate in the naive approach, 
finding the expected IR divergence in $q^{\star}$. 
For $q<q^{\star}$, one instead uses 
the HTL propagator for the mediator, yielding a 
UV-divergent soft contribution. 
The scale $q^{\star}$ 
drops out in the sum and one obtains a finite production rate.

The ``Strict LO'' moniker refers to 
the fact that, for $k\gtrsim T$, 
this scheme depends on the bath couplings 
as $g^a(\ln(1/g)+f(k/T))$ with $a>0$ the 
appropriate leading-order power of the coupling $g$. 
The non-analytical logarithmic 
dependence stems from the soft scale $gT$ curing the 
unphysical logarithmic divergence, 
while $f(k/T)$ is coupling-independent.\footnote{%
  Frequently, $f$ is referred to as the next-to-leading log 
  (NLL) ``constant''. 
} Differently from our two other schemes, no 
higher-order terms in $g$ are included.

In \myaut{} we do not determine the strict LO rate using 
the cutoff procedure, but rather by observing 
that the subtracted one, as given in Eq.~\eqref{eq:paramratesubtr}, 
is almost of the expected form. The terms 
in curly brackets depend on the couplings only 
multiplicatively through the coupling dependence of 
the $\tilde{a}_i$
and $\tilde{b}_i$ coefficients: 
they thus contribute to  $g^a\,f(k/T)$. 
The HTL term in $\Gamma^{(\opdim)\text{ HTL}}_\vartheta$, 
on the other hand, has a further dependence on the coupling through 
$\mD$ or $\masym$ in the logarithms in Eqs.~\eqref{eq:htlcontrd4a} 
and \eqref{eq:htlcontrd4a}. 
Let us note that 
\begin{equation}
  \ln\left(1+\frac{4k^2}{m_T^2}\right)
  \; \stackrel{k\gtrsim T\gg m_T\sim gT}{=} \;
  \ln\left(\frac{4k^2}{m_T^2}\right)
  \, +\, 
  \mathcal{O}\left(\frac{m_T^2}{k^2}\sim g^2\right)
  \; ,
  \label{eq:logseries}
\end{equation}
where $m_T$ is the thermal mass \textit{i.e.} 
the Debye mass or the asymptotic mass depending on the case. 
Hence, the strict LO scheme is simply obtained as
\begin{equation}
  \Gamma_\vartheta^{\text{strict LO}}
  \; = \;
  \Gamma_\vartheta^{\text{subtr}}
  \Bigg\vert_{\ln\left(1+\frac{4k^2}{m_T^2}\right)
           \to\ln\left(\frac{4k^2}{m_T^2}\right)   }
  \;.
  \label{eq:paramratestrict}
\end{equation}
Similarly, the logarithmic term therein  
--- often termed the \emph{leading-logarithmic term} --- reads
\begin{equation}
  \Gamma^{(\opdim)\text{ HTL strict LO}}_\vartheta
  \; = \;   
  \Gamma^{(\opdim)\text{ HTL}}_\vartheta
  \Bigg\vert_{\ln\left(1+\frac{4k^2}{m_T^2}\right)
           \to\ln\left(\frac{4k^2}{m_T^2}\right)   }
  \; .
  \label{eq:paramratestrictLL}
\end{equation}

The definition~\eqref{eq:paramratestrict}, 
together with Eq.~\eqref{eq:logseries}, implies that 
the difference between the rate for $k\gtrsim T$ in 
the  subtracted and strict LO schemes is of order $g^2$. 
We shall come back to this later. 

Strict LO rates obtained using Eq.~\eqref{eq:paramratestrict} 
have been found to agree in all cases with existing
calculations carried out by employing a cutoff, 
as described in the beginning of this Section. 
$\opdim=5$ examples include the case of axion production 
by comparing numerical results~\cite{Bouzoud:2024bom} by two of us 
with those of~\cite{Graf:2010tv}
and gravitino production: 
\myaut{}'s strict LO results in 
our companion paper~\cite{gravitinopaper} agree 
in their gauge-coupling component 
with those in \cite{Bolz:2000fu,Pradler:2006qh}. 
A $\opdim=4$ example is right-handed neutrino production: 
the results of~\cite{Ghiglieri:2016xye} in 
the subtracted scheme and of~\cite{Besak:2012qm} 
in the strict LO scheme agree in the $k\gg m_T$ limit.

%
\subsubsection{Tuned mass  scheme}
\label{sec:tuned}

In a nutshell, this scheme consists in introducing an effective 
mass in the propagator 
of the mediator, thus addressing IR finiteness.
From the arguments detailed in App.~\ref{sec:htl} 
this mass should be equal 
to the thermal mass of the exchanged particle times
a \emph{tuning factor} $\xi$ to be determined analytically.
This idea was first introduced in 
\cite{Peshier:2008bg,Gossiaux:2008jv,York:2014wja}; 
our implementation follows that of \cite{Kurkela:2018oqw} 
which uses the following replacement 
\begin{equation}
  \frac{u-s}{t}\to\frac{u-s}{t}f(q),
  \quad\text{with}\quad 
  f(q)=\frac{q^2}{q^2+\xi_{\opdim}^2 m_T^2}
  \; ,
  \label{eq:tuned}
\end{equation}
where $q$ is as usual the three-momentum of the exchanged mediator 
and $\xi^{ }_{\opdim}$ is a coefficient to be fixed 
as described below. 

To make contact with the chosen 
bases used for the matrix element squared in 
Eqs.~\eqref{eq:basis4}--\eqref{eq:basis5} 
we can use the following relations 
\begin{alignat}{2}
  & \opdim=4 \,: 
  && \qquad
  \frac{s}{t} \; = \; 
  -\frac{1}{2}-\frac{1}{2}\frac{u-s}{t} 
  \; ,
  \nonumber \\[2mm]
  & \opdim=5 \,:
  && \qquad
  \frac{s^2+u^2}{t}
  \; = \;
  \frac{t}{2}\left(1+\left(\frac{u-s}{t}\right)^2\right)
  \; ,
\end{alignat}
and then apply the substitution in Eq.~\eqref{eq:tuned}. 
We can then define 
\begin{equation}
  \Gamma_\vartheta^\text{tuned}
  \; = \;
  \frac{1}{4 k d_\vartheta}
  \sum_{\tau_1,\sigma_1,\sigma_2} 
  \left[
    \gamma^{t(\opdim)\text{ tun.}}_{\tau_1;\sigma_1,\sigma_2}
    (\tilde{a}_i)
    +
    \gamma^{s(\opdim)}_{\tau_1;\sigma_1,\sigma_2}(\tilde{b}_i)
  \right]
  \; .
  \label{eq:paramratetuned}
\end{equation}
From Eq.~\eqref{eq:tuned} we find for $\opdim=4$
\begin{equation}
  \gamma^{t(4)\text{ tun.}}_{\tau_1;-\tau_1,-\tau}(\tilde{a}_i)
  \; =\; 
  \tau_1
  \int{\rm d}\Omega_{2\to2} \,
  \mathcal{N}_{\tau_1;-\tau_1,-\tau}\,
  \biggl[
      \tilde{a}_1\left(\frac{s}{t}
    + \frac12\right)f(q)
    + \tilde{a}_2
    - \frac{\tilde{a}_1}{2}
  \biggr]
  \; .
  \label{eq:g4tuned}
\end{equation}
Similarly, for $\opdim=5$ we have
\begin{equation}
  \gamma^{t(5)\text{ tun.}}_{\tau_1;\tau_1,\tau}(\tilde{a}_i)
  \; = \;
  \tau_1
  \int{\rm d}\Omega_{2\to2} \,
  \mathcal{N}_{\tau_1;\tau_1,\tau}\,
  \biggl[
     \tilde{a}_1 \left( \frac{s^2+u^2}{t}
                        - \frac t2 
                  \right) f^2(q)
    + \left( \tilde{a}_2{+}\frac{\tilde{a}_1}{2} \right) t
  \biggr]
  \; .
  \label{eq:g5tuned}
\end{equation}
Non-divergent $\gamma^{t(\opdim)}$, 
e.g. $\gamma^{t(5)}_{\tau_1;-\tau_1,-\tau}$
are unmodified, 
$
  \gamma^{t(5)}_{\tau_1;-\tau_1,-\tau}
  =
  \gamma^{t(5)\text{ tun.}}_{\tau_1;-\tau_1,-\tau}
  \,
$.
The parametrizations in  App.~\ref{sec:phaseint} are easily adapted 
for the numerical two-dimensional 
integration of Eqs.~\eqref{eq:g4tuned}--\eqref{eq:g5tuned}.

The resulting rate~\eqref{eq:paramratetuned} is then finite. 
The final ingredient is the analytical determination 
of the tuning factor. 
By imposing that Eq.~\eqref{eq:paramratetuned} agrees 
with Eq.~\eqref{eq:paramratestrict} for $k\gg m_T$
up to higher orders in $g\sim m_T/k$, one finds
\begin{align}
  \xi_4
  \; =\; 
  \frac{{ e}}{2}
  \; , \qquad
  \xi_5
  \; = \;
  \frac{{ e}^{1/3}}{2}
  \; .
\end{align}
$\xi_4$   and  $\xi_5$ 
are due to~\cite{Kurkela:2018oqw} and 
\cite{Boguslavski:2023waw,Bouzoud:2024bom} respectively.

%
\subsubsection{Comparing the schemes}
\label{sub:compare}

We close out this discussion of the  schemes implemented in 
\myaut{} by reviewing their properties. 
First of all, 
for $k\gg m_T$ the difference between the  schemes 
in terms of $g$ is
\begin{align}
  & 
  \Gamma_{\vartheta}^\mathrm{subtr}
  \, -\, 
  \Gamma_{\vartheta}^\text{strict LO}
  \; = \; \mathcal{O}(g^2)
  \; , \nonumber
  \\[2mm]
  &
  \Gamma_\vartheta^\text{tuned}
  \, - \, 
  \Gamma_{\vartheta}^\text{strict LO}
  \; = \; \mathcal{O}(g)
  \; .
\end{align}
The first line was shown in Sec.~\ref{sec:strict}.
For what concerns the difference between the tuned 
and strict LO schemes, we refer the reader to \cite{Bouzoud:2024bom} 
for a discussion in the case of axion production 
($\opdim=5$ interaction) 
and to \cite{Ghiglieri:2018dib} for the case of 
a $\opdim=4$ interaction. 

Let us now examine the behavior of the production rate as 
a function of $k$. 
It is clear from Eq.~\eqref{eq:paramratestrict} 
that the rate in the strict LO scheme  can become 
unphysically negative for $k\ll m_T$. 
While the subtracted scheme alleviates this issue, 
the rate can still be negative for softer $k$. 
This is because for $m_T\gtrsim  T$ we are outside 
the validity range of the HTL theory.
The HTL-resummed contributions in Eqs.~\eqref{eq:htlcontrd5a} 
and \eqref{eq:htlcontrd4a} 
are not large enough to compensate for the subtraction of 
the divergent part. 
Hence, the  rate can become negative. 
Finally, the tuned mass scheme is positive-definite and 
does not suffer from such a problem. 

In other words, we have shown how the time-honored
strict LO \emph{assumes} the existence of  $q^\star$ 
to separate scales $g T \ll q^\star \ll T$, which gives 
the $\ln(T/(gT))$. When $T$ is not $\gg gT$, 
there is no ``large logarithm'', and worse: 
when $T \lesssim g T$ the logarithm becomes negative. 
A first idea to address this issue is then to replace 
$\ln(T/(gT)) \to 1/2 \ln(1+(T/(gT))^2)$, 
which is motivated from HTL sum rules: 
this is precisely the subtracted scheme. 
But, as we discussed, this turns out to not be enough, 
motivating the introduction of the tuned scheme, 
which is always positive by construction. 

This behavior of the various schemes in terms of $g$ and $k$ 
directed us in choosing them for \myaut{}, 
so as to showcase a conservative estimate of the theory uncertainty. 
We refer to Fig.~\ref{fig:gw} 
for an example of the spread between the three schemes. 
Note that this is not the only source of theory uncertainty; 
we refer the reader to \cite{Bouzoud:2026rur} for a more thorough, 
model-dependent analysis of the soft $k$ contribution 
in the case of axion production. 

%
\subsection{Automated determination of thermal masses}
\label{sec:masses}

As we have seen above, determining the thermal mass $m_T$ of 
the mediator is essential to computing 
the HTL contributions 
$\Gamma^{(\opdim)\text{ HTL}}_\vartheta$ and 
$\gamma^{t(\opdim)\text{ tun.}}$. 
We will now describe how thermal masses are computed by \myaut{}. 
The procedure and implemented routines 
are applicable to scalars, fermions and gauge bosons 
in thermal equilibrium and thus have a wider applicability 
than their use in the determination of the HTL contribution. 

At leading order in perturbation theory, 
the thermal dispersion relation of a massless particle 
$\extpart$ with 4-momentum $\mathcal{P}$ approaches 
$
  \mathcal{P}^2
  =
  p_0^2-p^2
  =
  m_{\rm asy}^2
$ 
at  momenta $p_0>p\gtrsim T$,
where $m_{\rm asy}$ is called the asymptotic mass. 
This is in principle not what we want, 
namely the matching coefficient for a gauge-boson or fermion HTL,
which is defined for \emph{soft} external momentum $\mathcal{P}$. 
However, it is well known that,
at leading order only, the HTL dispersion relations for transverse 
gauge bosons and for fermions with positive chirality 
to helicity ratios\footnote{%
  These are the propagating modes for a free theory in vacuum: 
  coherently, they are the only propagating modes in 
  the UV limit of the HTL theory. 
  For scalars the discussion is much simpler, 
  as the on-shell one-loop self-energy
  is momentum-independent for massless loop particles. 
  See~\cite{Biondini:2020ric} for the effect of masses.
} 
are determined in the $p\gg gT$ limit 
by \emph{the same} asymptotic masses that are obtained in 
the $p_0>p\gtrsim T$ regime. 
Hence, by determining the latter we do fix the needed HTL 
matching coefficients. 

%
\begin{figure}[t]

\begin{center}
    \includegraphics[width=0.7\linewidth]{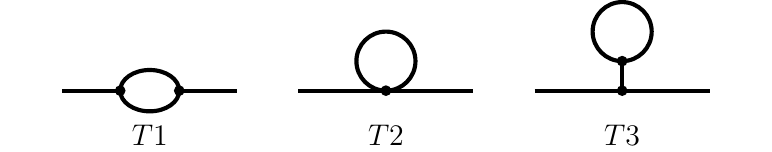}    
\end{center}

\vspace{-4mm}

\caption[a]{\small
  All possible one-loop topologies for the self-energy  
  in this study. 
  The graphical conventions
  follow those of Fig.~\ref{fig:top2to2}. 
}
\label{fig:topself}
\end{figure}
%

At leading order, 
$m_{\rm asy}$ emerges from the real part of the retarded one-loop 
self-energy of $\extpart$, evaluated at 
the argument $\mathcal{P}$ such that $\mathcal{P}^2=0$ 
in order to obtain the first perturbative correction to 
the dispersion relation. 
Let us then examine all possible type of diagrams for 
the one-loop self-energy, shown in Fig.~\ref{fig:topself}.
For our purposes, the $T3$ tadpole  will not contribute: 
if the vertical propagator is a non-abelian gauge boson, 
the diagram will vanish as ${\rm SU}(N)$ generators are traceless. 
In the abelian case, the particle in the loop must be charged; 
the sum of the particle 
and antiparticle contributions will cancel out at 
vanishing chemical potential following Furry's theorem; 
see~\cite{Notzold:1987ik} for the neutrino self-energy 
at non-vanishing chemical potentials. 

We  compute the retarded self-energy using 
the real-time formalism of TFT, 
choosing the $r/a$ basis for propagators for convenience 
(see \cite{Ghiglieri:2020dpq} for a review). 
This basis contains three non-zero propagators: 
the  causal retarded ($R$) and advanced ($A$) propagators 
and the \emph{symmetric} $rr$ propagator. 
The free causal 
propagators are identical to their zero-temperature counterparts.

%
\begin{figure}[t]

\begin{center}
    \includegraphics[width=0.75\linewidth]{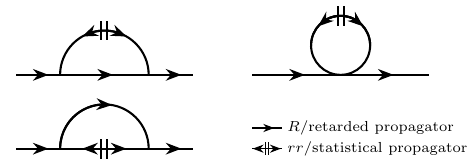}
\end{center}

\caption{
  Causality assignments for the one-loop retarded self-energy. 
  See main text for more details on the $R$ and $rr$ propagators.
}
\label{fig:1to1causal}
\end{figure}
%

All possible causality assignments for the propagators in 
the one-loop topologies are shown in Fig.~\ref{fig:1to1causal}.
We can see that every possibility contains one $rr$ propagator,
whose expression we list below 
\begin{alignat}{2}
  D_{rr}(\mathcal{Q})
  & \; = \; 
  \left[ \frac{1}{2} + n_{+}(|q_0|) \right]
  2\pi\delta(\mathcal{Q}^2)
  \; \equiv \; 
  D_{rr}^+(\mathcal{Q})
  \; , 
  && \qquad\text{[scalar]}
  \nonumber \\[2mm]
  S_{rr}(\mathcal{Q})
  & \; = \;
  \slashed{\mathcal{Q}}
  \left[ \frac{1}{2} + n_{-}(|q_0|) \right]
  2\pi\delta(\mathcal{Q}^2)
  \; \equiv \; 
  \slashed{\mathcal{Q}}D_{rr}^-(\mathcal{Q})
  \; , 
  && \qquad\text{[fermion]}
  \nonumber \\[2mm]
  G_{rr}^{\mu\nu}(\mathcal{Q})
  & \; = \; 
  - \eta^{\mu\nu}
  \left[ \frac{1}{2} + n_{+}(|q_0|) \right]
  2\pi\delta(\mathcal{Q}^2)
  \; = \;
  -\eta^{\mu\nu}D_{rr}^+(\mathcal{Q})
  \; . 
  && \qquad\text{[gauge]}
  \label{eq:rrprops}
\end{alignat}
The gauge boson $rr$ propagator was given in the Feynman gauge.
This choice does not affect the results, 
We address this later on and explain why Fadeev--Popov ghosts 
are not needed.

We can observe from Eq.~\eqref{eq:rrprops} that the $rr$ propagators
are proportional to 
$
  D_{rr}^{\pm}(\mathcal{Q})
  \propto
  \delta(\mathcal{Q}^2) \,
$.
Diagrammatically, this corresponds to cutting the $rr$ line and 
putting it on shell with frequency $q_0=\pm q$. This is depicted 
graphically in Fig.~\ref{fig:1to1causal} by 
the short double vertical lines. 
The  $1/2+n_{\pm}(\vert q_0\vert)$ factor 
in $D_{rr}^{\pm}(\mathcal{Q})$ is to be understood 
as a vacuum part, $1/2$, and a thermal part, $n_{\pm}(\vert q_0\vert)$. 
The thermal mass 
naturally arises from the latter contribution. 

\begin{figure}[t]

\begin{center}
    \includegraphics[width=0.7\linewidth]{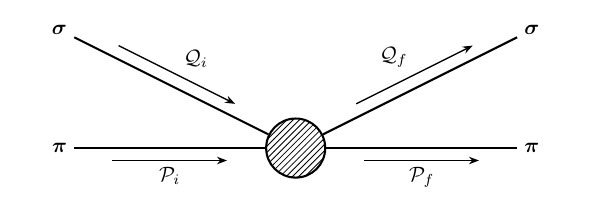}
\end{center}

\vspace{-6mm}

\caption[a]{\small
  Notation and labeling of momenta  
  for the computation of the 
  $\extpart+\bathpart\to \extpart+\bathpart$ amplitude.
  See main text for more details.
}
\label{fig:massampl}
\end{figure}

Cuts of $T1$ yield the $S$ and $U$ diagrams 
in Fig.~\ref{fig:top2to2}, according to the sign of $q_0$; 
the single cut of $T2$ yields the $X$ diagram. 
Therefore, the evaluation of the one-loop self-energy 
is equivalent to the convolution of the on-shell amplitude of 
$\twotwo$ processes of the form 
$\extpart+\bathpart\to \extpart+\bathpart$ 
with the statistical function $n_{\bathpart}(\vert q_0\vert)\,$. 
$\bathpart$ is the particle associated with the $rr$ propagator 
(see Fig.~\ref{fig:massampl}) and
$n_{\bathpart}(\vert q_0\vert)$ its corresponding statistical function.
This factorization is analogous to loop-tree 
duality~\cite{Catani:2008xa},
recently extended in the thermodynamical context to 
finite temperature and density in~\cite{Karkkainen:2025nkz}. 
This is particularly advantageous because, once again, 
at this order we shall need the on-shell zero-temperature 
tree-level $\twotwo$ amplitude, 
which is easily computed using the in-vacuum automation tools 
we adopt. 
Note that both $\extpart$ and $\bathpart$ are particles belonging 
to the thermal bath and therefore neither of them can be $\vartheta$. 

%
%

In Fig.~\ref{fig:massampl} we depict the kinematics of the cut process.
The  momenta of the incoming and outgoing 
$\extpart$ and $\bathpart$ are the same,  
$
  \mathcal{P}_i
  =
  \mathcal{P}_f
  \equiv
  \mathcal{P}
$ and
$
  \mathcal{Q}_i
  =
  \mathcal{Q}_f
  \equiv
  \mathcal{Q} \,
$.
We thus have $t=0$ and $s=-u$, since
we also previously imposed $\mathcal{P}^2=0$ and we have 
$\mathcal{Q}^2=0$ from the  $rr$ propagator. 
This is nothing but \emph{forward scattering}; see~\cite{Moore:2020wvy}
for an intuitive explanation of its connection to shifts in 
the dispersion relation.

As all external states are on shell, the total amplitude for a 
given $\extpart+\bathpart\to \extpart+\bathpart$ 
process is gauge invariant.
Our choice of not considering Fadeev--Popov ghosts stems from the known 
equivalence~\cite{Landshoff:1992ne,Landshoff:1993ag}
between assigning opposite or vanishing thermal parts 
(the $n_\pm$ part of $D_{rr}^\pm$) to 
unphysical gauge-boson polarizations and to ghosts. 
Our choice of a vanishing thermal part
is more easily handled by in-vacuum automation tools.

Under these kinematical constraints, 
the amplitudes simplify tremendously, becoming
independent on $\mathcal{P}$ or $\mathcal{Q}$.
This is easy to understand for $X$ diagrams.
For $S$ and $U$ diagrams, 
the  propagator of the intermediate particle will introduce an 
$s=-u=2\mathcal{P}\cdot\mathcal{Q}$ denominator.
In the numerator, once all spin and polarization sums or projections
have been carried out, the only nonzero Lorentz-invariant 
combination of $\mathcal{P}$ and $\mathcal{Q}$ 
is their product $\mathcal{P}\cdot\mathcal{Q}$,
compensating the structure in the denominator.

Quantitatively, let us define 
$i\mathcal{M}_{\extpart \bathpart\to \extpart \bathpart}$ as 
the connected, 
amputated amplitude traced over all degeneracies of 
the $\bathpart$ state 
--- cutting the $rr$ propagator implies tracing over the indices of 
the external $\bathpart$ legs --- 
and for a single degenerate $\extpart$ state, 
e.g. a single helicity and color state 
for a gluon. 
This corresponds to a diagonal self-energy in 
the space of these degeneracies. 

Let us note that, in the SM or extensions thereof, 
Yukawa couplings break the degeneracy of fermion generations. 
The resulting self-energies and thermal masses are 
thus in principle not diagonal in generation space, 
see~\cite{Hataei:2025lzz} for possible consequences. 
In its current form, \myaut{} can only deal with cases where 
the degeneracy, though broken, still yields a diagonal 
thermal mass matrix in generation space. 
This is for instance the case when only 
the top coupling is considered 
nonzero among the Yukawa couplings. 
We refer to App.~\ref{app:yukawa} for more details of our handling 
of this case. 

The asymptotic mass squared can be expressed as the sum over 
$\bathpart$ of these 
$
  i\mathcal{M}_{\extpart \bathpart\to \extpart \bathpart}
$
amplitudes multiplied by the  integral over the loop momentum 
$\mathcal{Q}$ of $D_{rr}^{\pm}(\mathcal{Q})\,$, 
which only depends on the nature of $\bathpart$. 
This can be seen in Refs.~\cite{Arnold:2002zm,Caron-Huot:2008vbk}, 
for example. 
In the notation of Figure \ref{fig:massampl} we have 
\begin{equation}
  m_{\rm asy}^2
  \; = \;
  - \sum_{\bathpart}
  \mathcal{M}_{\extpart \bathpart \to 
               \extpart \bathpart  }\,
  H_{\bathpart}\times
  \begin{cases}
    -1 & \text{if fermion loop cut} \\
    1 & \text{otherwise}
  \end{cases}
  \; ,
\end{equation}
The overall  sign is due to the self-energy being defined as 
minus the connected, amputated two-point function. 
We also account for the usual $-1$ factor when cutting a fermion loop. 
$H$ is a function of the statistics of $\bathpart$, 
\begin{equation}
  H_{\pm}
  \; = \;
  \frac12 \int\frac{\mathrm{d}^4\mathcal{Q}}{(2\pi)^4}
  D_{rr}^{\pm}(\mathcal{Q})
  \; = \;
  \pm \frac{1}{2\pi^2}
  \int_0^\infty {\rm d}q\,q^2\,
  \frac{f_\pm(q)}{2q}
  \; = \;
  (2^{\pm1}-1)\frac{T^2}{24}
  \; .
\end{equation}
The overall factor of 1/2 emerges from our 
accounting for a single solution
of the two $q_0=\pm q$ 
when carrying out the $q_0$ integral using 
$\delta(\mathcal{Q}^2)$ 
with our cutting procedure.
The $1/2$ vacuum part of $D_{rr}^{\pm}(\mathcal{Q})$ integrates 
to zero in dimensional regularization.

\myaut{} computes the amplitude for each relevant process 
automatically, by 
generating and evaluating  the diagrams using \myfa{} and \myfc{}. 
We use the \textsc{VecSet} module of \myfc{} to set 
the coordinates of $\mathcal{P}$ and $\mathcal{Q}$ 
and to  perform the spin/polarization sum of $\bathpart$ if needed. 
Summing over all $\bathpart$, 
we finally obtain the asymptotic mass $m_{\rm asy}^2\,$. 
For gauge bosons, we recall that at leading order 
$\mD^2=2m_{\rm asy}^2\,$. 

%
\section{Structure of \myaut{}}
\label{sec:struct}

%
\begin{figure}[t]
    \centering
    \includegraphics[width=\linewidth]{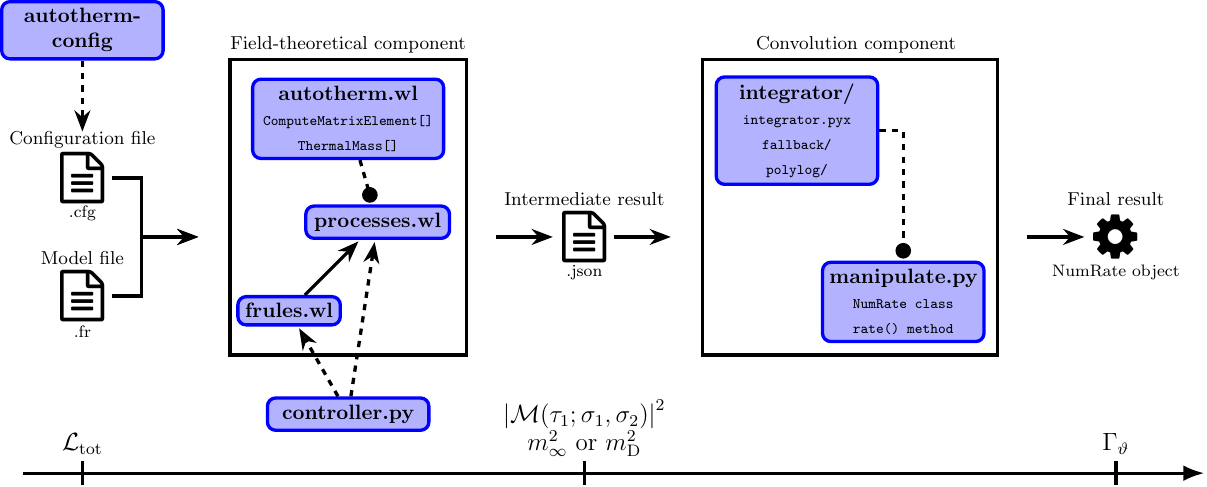}
    \caption[a]{\small
      Diagram showing the main components of 
      the program and at which stage certain 
      files and  functions  
      get used during the computation.
    }
    \label{fig:flowchart}
\end{figure}
%

The structure of this first release of \myaut{} is shown 
in Fig.~\ref{fig:flowchart}. 
Its modular design 
currently features two main components, 
the field-theoretic manipulations 
and the subsequent convolution with numerical integrators. 
The \anpart{} component is tasked with
the generation of tree-level Feynman diagrams for all 
$ab\to c\vartheta$ processes and the 
evaluation of the corresponding matrix elements squared 
using \myfa{} and \myfc{}
with the ad-hoc modifications mentioned in App.~\ref{app:fafc}. 
When the resulting naive rate 
is IR divergent, this component implements the procedure 
described in Sec.~\ref{sec:masses}
to determine the thermal mass for each particle mediating the 
would-be divergent exchange.

The main element of this \anpart{} component is 
the \texttt{autotherm.wl} \textsc{Wolfram} package, 
which implements all needed functions. Further information 
on its possible standalone usage 
will be given in Sec.~\ref{sec:runmath}. 
Its main use, though, is within the fully-automated 
pipeline going from $\mathcal{L}_\mathrm{tot}$ 
to $\Gamma_\vartheta$; in that context, 
the \texttt{controller.py} Python module will orchestrate 
the relevant steps by calling 
two \textsc{Wolfram} language scripts. 
\texttt{frules.wl} takes care of using \myfr{} to derive the 
Feynman rules, 
whereas \texttt{processes.wl} takes care of the determination of 
$\left\vert \mathcal{M}(\tau_1;\sigma_1,\sigma_2)\right\vert^2$ 
and of the thermal masses 
by calling the appropriate functions in \texttt{autotherm.wl}. 
Usage instructions will be provided in Sec.~\ref{sec:runpython}. 

The \numpart{} component takes as input the 
$\left\vert \mathcal{M}(\tau_1;\sigma_1,\sigma_2)\right\vert^2$ 
and any relevant thermal mass; 
it first proceeds to project the former onto
our bases in Eqs.~\eqref{eq:basis4}--\eqref{eq:basis5} 
or onto the fallback basis in App.~\ref{sec_fallback}. 
It then performs the numerical integrations and possible subtractions 
following the schemes presented in Sec.~\ref{sub:schemes}, 
finally outputting the $\Gamma_\vartheta$ rate, 
in a tabulated form. 

The main workhorse of this component is 
the \texttt{manipulate.py} Python module. 
It relies on SymPy \cite{10.7717/peerj-cs.103} for 
the symbolic manipulations needed for projecting matrix elements 
squared onto bases. 
Two-dimensional numerical integrations, as in App.~\ref{sec:phaseint}, 
are carried out using SciPy\cite{2020SciPy-NMeth}, with fast, 
compiled Cython \cite{behnel2010cython} 
functions for the integrands implemented in 
our \texttt{integrator} module. 
Four-dimensional fallback integrations, 
as per App.~\ref{sec_fallback}, are carried out in C 
with the Cubature package~\cite{cubature}. 

The end result, the rate for user-provided values 
of the momentum of $\vartheta$ and of the relevant coupling constants, 
is obtained from methods of the \texttt{NumRate} class 
that will be described in Sec.~\ref{sec:runpython}. 
Possible standalone usage of the \texttt{manipulate.py} module, 
without any prior 
execution of the \anpart{} component, will be illustrated 
in Sec.~\ref{sec:runpythonpackage}.

%
\section{Using the program}
\label{sec:tuto}

\myaut{} can be downloaded from \url{https://autotherm.in2p3.fr}; 
documentation, tutorials and examples are also available there.

%
\subsection{Installing \myaut{}}

\myaut{} has been developed and tested on Linux and  macOS. 
Installation on Windows is left to 
the courageous reader. 
\myaut{} requires a working installation of \mymath{}; 
it is tested on versions 11 and greater. 
The 
\myfr{} \cite{Alloul:2013bka},  
\myfa{} \cite{Hahn:2000kx} and 
\myfc{} \cite{Hahn:2016ebn}
\mymath{} packages need to be installed on your system. 
See their websites and documentation for further instructions.

Python packages 
NumPy~\cite{harris2020array}, 
SciPy~\cite{2020SciPy-NMeth},
SymPy~\cite{10.7717/peerj-cs.103} and 
Cython~\cite{behnel2010cython} 
will be automatically installed by the setup procedure.
Finally, the Polylogarithm~\cite{Voigt_Polylogarithm_2024} and 
Cubature~\cite{cubature} C libraries 
are bundled within our source code.

Once the code has been unpacked in the \texttt{\myautdir{}} folder, 
we recommend  installing 
\myaut{} within a virtual environment.
This can be done with the following 
commands 
(in the bash or zsh shells): 

\begin{mdframed}[style=mylststyle]
\begin{lstlisting}[language=myCMD,style=CMDstyle,escapeinside={(*}{*)}]
cd (*\myautdir{} *)
python3 -m venv .venv
source .venv/bin/activate
pip install .
\end{lstlisting}
\end{mdframed}

The second line above requires the \texttt{ensurepip} package 
to be installed.\footnote{\label{foot_venv}%
  On the \texttt{csh/tcsh} and \texttt{fish} shells, 
  the \texttt{source .venv/bin/activate} command should be replaced 
  by \texttt{source .venv/bin/activate.csh} and 
  \texttt{source .venv/bin/activate.fish} respectively.
  In all cases, a successful creation of a virtual environment will 
  cause the command prompt to be prefixed by \texttt{(.venv)}. 
  From now on, all calls to \texttt{python} and \texttt{pip} 
  will refer to the versions of those tools installed inside 
  the virtual environment. 
  This prevents interferences with system-wide installations. 
  To leave the virtual environment, just run \texttt{deactivate}. 
  Conversely, to activate it in a new terminal, 
  \texttt{source .venv/bin/activate} will then need to be executed 
  again before using \myaut{}.
}
The final command will install the Python dependencies of \myaut{} as 
well as compile the Cython/C code used by the \numpart{} component.
The commands \texttt{autotherm-config} 
and \texttt{autotherm-run}, whose usage will be discussed in 
Sections \ref{sec:conf} and \ref{sec:runpython} respectively, 
will also be made available.
Finally, 
we note that the fallback integration routines support parallelism 
through \textsc{OpenMP}. 
See the documentation for activation instructions.

%
\subsection{Writing a configuration file for \myaut{}}
\label{sec:conf}

\myaut{} requires two files to run:
\begin{itemize}
  \item A \textit{model file} in the \texttt{.fr} \myfr{}  format; 
  \item A \textit{configuration file} in the \texttt{.cfg} format. 
  This file contains required  information 
  such as paths to the tools used in the code 
  and model-specific settings. 
\end{itemize}

For what concerns the \emph{model file}, 
we refer the reader to~\cite{Alloul:2013bka} for more information. 
Here we just point out a few \myaut{}-specific aspects: 
\begin{itemize}
  \item 
  Fermions are treated by \myfa{} as Dirac or Majorana fermions.
  When single-chirality Weyl fermions are required, 
  e.g. when studying the Universe before electroweak symmetry breaking,
  they should be implemented as massless Dirac fermions; 
  the Lagrangian terms describing them 
  should be written including the chirality projectors
  \texttt{ProjP} and \texttt{ProjM} 
  --- see \cite{Alloul:2013bka} for more information. 
  \item 
  Models extending the SM can profit from our provided implementation 
  of the sym\-met\-ric-phase SM and from \myfr{}'s ability 
  to load multiple files. 
  One can then write a model file encoding only the BSM fields 
  and couplings. 
  We refer to     Sec.~\ref{sec:validation} for several examples 
  and to App.~\ref{app_sm} for our conventions 
  for the SM field content. 
  \texttt{includeSM} needs to be 
  set to \texttt{yes} in the configuration file, 
  as illustrated in App.~\ref{app:cfg}. 
  \item 
  The implementation of couplings having vector or tensor form 
  in indices such as generation ones 
  --- the prime example being the Yukawa coupling matrices 
  in generation space in the SM --- 
  requires special care and has consequences on thermal masses 
  and HTL resummation. 
  Details are illustrated in App.~\ref{app:yukawa}. 
  \item
  \texttt{Mass} and \texttt{Width} should be set to \texttt{0} 
  for each field, otherwise \myfa{} will automatically 
  generate symbols that might cause errors elsewhere.
\end{itemize}
Our online documentation and packaged model files, 
as discussed later in Sec.~\ref{sec:validation}, 
can also provide additional guidance.

The \textit{configuration file} is documented in 
App.~\ref{app:cfg} and online; 
the model-dependent section of the file is provided with 
the example files discussed in Sec.~\ref{sec:validation}. 
We remark that the \texttt{autotherm-config} command generates 
a template for the configuration file from user input. 

%
\subsection{From \texorpdfstring{$\mathcal{L}_\mathrm{tot}$}{Ltot} to \texorpdfstring{$\Gamma_\vartheta$}{Γθ} with \myaut{}}
\label{sec:runpython}

Once a proper configuration file, 
henceforth called  \texttt{config.cfg}, 
has been prepared, you can run the \myaut{} code. 
It is advised to run the \anpart{} and \numpart{} components 
separately, 
so as to avoid re-computing all matrix elements squared 
and thermal masses 
if the model file has not been changed. 
To run the \anpart{} component by itself, 
you can either just execute in a terminal

\begin{mdframed}[style=mylststyle]
\begin{lstlisting}[language=myCMD,style=CMDstyle]
autotherm-run config.cfg
\end{lstlisting}
\end{mdframed}
or, in a Python script or notebook,
\begin{mdframed}[style=mylststyle]
\begin{lstlisting}[language=Python,style=pystyle]
from analytical.controller import *
analytical_res = analytical_pipeline("config.cfg")
\end{lstlisting}
\end{mdframed}

In either case, the \texttt{controller} module acts
as an orchestrator for the \textsc{Wolfram} language routines 
taking care of running \myfr{} to derive Feynman rules 
--- if \texttt{rules} is toggled --- and of 
generating and squaring matrix elements and computing thermal masses 
if \texttt{proc} is toggled. 

For a first execution, you should run all steps of 
the \anpart{} pipeline, that is to say 
set the \texttt{conf}, \texttt{rules} and \texttt{proc} keys
in the configuration file to \texttt{yes}.
Should no errors arise, you will notice \myaut{} generated 
files and directories illustrated in App.~\ref{app:json}, 
with particular 
attention to \texttt{config\_result.json}, 
which contains the final result 
for $\left\vert \mathcal{M}(\tau_1;\sigma_1,\sigma_2)\right\vert^2$ 
and,
when needed, the thermal masses, in a human-readable format.

Once these analytical results  have been obtained, 
one can move on to the numerical evaluation 
of the production rate. 
A few lines of Python are enough to perform this task, i.e.
\begin{mdframed}[style=mylststyle]
\begin{lstlisting}[language=Python,style=pystyle]
from analytical.controller import *
from numerical.manipulate import *

# If we already ran the field-theoretical pipeline
# set "conf", "rules" and "proc" in config.cfg to "no"
analytical_res = analytical_pipeline("config.cfg")

rateobj = NumRate(*analytical_res,deg)
rateres = rateobj.rate(k,noneqnum,numcouplings,mode)
\end{lstlisting}
\end{mdframed}

What is happening here is that \texttt{analytical\_pipeline} can
optionally run the routines dealing with configuration generation, 
Feynman rules derivation
and evaluation of matrix elements squared and thermal masses, 
according to the values 
of the corresponding configuration keys. 
In all cases it will then parse the resulting 
\texttt{config\_result.json}
output file and return three Python structures that serve as input 
to the \texttt{NumRate} class. 
We shall return to this 
in Sec.~\ref{sec:runpythonpackage}.\footnote{\label{foot_fermionflag}%
  The handling of the interface between the two components 
  is different in cases where a fermionic $t$- or $u$-channel 
  mediator has a generation-dependent thermal mass. 
  We refer to App.~\ref{app:yukawa} and to 
  the provided example reproducing the results of 
  Ref.~\cite{Biondini:2020ric} for more details.
} 
Finally, its \texttt{rate} method performs the numerical
integrations and returns numerical values for the rate.

There are a few  input parameters which  need to be provided by 
the user: 
\begin{itemize}
  \item 
  \texttt{deg} is  $d_\vartheta$.\footnote{\label{foot:weyl}%
    Note that if a Weyl fermion is implemented as a Dirac fermion 
    \texttt{F[i]} with chirality projectors, 
    its two helicity degrees of freedom are treated by 
    \myfa{}, \myfc{} and \myaut{} 
    as particle and antiparticle modes 
    \texttt{F[i]} and \texttt{-F[i]}. Hence, 
    if $\vartheta$ is set to \texttt{F[i]} in the configuration file,
    \myaut{} will compute the production rate for 
    a single helicity component, lifting the degeneracy.
  }
  \item
  \texttt{k} is the $\vartheta$ momentum, normalized by 
  the temperature ($k/T$). 
  This variable can either be a float or a \texttt{NumPy} array. 
  \item
  \texttt{noneqnum} is the numerical value of 
  the \texttt{noneq} bath-$\vartheta$ coupling. 
  Setting it to $1$ is equivalent to normalizing 
  the rate by the coupling squared. 
  For  $\opdim\ge 5$, \texttt{noneqnum} is dimensionful. 
  As all other dimensionful quantities, including $\Gamma_\vartheta$, 
  are rescaled by appropriate powers of $T$, it is advised to set 
  this quantity in the same units. 
  \item 
  \texttt{numcouplings} is a tuple containing the numerical values 
  of the other couplings.\\ \texttt{analytical\_res[1]}, 
  the second component of the list returned by 
  \texttt{analytical-}\\ \texttt{\_pipeline}\footnote{%
    Equivalently,
    this is also given by the final result file 
    \texttt{config\_result.json}, see App.~\ref{app:json}.
  } 
  is a dict whose structure is to be followed 
  to order the elements of \\ \texttt{numcouplings}, \textit{i.e.} 
  first the gauge couplings in the order they are listed in 
  the \texttt{gauge} dict key, then the other parameters in 
  the order that are listed in the \texttt{others} dict key. 
  \item 
  The \texttt{mode} integer variable determines how 
  $\Gamma_\vartheta$ is computed.
  \begin{itemize}
    \item 
    For a value of $1$, 
    the strict LO scheme of Sec.~\ref{sec:strict} is used.
    \item 
    For a value of $2$, 
    the subtracted scheme of Sec.~\ref{sec:subtr} is used.
    \item 
    For a value of $3$, 
    the tuned mass scheme of Sec.~\ref{sec:tuned} is used.
    \item 
    For a value of $0$, all three schemes above are used.
    \item 
    For any \texttt{mode}$<0$, 
    the fallback integration in App.~\ref{sec_fallback} is used.
  \end{itemize}
  In case of a finite naive rate, 
  the three schemes are equivalent and any non-negative choice
  is ignored. Negative ones still use the fallback integration, 
  which is also automatically activated for $\opdim>5$ couplings.
\end{itemize}

The \texttt{rate} method returns a list. The first item is 
a \texttt{NumPy.ndarray} for the dimensionless momenta $k/T$,
regardless of the type of \texttt{k}.
Subsequent items contain 
$\Gamma_\vartheta/T$ for each value of $k/T$.\footnote{%
  If \texttt{noneqnum} is set to units other than $T$ for $\opdim>4$, 
  then this rescaling will be reflected here. For instance, setting 
  \texttt{noneqnum}$=1$ implies that 
  the returned quantity has dimension $\Gamma_\vartheta/T^{2\opdim-7}$. 
}
If \texttt{mode} was set to $0$ and the naive rate is divergent, 
the last 3 elements of the list are \texttt{NumPy.ndarray} 
for $\Gamma_\vartheta/T$ in the order strict LO, subtracted and tuned. 
For any other value of \texttt{mode}, the resulting list 
contains just the arrays for the momentum and 
the rate computed using the selected scheme. 

To summarize this Section, \myaut{} is divided into 
two main components. 
The \anpart{} component processes the model file 
and the configuration and outputs 
a result file containing the matrix elements squared 
for all relevant processes as 
well as the thermal masses of the mediators in processes 
that would naively lead to a divergence of the production rate. 
The \numpart{} component then processes this result file 
and returns a \texttt{NumRate} object. 
This object possesses a \texttt{rate} method, 
allowing the user to get numerical 
results for the $\vartheta$ production rate. 

A \texttt{NumRate} object also has the methods \texttt{get\_coeffs} 
and \texttt{get\_leadlog}. 
The former returns the symbolic form of 
the $\tilde{a}_i$ and $\tilde{b}_i$ coefficients used in 
the decomposition of 
$
  \left\vert \mathcal{M}(\tau_1;\sigma_1,\sigma_2)\right\vert^2
$,
as explained in Sec.~\ref{sub:naive}.
\texttt{get\_leadlog} returns the symbolic form of 
the leading-logarithmic part of the production rate, 
which corresponds to Eq.~\eqref{eq:paramratestrictLL} 
--- see Sec.~\ref{sec:strict} for more details. 
Both of those methods do not take any argument. 
We refer to the online 
documentation for other methods of the \texttt{NumRate} class.

%
\subsection{\textsc{AutoTherm} as a standalone Python package}
\label{sec:runpythonpackage}

We have just seen how the \anpart{} and 
\numpart{} components are interfaced by having the main 
component of the latter, the \texttt{NumRate} class, receive as input
three structures generated by the former \anpart{} component. 
This intentionally modular design allows the user to run 
either component independently of the other. 

Imagine for instance that one is interested in performing 
the phase-space convolutions 
to obtain the rates in a model where all needed 
matrix elements squared and thermal masses are well known: 
there would be little point, then, in taking 
the time to write a \myfr{} model file and have 
the \anpart{} component 
regenerate all matrix elements and thermal masses. 

In this case one can simply provide the expressions for 
matrix elements squared 
and masses as input to the call to \texttt{NumRate}, 
e.g. for thermal production
of the KSVZ axion~\cite{Graf:2010tv,Bouzoud:2024bom} 
--- see Sec.~\ref{sub:d5val}:
\begin{mdframed}[style=mylststyle]
\begin{lstlisting}[language=Python,style=pystyle]
from numerical.manipulate import *

# factor out the couplings in the |M|^2 for readability
# here g=g3 is the QCD gauge coupling, fa is the axion scale
# Nf is the number of light quarks
strconst = "g^6/(2*fa*pi*pi)^2"
msqdict = {(1,1,1):strconst+"*3*(s*u/t+t*u/s+s*t/u)",\
            (-1,-1,1):strconst+"*(s^2+u^2)/t*(-Nf/2)",\
            (-1,1,-1):strconst+"*(s^2+t^2)/u*(-Nf/2)",\
            (1,-1,-1):strconst+"*(u^2+t^2)/s*Nf/2"}
coupsdict = {"noneq":("fa",),"gauge":("g",),"others":("Nf",)}
masstuple = ("g^2*(1+Nf/6)",)
axionclass = NumRate(msqdict,coupsdict,masstuple,1)

\end{lstlisting}
\end{mdframed}
Here \texttt{msqdict} is a dict with $(\tau_1,\sigma_1,\sigma_2)$ keys
and the corresponding string format expressions for 
$\left\vert \mathcal{M}(\tau_1;\sigma_1,\sigma_2)\right\vert^2$ 
as entries. 
The constructor of the \texttt{NumRate} class will then parse 
these expressions through SymPy. 
\texttt{coupsdict} is a dict with 
entries in the form of Python tuples 
--- hence the \texttt{("fa",)} for one-element tuples; 
the keys correspond to the \texttt{couplings} section in 
the .json file described in App.~\ref{app:json}. 
Finally, 
\texttt{masstuple} contains expressions for the needed thermal masses. 
It must be  a tuple of the same length of 
the \texttt{gauge} coupling tuple for $\opdim=5$. 
In the $\opdim=4$ case it must instead be 
a simple string rather than a tuple, as a single mass is expected. 
See App.~\ref{app:yukawa} for more details in 
the case of generation-dependent fermionic thermal masses. 

The \texttt{rate} method of \texttt{NumRate} can then be called 
to obtain numerical values of the rate, 
as shown in Sec.~\ref{sec:runpython}. 

%
\subsection{\textsc{AutoTherm} as a standalone \textsc{Mathematica} package}
\label{sec:runmath}

The \textsc{Wolfram} language components of \myaut{} can also be used 
as a standalone \mymath{} package, in particularly
to compute the matrix element squared of a given process or 
the thermal mass of a given particle.
Contrary to the previous sections, 
the processes do not have to contain $\vartheta$.

In order to use this package, a configuration file 
(let us call it \texttt{config.cfg}) must be 
created according to App.~\ref{app:cfg}. 
\texttt{conf} should be set to \texttt{yes} therein to create 
the needed configuration files in \mymath{} format, 
which are created by executing \texttt{autotherm-run config.cfg}. 
You can similarly use the \texttt{rules} flag 
to generate Feynman rules from a model file. 

Once this is done, 
you can import the configuration and required tools:

\begin{mdframed}[style=mylststyle]
\begin{lstlisting}[language=mymath,style=mathstyle]
conffilename = "config.m";
conffile = Import[conffilename];

Get[conffile[["feynarts"]]];
Get[conffile[["formcalc"]]];
Get[FileNameJoin[Join[Drop[FileNameSplit[conffile[["formcalc"]]], -1],{"tools","VecSet.m"}]]];
Get[FileNameJoin[Join[FileNameSplit[conffile[["autothermdir"]]],{"analytical", "autotherm.wl"}]]];
\end{lstlisting}
\end{mdframed}

Then, parse the configuration file to initialize internal variables.

\begin{mdframed}[style=mylststyle]
\begin{lstlisting}[language=mymath,style=mathstyle]
ConfigParse[conffilename];
\end{lstlisting}
\end{mdframed}

You are now ready to use the \myaut{} package. 
As previously discussed, 
its two main uses are computing matrix elements squared and 
thermal masses. 
To compute the matrix element squared of a process of the form 
$ab\to cd$, where the names of $a,b,c,d$ within \myfr{} 
are respectively \texttt{X[1],X[2],X[3],X[4]}, 
run the following code:

\begin{mdframed}[style=mylststyle]
\begin{lstlisting}[language=mymath,style=mathstyle]
mat2 = ComputeMatrixElement[{X[1],X[2]}->{X[3],X[4]}]
mat2 = mat2/.{AUThast[_]:>1,AUTstatspart[__]:>1}
\end{lstlisting}
\end{mdframed}

The \texttt{ComputeMatrixElement} function returns 
the analytical expression 
for the tree-level matrix element squared of the given process, 
traced over all indices for both initial \emph{and} final states.
The  result  contains helper functions used by \myaut{}, 
as discussed in App.~\ref{app:json}. 
They can be removed using the replacement rule on the second line.
As with a full run of the \anpart{} component, 
\myaut{} will also create a \texttt{Diagrams/} directory and 
place \textsc{PostScript} files for the Feynman diagrams in it. 
In the case that a particle in \texttt{flavorexpand} appears 
as a mediator in the process,
\texttt{ComputeMatrixElement} will return a list, 
see App.~\ref{app:yukawa}. 

To compute the thermal mass of a particle \texttt{X[1]} run 
the following code: 

\begin{mdframed}[style=mylststyle]
\begin{lstlisting}[language=mymath,style=mathstyle]
mass = ThermalMass[X[1]]
\end{lstlisting}
\end{mdframed} 

The \texttt{ThermalMass} function returns the analytical expression 
for the dimensionless thermal mass squared of the particle 
--- $\mD^2/T^2$ for vector bosons, $\masym^2/T^2$ otherwise. 
If the fermionic or scalar particle has 
a generation-dependent Yukawa coupling, the result will 
be a list, as shown in App.~\ref{app:yukawa}. 

To get the thermal masses of \emph{all} particles in the thermal bath, 
the \texttt{AllMasses} function can be used:

\begin{mdframed}[style=mylststyle]
\begin{lstlisting}[language=mymath,style=mathstyle]
res = AllMasses
\end{lstlisting}
\end{mdframed}

This function returns a list of two-element lists containing 
the particle's name and its thermal mass. 
Note that \texttt{AllMasses} only computes 
the thermal masses of particles, not antiparticles. 

To get more information about the \mymath{} functions, 
you can type the function name preceded by a question mark 
(e.g. \texttt{?ConfigParse}) or refer to the documentation website.

%
\section{Validation with existing results}
\label{sec:validation}

We have  used \myaut{} to successfully reproduce and, when necessary, 
amend, a host of results from the literature 
in a broad range of models. 
We dedicate this Section to discussing their physical context 
and results. 
The corresponding model files and example jupyter notebooks 
provided with our release are summarized in table~\ref{tab:allmodels}. 
Consideration on performance and execution times 
are to be found in Sec.~\ref{sub:perf}. 

%
\begin{table}[!ht]
\begin{center}
\begin{tabular}{|C{5cm}|C{0.6cm}|C{4.2cm}|C{2.cm}|}
  \hline
  \rule{0pt}{3ex} Model & $\opdim$&File names & \texttt{includeSM}
  \\[4pt]
  \hline
  \rule{0pt}{3ex} SPSM & --  & \texttt{symmetric.fr} & -- 
  \\[4pt]
  \hline
  \rule{0pt}{3ex} Right-handed neutrino~\cite{Besak:2012qm} & 4 & \texttt{rhn.fr}, \texttt{rhn.ipynb}  & yes  
  \\[4pt]
  \hline
  \rule{0pt}{3ex} Majorana fermion DM~\cite{Biondini:2020ric}& 4 & \texttt{chiDM.fr}, \texttt{chiDM.ipynb} & yes 
  \\[4pt]
  \hline
  \rule{0pt}{3ex} QGP + photon~\cite{Baier:1991em,Kapusta:1991qp} & 4 & \texttt{QGPphoton.fr} 
  & no 
  \\[4pt]
  \hline
  \rule{0pt}{3ex} Axion~\cite{Graf:2010tv,Salvio:2013iaa,Bouzoud:2024bom} & 5 & \texttt{axion.fr}, \texttt{axion.ipynb} & yes 
  \\[4pt]
  \hline
  \rule{0pt}{3ex} SPSM + graviton~\cite{Ghiglieri:2020mhm} & 5 & \texttt{SPSM\_GW.fr}, \texttt{GWs.ipynb} & yes 
  \\[4pt]
  \hline
  \rule{0pt}{3ex} Symmetric-phase SMASH  + graviton~\cite{Ringwald:2020ist} & 5  & \texttt{SMASH\_GW.fr},  \texttt{GWs.ipynb} & yes 
  \\[4pt]
  \hline
  \rule{0pt}{3ex} SUSY-phase MSSM + graviton \& gravitino~\cite{Bolz:2000fu,Pradler:2006qh,Rychkov:2007uq,Ringwald:2020ist,gravitinopaper} 
  & 5 &  \texttt{MSSM.fr},  \texttt{MSSM.ipynb} & no 
  \\[4pt]
  \hline
  \rule{0pt}{3ex} Dark photon~\cite{Salvio:2022hfa} & 6 & \texttt{darkphoton.fr}, \texttt{darkphoton.ipynb} & yes 
  \\[4pt]
  \hline
\end{tabular}
\end{center}

\caption[a]{\small
  List of all included model files and example notebooks. 
  The former (\texttt{*.fr} files) 
  are found in the \texttt{analytical/models/} directory, 
  the latter (\texttt{*.ipynb} files) in 
  \texttt{doc/examples/} and in the online documentation. 
  SPSM stands for Symmetric-Phase Standard Model. 
  Models having \texttt{includeSM} set to yes 
  inherit from the \texttt{symmetric.fr} SPSM model file. 
}
\label{tab:allmodels}
\end{table}
%

%
\subsection{\texorpdfstring{$\opdim=4$}{d=4} models}
\label{sub:d4val}

For what concerns renormalizeable, marginal $\vartheta$-bath 
couplings we considered two models which share 
some similarities. 
The first one extends the SM with right-handed 
``sterile'' neutrinos $N$ with Majorana masses $M_N$. 
These have Yukawa couplings with the Higgs field $\phi$ and 
the lepton doublets $\ell_\mathrm{L}$ 
--- see App.~\ref{app_sm} for our notation --- 
giving mass to active neutrinos through the see-saw 
mechanism~\cite{Minkowski:1977sc,GellMann:1980vs,Yanagida:1980xy}. 
As pointed out in~\cite{Fukugita:1986hr}, baryogenesis 
via leptogenesis is a viable path for addressing the BAU
within this class of models. 
In the context of  ``ARS leptogenesis'' \cite{Akhmedov:1998qx} 
--- see also~\cite{Klaric:2020phc,Klaric:2021cpi} --- 
the dynamics 
of these right-handed neutrinos in the ultrarelativistic regime, 
before sphaleron freezeout, is very important, 
thus motivating precise determinations of their UR production rate. 
We refer to~\cite{Laine:2022pgk}
for a review on determinations of the production rate 
across all kinematical regimes. 

The complete leading-order $\Gamma_N$ for $M_N\ll T$ 
was first obtained in~\cite{Besak:2012qm} 
in the symmetric-phase SM by complementing 
the $1\leftrightarrow 2$ component from effective 
$N\leftrightarrow \phi \ell_\mathrm{L}$ processes determined 
in \cite{Anisimov:2010gy} with the $\twotwo$ component. 
Only the latter  can be determined through the current release 
of \myaut{}; these processes feature would-be IR divergences 
from $t$-channel $\ell_\mathrm{L}$ exchange, 
as shown in Fig.~\ref{fig:d4diags}.
This requires resumming their HTL. 
Our automated determination is in perfect agreement with 
the results in~\cite{Besak:2012qm}, 
as shown in the provided notebook.

These fermion-exchange $t$-channel models are also used 
to provide Majorana DM candidates not in 
the form of right-handed neutrinos 
--- see e.g.~\cite{Hall:2009bx,Garny:2018ali,Arina:2025zpi}. 
\cite{Biondini:2020ric} explored in detail the UR regime 
of a model where a Majorana fermion $\chi$ is coupled 
to an up-type right-handed quark and to a colored scalar, precisely
determining the $\Gamma_\chi$ production rate. 
Once again we can currently only determine the $\twotwo$ 
component thereof with \myaut{}. 
Our automated implementation allowed us to find a minor, 
unfortunate sign mistake in the 
numerical implementation of a part of the $s$ parametrization 
in~\cite{Biondini:2020ric}.\footnote{\label{foot_oops}%
  This is to be entirely attributed to the one element of 
  the intersection between the authors 
  of~\cite{Biondini:2020ric} and the present ones.
} 
Hence, results obtained through \myaut{} supersede 
those in~\cite{Biondini:2020ric}, though at the practical level 
the main results of that paper remain unchanged: 
in scenario ``P3'' in~\cite{Biondini:2020ric} 
--- the most sensitive to this UR $\twotwo$ component --- 
our updated numerical values for the rate change by 1\% only 
the determination of the Yukawa 
coupling yielding the observed $\Omega_\mathrm{DM}$. 

Furthermore, this model features would-be divergences mediated 
by up-type right-handed quarks; 
it is thus an ideal testing ground for our handling 
of generation-dependent contributions to fermionic 
thermal masses, as described in more detail in App.~\ref{app:yukawa}. 

We also provide a simple model file coupling QCD 
with three light flavors to the photon: 
this is the relevant Lagrangian to describe 
thermal photon production from a quark-gluon plasma. 
Though we do not provide an example notebook, 
it is easy to obtain the $\twotwo$ component 
of the photon production rate, 
as obtained in Refs.~\cite{Baier:1991em,Kapusta:1991qp}. 
This is to be complemented by the $1\leftrightarrow 2$ component, 
as derived in Refs.~\cite{Arnold:2001ba,Arnold:2001ms}.

%
\subsection{\texorpdfstring{$\opdim=5$}{d=5} models}
\label{sub:d5val}

Our provided model file \texttt{axion.fr} implements 
the most generic $\opdim=5$ Lagrangian coupling an 
axion~\cite{pecceiquinn,weinberg,wilczek}
or axion-like particle to the SM, as written in~\cite{Salvio:2013iaa}. 
If all Wilson 
coefficients except the one multiplying the axion coupling to 
the QCD topological charge
are set to zero, the results of~\cite{Bouzoud:2024bom} for 
the so-called KSVZ axion~\cite{Kim:1979if,Shifman:1979if}
are automatically obtained by \myaut{}. 
As shown in~\cite{Bouzoud:2024bom} by two of us,
the strict LO rate agrees with that determined in~\cite{Graf:2010tv}. 

We also examined  the axion $a$ coupling to the third-generation
quarks and Higgs field, 
$
  \mathcal{L}_\mathrm{int}
  \propto 
  a h_t
  \bar{Q}_{\mathrm{L}\,3} 
  u_{\mathrm{R}\,3}
  \tilde{\phi}
$
--- see App.~\ref{app_sm} for our conventions. 
Our numerical result for the integral 
$
  \int_0^\infty \mathrm{d} k\,k^2\,
  n_+^{ }(k) \Gamma_a(k)
$
turns out to be 2\% larger than what is given in Eq.~(3.20) 
of \cite{Salvio:2013iaa}: our numerical 
factor multiplying the analytical result obtained in 
the Maxwell--Boltzmann approximation reads 0.96 rather than their 0.94.
We refer to the bundled \texttt{axion.ipynb} notebook for details and 
to our companion paper~\cite{gravitinopaper} for a discussion of 
a similar small discrepancy in the 
related top Yukawa contribution to gravitino production. 

%
\begin{figure}[t]

\begin{center}
    \includegraphics[width=\linewidth]{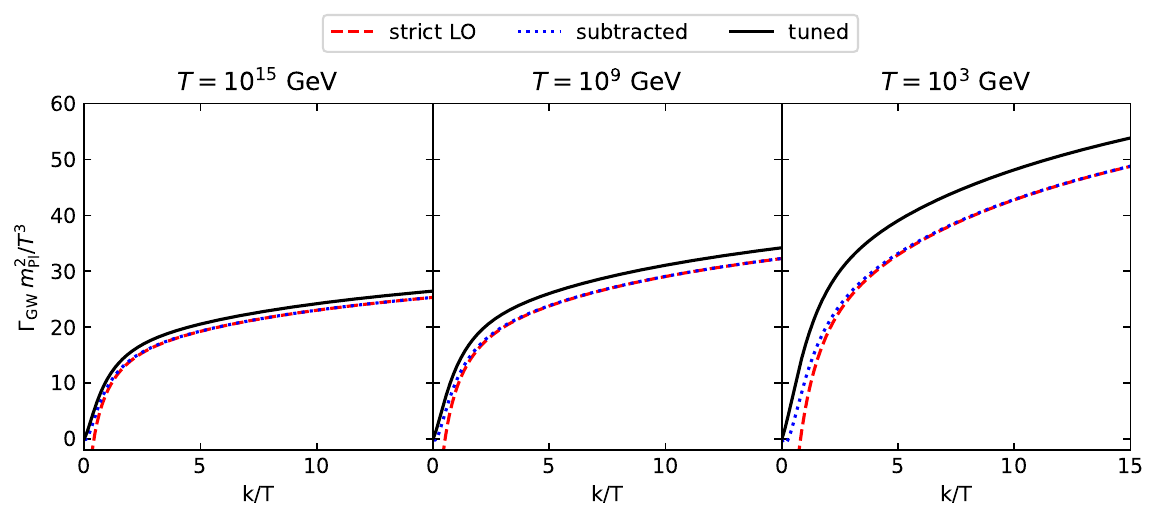}
\end{center}

\vspace{-5mm}

\caption[a]{\small
  The GW rate from a SM plasma, as in Fig.~3 (left) 
  of~\cite{Ghiglieri:2020mhm}. 
  Figure obtained from \myaut{}, see the bundled \texttt{GWs.ipynb}. 
  Our couplings run with the temperature as described 
  in~\cite{Ghiglieri:2020mhm}; we differ in 
  our value for the Debye masses for the three SM gauge groups. 
  They are implemented at LO 
  in \myaut{} and at two-loop order in~\cite{Ghiglieri:2020mhm}, 
  following~\cite{Laine:2019uua}.
}
\label{fig:gw}
\end{figure}
%

We have also successfully employed \myaut{} to reproduce LO results 
for gravitational-wave (GW) production from equilibrated plasmas. 
As pointed out in~\cite{Ghiglieri:2015nfa}, this represents an 
inevitable GW background from the radiation epoch. 
Its contribution to the current energy density of 
the universe is approximately proportional to 
$T_\mathrm{max}/m_\mathrm{Pl}$, 
where $T_\mathrm{max}$ is the maximal temperature of 
the radiation epoch and $m_\mathrm{Pl}\equiv G^{-1/2}$ 
is the Planck mass. 

In more detail, as illustrated in the accompanying notebook, 
we are able to reproduce the LO GW 
rate $\Gamma_\mathrm{GW}$ from the symmetric-phase SM, 
obtained in~\cite{Ghiglieri:2020mhm}, 
as well as that arising from a 
Standard Model--axion--seesaw--Higgs portal inflation (SMASH) 
\cite{Ballesteros:2016euj,Ballesteros:2016xej} bath, 
obtained in~\cite{Ringwald:2020ist}. 
Both references~\cite{Ghiglieri:2020mhm,Ringwald:2020ist} adopted 
the subtracted scheme only. 
In Fig.~\ref{fig:gw} we plot the obtained GW rate in the SM, 
highlighting the smaller spread between the strict LO 
and subtracted scheme than between any of those two and 
the tuned one, in agreement with 
the theoretical expectations outlined in Sec.~\ref{sub:compare}. 

Our successful reproduction of these results is 
a highly non-trivial test of \myaut{}, 
given the large number of processes and diagrams to be evaluated, 
as expected from the universal coupling of gravity. 
We also remark that \cite{Ghiglieri:2020mhm} carried out 
its LO calculation for the SM using two independent methods, 
one of them based on an early version of our 
\myfr{}--\myfa{}--\myfc{} pipeline but with manual handling of 
HTL resummation, 
thus representing a proof-of-concept of 
the viability of automation techniques.

Finally, we also provide a working implementation of 
the MSSM at temperatures much larger 
than any superpartner mass, 
so that both the electroweak symmetry and supersymmetry 
are fully realized at the Lagrangian level, 
the latter being broken only by thermal effects. 
We couple the MSSM to gravitons and gravitinos 
--- technically to its spin one-half Goldstino component --- 
and compute the production rates for these particles. 
We refer to our companion paper~\cite{gravitinopaper} for 
an extensive illustration of our results. 
Here we anticipate that our graviton rate agrees with 
the one in~\cite{Ringwald:2020ist}. 
For what concerns the gravitino, 
the gauge component of our strict LO  rate agrees 
with~\cite{Bolz:2000fu,Pradler:2006qh,Pradler:2006hh,Pradler:2006tpx}. 
Our top Yukawa contribution is in reasonable agreement 
with~\cite{Rychkov:2007uq}, 
while it disagrees with that of~\cite{Eberl:2024pxr}. 
The discrepancy likely arises from an issue in 
the derivation of Feynman rules
and matrix elements squared in~\cite{Eberl:2024pxr}. 
We further discuss why gauge-dependent, non HTL resummations, 
as carried out 
in~\cite{Rychkov:2007uq,Strumia:2010aa,Eberl:2020fml,Eberl:2024pxr}, 
give rise to a pathologically divergent rate. 
We argue that the finite numerical results 
in~\cite{Rychkov:2007uq,Strumia:2010aa,Eberl:2020fml,Eberl:2024pxr} 
must follow from 
an accidental, artificial numerical regularization of this divergence.
Finally, we provide a handy parametrization of our results in a form 
similar to that in~\cite{Ellis:2015jpg}. 
This is an excellent illustration of the 
potential of \myaut{} in automating and streamlining intricate 
calculations that are prone to 
technical and methodological blunders, 
such as the determination of the Yukawa matrix elements squared 
and the use of gauge-dependent resummations respectively.

%
\subsection{\texorpdfstring{$\opdim=6$}{d=6} models}
\label{sub:d6val}

We provide an implementation of 
the generic effective Lagrangian in~\cite{Salvio:2022hfa}, 
coupling a dark photon to the SM through dimension-six operators. 
As we show in the accompanying 
notebook, we are able to  automatically compute 
single dark-photon production. 
Furthermore, 
using preliminary features hinted at after Eq.~\eqref{eq:lintJ}, 
we can determine semi-automatically 
the double-production contribution. 

As we show in the accompanying notebook, our results, 
both for what concerns the evaluation of the matrix elements 
and for the numerical integration, are in 
agreement with those of~\cite{Salvio:2022hfa}.\footnote{%
  In the case of single production, 
  our result for the $W$-Higgs scattering contribution 
  is a factor of 3/4 smaller than Eq.~(3.19) 
  in Ref.~\cite{Salvio:2022hfa}. 
  This apparent discrepancy can be understood 
  by Eq.~(3.19) implicitly defining a somewhat non-standard
  normalization for the SU(2) generators and gauge coupling 
  that enter in the $W^{\mu\nu}$ and $\tilde{W}^{\mu\nu}$ 
  SU(2) field-strength tensors in the Lagrangian in Eq.~(2.1) 
  in~\cite{Salvio:2022hfa}. 
  These are shorthand  for 
  $W^{\mu\nu\,a}T^a$ and $\tilde{W}^{\mu\nu\,a}T^a$ respectively, 
  with $a$ the adjoint SU(2) index and $T^a$
  the fundamental-representation generator. 
  We choose the standard form $T^a=\sigma^a/2$, 
  with $\sigma^a$ a Pauli matrix. 
  Equation~(3.19) in~\cite{Salvio:2022hfa} instead implicitly 
  fixes the generators such that 
  $\mathrm{Tr}[T^a T^a]=\mathrm{Tr}[\mathbb{I}]$, 
  while ours implies 
  $\mathrm{Tr}[T^a T^a]=3/4\mathrm{Tr}[\mathbb{I}]$ for SU(2). 
  A similar rescaling is then needed for the gauge coupling as well. 
  The SU(2) Yang-Mills term in the SM Lagrangian, 
  not given explicitly in~\cite{Salvio:2022hfa}, 
  would then differ from the 
  $-1/2 \, \mathrm{Tr}[W_{\mu\nu}W^{\mu\nu}]$ implemented in our 
  SPSM model file.
  We are grateful to
  Alberto Salvio for communications on this matter.
}

%
\subsection{Performance}
\label{sub:perf}

The execution times of the \anpart{} component 
are largely model dependent. In cases 
where $\vartheta$ has a small set of couplings to the bath, 
such as in the right-handed neutrino or axion models, its execution, 
from Feynman-rule derivation to the determination of 
$
  \left\vert \mathcal{M}(\tau_1;\sigma_1,\sigma_2)\right\vert^2
$ 
and of the needed  thermal masses takes 
about 20 seconds on an Apple M3 Pro MacBook. 
Models with more $\vartheta$-bath couplings and/or needing 
our ad-hoc implementations of tensor polarization and SU(2) algebra 
--- see App.~\ref{app:fafc} --- do require longer execution times, 
with the record holder being graviton production in the MSSM, 
where the \anpart{} component takes 
about 8 minutes on the same machine. 
Note that \myfr{} takes about one minute 
to generate the supersymmetric Lagrangian and its Feynman rules from 
the definition of the superfields. 
On an 11th-generation Intel i5 Linux machine 
all execution times are approximately twice as long. 

The \numpart{} component is in general quite rapid. 
Creating an instance 
of the \texttt{NumRate} class take $\mathcal{O}$(s) on 
an Apple M3 Pro MacBook, 
due to the symbolic work in projecting 
$
  \left\vert \mathcal{M}(\tau_1;\sigma_1,\sigma_2)\right\vert^2
$
onto the appropriate basis. 
On that machine, calling the \texttt{rate} method to 
generate Fig.~\ref{fig:gw} took 14s using a single core. 
It required evaluating the 
rate in all schemes at three different values of 
the couplings for $k/T$ on a geometric progression 
of 200  values,  $0.01<k/T< 15$. 
On that same grid, the four-dimensional fallback integration 
took approximately 120s using six cores to evaluate 
the integrand at multiple points in parallel.

%
\section{Conclusions and outlook}
\label{sec:concl}

In this paper we presented the first release of \myaut{}, 
a new program geared towards 
the automated determination of the $\twotwo$ component of 
the thermal production rate $\Gamma_\vartheta$ 
at leading order in the ultrarelativistic regime $T\gg M$, 
where $M$ labels the mass of any particle in the process. 
This release can thus be used for any model where 
a light $\vartheta$ species is coupled to a thermal bath via 
operators of $\opdim\ge 4$ mass dimension. 

The main strength of our approach is the consistent, 
automated implementation of 
collective effects through Hard Thermal Loop resummation. 
This is an integral part of the LO rate whenever a massless 
particle mediates a $t$-channel $\twotwo$ process of 
the form $ab\to c\vartheta$, with $a,b,c$ equilibrated particles. 
As we discuss at length in Sec.~\ref{sec:soft}, 
this relies on the known universality of HTLs, 
which factorize in a  thermal mass times a kinematical
function of the energy over momentum of the exchanged particle. 
While the thermal mass differs for each thermal particle, 
the kinematical functions depend only on spin; 
as we show, we only need the  fermionic and gauge-boson cases. 

At the practical level, 
the $\twotwo$ rate is at naive leading order a convolution 
of tree-level matrix elements squared for all possible 
$ab\to c\vartheta$ processes with the 
statistical functions of the thermalized $a,b,c$ particles. 
In a nutshell, \myaut{} takes as input the Lagrangian of 
the model in the form of a \myfr{} model file; 
Feynman rules are then derived by this tool. 
Our routines then use \myfa{} and \myfc{} to compute 
all tree-level matrix elements squared. 
If a sensitivity to collective effects is present 
for the model being studied, 
it shows up as $\propto 1/t$ and $\propto 1/u$ terms 
in these matrix elements squared. 
In these cases, we exploit the aforementioned universality 
to perform HTL resummation analytically. 
To this end, we describe in Sec.~\ref{sec:masses}
how the thermal masses for the particles mediating 
the $\propto 1/t$ and $\propto 1/u$ terms 
can be determined automatically. 
This is achieved by exploiting the cutting rules of 
Thermal Field Theory to reduce the evaluation of 
thermal loop diagrams to a convolution of tree-level in-vacuum
$\twotwo$ amplitudes and statistical factors. 

Another key ingredient in our automation algorithms for 
HTL resummation is provided 
by analyticity-based \emph{light-cone techniques}
~\cite{Aurenche:2002pd,CaronHuot:2008ni,Besak:2012qm,Ghiglieri:2013gia},
which allow compact, closed-form expressions for 
the HTL-resummed soft-mediator contribution 
to the rate $\Gamma_\vartheta$. 
This furthermore allows us to provide three LO-equivalent 
implementations of HTL resummation. 
As their difference can be of relative order $g$, 
which labels the gauge and Yukawa couplings between 
the equilibrated particles, 
their spread represents a reliable proxy for 
the theory uncertainty stemming from 
unknown, model-dependent higher-order corrections. 
We refer to~\cite{Jackson:2019tnr,Bouzoud:2024bom,Bouzoud:2026rur} 
for models where such potentially large corrections 
have been determined 
and to~\cite{Becker:2023vwd,Becker:2025yvb,Becker:2025lkc} 
for an alternative, 1PI-based approach 
to resummation and its comparison with 
the HTL-based schemes we provide.

In Secs.~\ref{sec:struct}--\ref{sec:tuto} we provide an overview of 
the structure and usage of \myaut{}. 
As we show, our \textsc{Mathematica}-based \anpart{} routines 
and Python-based \numpart{} routines fully automate 
the procedure we sketched. 
Each component can be used independently to perform specific tasks 
--- such as evaluating matrix elements and thermal masses, 
or calculating rates from known 
matrix element squared and thermal masses. 
Double-production of $\vartheta$ or of $\vartheta$ and 
its conjugate are also supported in a semi-automated fashion. 

Sec.~\ref{sec:validation} is then dedicated to applying \myaut{} 
to a multitude of results from the literature; 
the relevant model files and example notebooks 
are bundled with this release.
As we show, these results can be successfully reproduced 
--- and in a few cases corrected --- by this release. 
The case of gravitino production is handled 
in extensive detail in our companion paper~\cite{gravitinopaper}. 
Due to the large number of equilibrated particles and 
their many couplings in the MSSM at very high temperatures, 
this is a tedious and very error-prone calculation. 
Our companion paper showcases the strengths of \myaut{} 
and shows how the one-loop gauge-dependent 
resummation scheme introduced in~\cite{Rychkov:2007uq} 
is incompatible with non-abelian gauge theories and 
gives rise to a pathologically divergent rate. 

Our first release is then complementary to existing tools
such as micrOMEGAs~\cite{Belanger:2018ccd}, 
which is mostly geared towards \emph{massive} relics. 
Would-be IR divergences from massless 
$t$-channel mediators are handled in its latest 
release~\cite{Alguero:2023zol} only 
and through a phase-space cut. 

In future releases we plan to, first, automate the effective, 
LPM-resummed $1\leftrightarrow 2$ processes that contribute to 
the leading-order rate for $\opdim=4$ models in the UR regime. 
In a later stage we also plan 
to extend our framework to the $M\gtrsim T$ regime with full 
accounting of thermal effects at NLO: 
in models where the leading order is given, e.g., 
by a $1\to 2$ decay, NLO corrections arise from \emph{real} $\twotwo$ 
and $1\leftrightarrow3$ processes and \emph{virtual}, 
thermal correction to the $1\to 2$ process. 
The latter cannot be determined by in-vacuum automation tools, 
but they are known to cancel
would-be soft and/or collinear divergences from the real processes. 
The method 
proposed in~\cite{Jackson:2021dza} exploits this fact 
to extract virtual processes from the real amplitudes.

%
\section*{Acknowledgements}

We thank Mikko Laine and Alberto Salvio for useful discussions. 
J.G.\ and G.J.\ are partly funded by the Agence 
Nationale de la Recherche under grant ANR-22-CE31-0018 (AUTOTHERM). 

%
\appendix

\gdef\thesection{\Alph{section}} 
\makeatletter
\renewcommand\@seccntformat[1]{\appendixname\ \csname the#1\endcsname.\hspace{0.5em}}

\makeatother

%
\section{Phase space integrals}
\label{sec:phaseint}

The phase  space for $2\leftrightarrow 2$ processes is, 
using the notation of Figure \ref{fig:kinematics}
\begin{equation}
  \int{\rm d}\Omega_{2\to2}\equiv
  \int\frac{{\rm d}^3\mathbf{p_1}}{(2\pi)^3 2p_1}
  \int\frac{{\rm d}^3\mathbf{p_2}}{(2\pi)^3 2p_2}
  \int\frac{{\rm d}^3\mathbf{k_1}}{(2\pi)^3 2k_1}
  (2\pi)^4 \delta^{(4)}
  (\mathcal{P}_1+\mathcal{P}_2-\mathcal{K}_1-\mathcal{K}).
  \label{eq:phaselem}
\end{equation}
This  integral can be reduced to a four-dimensional one 
using the $\delta$ function and rotational invariance. 
In the \emph{$t$ parametrization}, 
the four remaining integration variables 
are handled by letting 
$
  \mathcal{Q}
  \equiv 
  \mathcal{P}_1-\mathcal{K}_1
  =
  \mathcal{K}-\mathcal{P}_2
  =
  (q_0,\mathbf{q})
$.
Then it follows that $t=q_0^2-q^2$ (where $q=\vert\mathbf{q}\vert$),
\begin{align}
  s = & 
  - \frac{t}{2q^2}
  \bigg[
    (2k{-}q_0)(2p_1{-}q_0)+q^2 
    - \cos(\varphi)
    \sqrt{(2k{-}q_0)^2{-}q^2}\sqrt{(2p_1{-}q_0)^2{-}q^2}
  \bigg]
  \; ,
  \label{stpar}
\end{align}
where $\varphi$ is the azimuthal angle between the $3$-vectors 
$\mathbf{p_1}$ and $\mathbf{k}$. 
$u$ can be obtained through the relation $u=-t-s$.
We then have
\begin{equation}
  \gamma^{t(\opdim)}_{\tau_1;\sigma_1,\sigma_2}
  \; = \;
  \frac{\tau_1}{(4\pi)^3 k}
  \int_{-\infty}^k {\rm d}q_0
  \int_{|q_0|}^{2k-q_0}{\rm d}q
  \int_{\qp}^\infty {\rm d}p_1
  \int_{-\pi}^{\pi}\frac{{\rm d}\varphi}{2\pi}
  \mathcal{N}_{\tau_1;\sigma_1,\sigma_2}\,f(s,t,u)
  \;,
  \label{eq:gammatstart}
\end{equation}
where we introduced
\begin{equation}
  \qpm\equiv\frac{q_0\pm q}{2}
  \; .
  \label{eq:qpqm}
\end{equation}

This parametrization is inconvenient wherever $s$ is at 
the denominator, due to $\cos(\varphi)$ in Eq.~\eqref{stpar}.
For this reason, we introduce  the \emph{$s$ parametrization}, 
whereby we let 
$\mathcal{Q}=\mathcal{P}_1+\mathcal{P}_2=\mathcal{K}_1+\mathcal{K}$.
This means $s=q_0^2-q^2$,
\begin{align}
  t = & 
  \frac{s}{2q^2} 
  \bigg[
    (2k{-}q_0)(2p_2{-}q_0)-q^2
    +\cos(\varphi)
    \sqrt{q^2{-}(2k{-}q_0)^2}\sqrt{q^2{-}(2p_2{-}q_0)^2}
  \bigg]
  \; . 
\end{align}
where $\varphi$ is now the azimuthal angle between 
$\mathbf{p_2}$ and $\mathbf{k}$. 
We can then write 
\begin{equation}
  \gamma^{s(\opdim)}_{\tau_1;\sigma_1,\sigma_2}
  \; = \;
  \frac{\tau_1}{(4\pi)^3 k}
  \int_k^{\infty} {\rm d}q_0 
  \int_{|2k-q_0|}^{q_0}{\rm d}q
  \int_{\qm}^{\qp} {\rm d}p_2
  \int_{-\pi}^{\pi}\frac{{\rm d}\varphi}{2\pi}
  \mathcal{N}_{\tau_1;\sigma_1,\sigma_2}\,g(s,t,u)
  \; .
  \label{eq:gammasstart}
\end{equation}

As it turns out, for the bases chosen in Eqs.~\eqref{eq:basis4} 
and \eqref{eq:basis5} the $p_{i=1,2}$ and $\varphi$ 
integrals in Eqs.~\eqref{eq:gammatstart} and \eqref{eq:gammasstart} 
can be performed analytically. 
This requires two identities following from the explicit form of 
the Bose--Einstein and Fermi--Dirac distributions, i.e. 
\begin{align}
 \mathcal{N}^{ }_{\tau_1;\sigma_1,\sigma_2} 
 \; = \; & 
 \bigl[
  1 + n^{ }_{\tau_1\sigma_1}(q^{ }_0) 
    + n^{ }_{\sigma_2}(k - q^{ }_0) 
 \bigr]
 \bigl[
  n^{ }_{\tau_1}(p^{ }_1 - q^{ }_0) - n^{ }_{\sigma_1}(p^{ }_1) 
 \bigr]\label{calnt} \\[2mm]
 \; = \; & 
 \bigl[
  1 + n^{ }_{\sigma_1}(q^{ }_0 - p^{ }_2) 
    + n^{ }_{\sigma_2}(p^{ }_2) 
 \bigr]
 \bigl[
  n^{ }_{\tau_1}(q^{ }_0 - k) - n^{ }_{\sigma_1\sigma_2}(q^{ }_0) 
 \bigr]
 \;.
 \label{calns}
\end{align}
The first (second) identity is employed for $t$ ($s$) parametrization.
Hence, the computation of $\gamma^t$ and $\gamma^s$ is reduced 
to a double numerical integral, on $q_0$ and $q$. 
For the fallback integration, 
all four integrals are computed numerically. 

We can thus write for the $\opdim=4$ and $\opdim=5$ rates
\begin{align}
  \gamma^{t(\opdim)}_{\tau_1;\sigma_1,\sigma_2}
  \; = \;
  \frac{\tau_1}{(4\pi)^3 k}
  \int_{-\infty}^k {\rm d}q_0\int_{|q_0|}^{2k-q_0}{\rm d}q\, 
  {\rm I}^{t(\opdim)}_{\tau_1;\sigma_1,\sigma_2}(\tilde{a}_i)
  \; ,
  \nonumber \\
  \gamma^{s(\opdim)}_{\tau_1;\sigma_1,\sigma_2}
  \; = \;
  \frac{\tau_1}{(4\pi)^3 k}
  \int_k^{\infty} {\rm d}q_0 \int_{|2k-q_0|}^{q_0}{\rm d}q\,
  {\rm I}^{s(\opdim)}_{\tau_1;\sigma_1,\sigma_2}(\tilde{b}_i)
  \; ,
  \label{eq:gammatsbase}
\end{align}
where the functions ${\rm I}^{t(\opdim)}$ and ${\rm I}^{s(\opdim)}$ 
will be presented in Apps.~\ref{sec_dim4}--\ref{sec_dim5}. 
For their numerical integration, it is convenient to switch to 
$(\qp,\qm)$ variables to have independent integration ranges, i.e. 
\begin{align}
  & \int_{-\infty}^k {\rm d}q_0\int_{|q_0|}^{2k-q_0}{\rm d}q 
  \; = \; 
  2\int_{-\infty}^{0}{\rm d}\qm\int_{0}^{k}{\rm d}\qp 
  \; , 
  \nonumber \\[1mm]
  & \int_k^{\infty} {\rm d}q_0 \int_{|2k-q_0|}^{q_0}{\rm d}q 
  \; = \; 
  2\int_{0}^{k}{\rm d}\qm\int_{k}^{+\infty}{\rm d}\qp 
  \; ,
\end{align}
where the factor of two is the Jacobian from the variable change.

Before we list the results for $\opdim=4$ and $\opdim=5$, 
it is useful to introduce 
\begin{align}
    &L_n\equiv T^n\Bigl[{\rm Li}_n\Bigl(\tau_1 { e}^{\qm/T}\Bigr)
    -{\rm Li}_n\Bigl(\sigma_1 { e}^{-\qp/T}\Bigr)\Bigr],\nonumber
    \\
    &L^{\pm}_{n}\equiv T^n\Bigl[{\rm Li}_n\Bigl(\sigma_2 { e}^{-\qpm/T}\Bigr)
    +(-1)^n{\rm Li}_n\Bigl(\sigma_1 { e}^{-\qmp/T}\Bigr)\Bigr].
\end{align}
We recall that ${\rm Li}_1(x)=-\ln(1-x)$.

%
\subsection{Dimension-four operator}
\label{sec_dim4}

From Eqs.~\eqref{eq:prodrate}, 
\eqref{eq:basis4}, 
\eqref{eq:paramrate}, and 
\eqref{eq:gammatsbase} we then have
\begin{align}
  {\rm I}^{t(4)}_{\tau_1;\sigma_1,\sigma_2}
  (\tilde{a}_1,\tilde{a}_2)
  & \; = \;
  \int_{\qp}^\infty{\rm d}p_1
  \int_{-\pi}^{\pi}\frac{{\rm d}\varphi}{2\pi}
  \mathcal{N}_{\tau_1;\sigma_1,\sigma_2}
  \left[ \tilde{a}_1\frac{s}{t} + \tilde{a}_2 \right]
  \; ,
  \nonumber \\[1mm]
  {\rm I}^{s(4)}_{\tau_1;\sigma_1,\sigma_2}(\tilde{b}_1)
  & \; = \;
  \int_{\qm}^{\qp} {\rm d}p^{ }_2
  \int_{-\pi}^{\pi}\frac{{\rm d}\varphi}{2\pi}
  \mathcal{N}_{\tau_1;\sigma_1,\sigma_2}\;
  \tilde{b}_1\frac{u}{s}
  \; .
  \label{eq:idim4}
\end{align}

Using the results of \cite{Besak:2012qm,Ghiglieri:2016xye}, 
the integrals in Eq.~\eqref{eq:idim4} can be computed analytically. 
This gives for the $t$ parametrization 
\begin{align}
  {\rm I}^{t(4)}_{\tau_1;\sigma_1,\sigma_2}
  (\tilde{a}_1,\tilde{a}_2)
  & \; = \; 
  \bigl[1+n_{\tau_1\sigma_1}(q_0)+n_{\sigma_2}(k-q_0)\bigr]
  \nonumber \\[2mm]
  & \; \times \;
  \biggl[
    \biggl(-\tilde{a}_1\frac{k-\qm}{q}+\tilde{a}_2\biggr)L_1
    \, + \,
    \tilde{a}_1\frac{-2k+q_0}{q^2}L_2
  \biggr]
  \; ,
\end{align}
and for the $s$ parametrization
\begin{align}
  {\rm I}^{s(4)}_{\tau_1;\sigma_1,\sigma_2}(\tilde{b}_1)
  & \; = \;
  \tilde{b}_1\,
  \bigl[n_{\tau_1}(q_0-k)-n_{\sigma_1\sigma_2}(q_0)\bigr]
  \nonumber \\[2mm]
  & \; \times \; 
  \biggl[
    -\frac{q}{2}+\frac{k-\qm}{q}L_1^{+}+\frac{k-\qp}{q}L_1^{-}
    +\frac{2k-q_0}{q^2}L_2^{+}+\frac{-2k+q_0}{q^2}L_2^{-} 
  \biggr]
  \; .
\end{align}

%
\subsection{Dimension-five operator}
\label{sec_dim5}

In this case, Eqs.~\eqref{eq:prodrate}, 
\eqref{eq:basis5}, 
\eqref{eq:paramrate} and 
\eqref{eq:gammatsbase} yield
\begin{align}
  & {\rm I}^{t(5)}_{\tau_1;\sigma_1,\sigma_2}
  (\tilde{a}_1,\tilde{a}_2)
  \; = \;
  \int_{\qp}^\infty {\rm d}p_1
  \int_{-\pi}^{\pi}\frac{{\rm d}\varphi}{2\pi}
  \mathcal{N}_{\tau_1;\sigma_1,\sigma_2}
  \left[\tilde{a}_1\frac{s^2+u^2}{t}+\tilde{a}_2t\right]
  \; , 
  \nonumber \\[1mm]
  & {\rm I}^{s(5)}_{\tau_1;\sigma_1,\sigma_2}
  (\tilde{b}_1,\tilde{b}_2)
  \; = \;
  \int_{\qm}^{\qp} {\rm d}p_2
  \int_{-\pi}^{\pi} \frac{{\rm d}\varphi}{2\pi}
  \mathcal{N}_{\tau_1;\sigma_1,\sigma_2}
  \left[\tilde{b}_1\frac{t^2}{s}+\tilde{b}_2s\right]
  \; .
  \label{eq:idim5}
\end{align}

The integrals can be evaluated analytically as in App.~\ref{sec_dim4}.
Using the results from \cite{Ghiglieri:2020mhm}, 
we get for the $t$ parametrization
\begin{align}
  {\rm I}^{t(5)}_{\tau_1;\sigma_1,\sigma_2}
  (\tilde{a}_1,\tilde{a}_2)
  & \; = \;
  \bigl[1+n_{\tau_1\sigma_1}(q_0)+n_{\sigma_2}(k-q_0)\bigr]
  (q^2 - q_0^2)
  \biggl\{
    -\biggl(\tilde{a}_2+\frac{2 \tilde{a}_1}{3}\biggr)L_1
    \nonumber \\[1mm]
  & \hspace{1cm}
  + \frac{
      \tilde{a}_1[q^2-3(q_0-2k)^2][12 L_3+6q L_2+q^2 L_1 ]
    }{
      6 q^4
    }
  \biggr\}
  \; ,
\end{align}
and for the $s$ parametrization
\begin{align}
  {\rm I}^{s(5)}_{\tau_1;\sigma_1,\sigma_2}
  (\tilde{b}_1,\tilde{b}_2)
  & \; = \;
  \bigl[ 
    n_{\tau_1}(q_0-k)-n_{\sigma_1\sigma_2}(q_0)
  \bigr](q^2-q_0^2)
  \nonumber \\[2mm]
  & \; \times \;
  \biggl\{
    \frac{
    \tilde{b}_1 [q^2{-}3(q_0{-}2k)^2]
    [  12(L^{-}_3{-}L^{+}_3)
    {-}6q(L^{-}_2{+}L^{+}_2)
    {+}q^2(L^{-}_1{-}L^{+}_1) ]
    }{
      12 q^4
    }
    \nonumber \; \\ 
    & \hspace{8mm}
    - \frac{
        \tilde{b}_1(q_0{-}2k)
        [2(L^{-}_2{-}L^{+}_2)-q(L^{-}_1{+}L^{+}_1)]
      }{
        2 q^2
      }
      \nonumber \\
    & \hspace{8mm}
    - \biggl(
        \frac{\tilde{b}_1}{3}+\tilde{b}_2
      \biggr)
      (L^{-}_1{-}L^{+}_1+q)
    \biggr\}
    \; .
\end{align}

%
\subsection{Fallback integration}
\label{sec_fallback}

This method performs the four-dimensional integrations in 
Eqs.~\eqref{eq:gammatstart} and \eqref{eq:gammasstart} numerically. 
In the case of $\opdim=4,5$ models with would-be divergent matrix 
elements squared $\propto 1/t$, 
the tuned scheme of Sec.~\ref{sec:tuned} is used. 
In all other cases, 
i.e. when the matrix element squared is at largest a constant 
for $t\to 0$, 
the $s$ parametrization of Eq.~\eqref{eq:gammasstart} is used and 
the matrix elements squared are decomposed as
\begin{equation}
  \label{eq:fallback}
  \left\vert\mathcal{M}\right\vert^2
  \; = \; 
  \sum_i c_i s^{m_i} t^{l_i}
  \,,
\end{equation}
with $m_i+l_i=\opdim -4$  and $l_i\ge0$.

In summary, this method represents a numerical cross-check of 
the two-dimensional numerics for the 
tuned scheme~\eqref{eq:paramratetuned} in 
$\opdim=4,5$ models with would-be divergent matrix elements squared. 
In $\opdim=4,5$ 
without would-be divergences it provides an 
independent implementation of the LO-correct naive rate 
in Eq.~\eqref{eq:paramrate}. 
For models with $\opdim>5$ and IR-finite matrix elements squared, 
it becomes our only 
implementation of the LO-accurate naive rate. 

%
\section{Hard Thermal Loop resummation}
\label{sec:htl}

In this Appendix we provide further technical details 
backing up results presented in Sec.~\ref{sec:soft}. 

%
\subsection{HTL propagators}
\label{sec:htlprops}

We start by listing the 
expressions for the  retarded HTL propagators of gauge bosons 
and fermions. 
For what concerns the former, as we shall 
make use of the techniques provided in~\cite{Ghiglieri:2015ala}, 
we follow their choice and adopt the Coulomb gauge. 
For the \textit{longitudinal} propagator, we then have 
\begin{align}
  & 
  G^L_R(\mathcal{Q})
  \; \equiv \;
  G^{00}_R(\mathcal{Q})
  \; = \;
  \frac{i}{q^2+\Pi_R^L}
  \,,\quad\text{with}\quad
  \Pi^L_R
  \; = \; 
  \mD^2
  \left[
    1-\frac{q_0}{2q}\ln\left(\frac{q_0+q}{q_0-q}\right)
  \right]
  \; .
  \label{htlL}
\end{align}
For the \textit{transverse} propagator, we have
\begin{align}
  G^{ij}_R(\mathcal{Q})
  \; \equiv \; & 
  \biggl(\delta_{ij}-\frac{q_iq_j}{q^2}\biggr)G^T_R(\mathcal{Q})
  \;, \quad\text{where}\quad
  G^T_R(\mathcal{Q})=\frac{i}{q_0^2-q^2-\Pi^T_R}
  \; ,\nonumber \\[1mm]
  \Pi^T_R 
  \; = \; &
  \frac{\mD^2}{2}
  \left[
    \frac{q_0^2}{q^2}
    -
    \left(\frac{q_0^2}{q^2}-1\right)
    \frac{q_0}{2q}\ln\left(\frac{q_0+q}{q_0-q}\right)
  \right]
  \; .
  \label{htlT}
\end{align}

For fermions, we  decompose the propagator in terms of 
the positive and negative helicity-to-chirality components 
\begin{align}
  S_R(\mathcal{Q})
  \; \equiv \; &
    h_{\mathbf{q}}^{+}S_R^{+}(\mathcal{Q})
  + h_{\mathbf{q}}^{-}S_R^{-}(\mathcal{Q})\,,
  \quad\text{where}\quad
  S_R^{\pm}(\mathcal{Q})
  \; = \;
  \frac{i}{q_0\mp\left(q+\Sigma^{\pm}_R\right)}
  \,,
  \nonumber \\[1mm]
  \Sigma^{\pm}_R 
  \; = \; &
  \frac{\masym^2}{2q}
  \left[
    1-\frac{q_0\mp q}{2q}\ln\left(\frac{q_0+q}{q_0-q}\right)
  \right]
  \; ,
  \label{htlS}
\end{align}
where 
$
  h^{\pm}_{\mathbf{q}}
  =
  \left(\gamma^0\mp\bm{\gamma}\cdot\mathbf{\hat{q}}\right)/2
$ 
are the helicity projectors.

In all those expressions, $q_0$ should be understood 
to be $q_0+i\varepsilon$ where $1\gg\varepsilon>0$. 
The advanced propagators can be obtained 
by instead substituting $q_0$ for $q_0-i\varepsilon$.

%
\subsection{\texorpdfstring{$\opdim=5$}{d=5} soft exchange}
\label{sec:dim5app}

Our starting point is  Eq.~\eqref{eq:htlcontrstart}. 
By taking the $t$ parametrization it turns into
\begin{align}
  \Gamma^{(5)\text{ HTL}}_\vartheta
  \; = \; &\, 
  \frac{1}{4(4\pi)^3d_\vartheta k^2}\sum_{\tau_1=\pm 1}
  \tau_1\,C^{(5)}_{\tau_1} 
  \int_{-\infty}^k {\rm d}q_0
  \int_{|q_0|}^{2k-q_0}{\rm d}q
  \int_{\qp}^\infty {\rm d}p
  \int_{-\pi}^{\pi}\frac{{\rm d}\varphi}{2\pi}
  \nonumber \\[1mm]
  & \; \times \; 
  \frac{
    n_{\tau_1}(p)n_{\tau}(k{-}q_0){\bar n}_{\tau_1}(p{-}q_0)
  }{
    n_\tau(k)
  }\,
  \left\vert \mathcal{M}^{(5)}_\mathrm{HTL} \right\vert^2
  \; .
  \label{eq:htlcontrstartt}
\end{align}
Plugging the propagators in Eqs.~\eqref{htlL} and \eqref{htlT} 
in the matrix element squared~\eqref{msqdim5} and expanding for 
$\mathcal{P},\mathcal{K}\gg\mathcal{Q}$, 
Eq.~\eqref{eq:htlcontrstartt} becomes, 
\begin{align}
  \Gamma^{(5)\text{ HTL}}_\vartheta
  \; = \; & 
  \frac{g^2 g_\vartheta^2}{16\pi^3d_\vartheta }
  \sum_{\tau_1=\pm 1} C_{\tau_1}
  \int_{-\infty}^k {\rm d}q_0
  \int_{|q_0|}^{2k-q_0}{\rm d}q
  \underbrace{
    \int_{0}^\infty {\rm d}p\, p^2 f_{\tau_1}(p){\bar n}_{\tau_1}(p)
  }_{\equiv I_{\tau_1}}
  \nonumber \\
  & \; \times \; 
  (q^2-q_0^2)
  \left(
    \left\vert G_L^R(\mathcal{Q})\right\vert^2
    +
    \frac{1}{2}\left(1-\frac{q_0^2}{q^2}\right)^2
    \left\vert G_T^R(\mathcal{Q})\right\vert^2
  \right)
  \; ,
  \label{eq:htlcontrdmid}
\end{align}
where we have used the softness of $\mathcal{Q}$ to extend 
the $p$ integration range down to $0$. 
With a slight abuse of notation, 
we dropped this and other subleading terms in 
the $\mathcal{Q}\ll \mathcal{K},\mathcal{P}$ expansion 
in going from Eq.~\eqref{eq:htlcontrstartt} to \eqref{eq:htlcontrdmid}.
The $p$ integration yields
\begin{align}
  \label{eq:IandJ}
  I_{\tau_1}
  \; \stackrel{\mathrm{IBP}}{=} \; 
  2T \int_{0}^\infty {\rm d}p\, p\, f_{\tau_1}(p)
  \; \equiv \;
  2T J_{\tau_1}\, ,
  \quad\text{where}\quad 
  J_{\tau_1}
  \; = \;
  \bigl\vert2^{\tau_1}-1\bigr\vert\frac{\pi^2 T^2}{6}
  \; ,
\end{align}
where for future convenience we used integration by parts (IBP) 
to introduce the $J_{\tau}$ function.

The explicit forms of the retarded propagators 
in Eqs.~\eqref{htlL}--\eqref{htlT} 
can be used to rewrite the second line in Eq.~\eqref{eq:htlcontrdmid} 
in terms of a spectral function $\rho^i\equiv G_R^i-G^i_A$. 
One then finds, see e.g.~\cite{Ghiglieri:2015ala} 
\begin{align}
  \left \vert G^L_R(\mathcal{Q})\right\vert^2
  \; = \;
  \frac{1}{\pi\mD^2}\frac{q}{q_0}\rho_L(\mathcal{Q})
  \; , \qquad
  \left \vert G^T_R(\mathcal{Q})\right\vert^2
  \; = \;
  \frac{2}{\pi\mD^2}\, 
  \frac{q}{q_0}\,
  \frac{q^2}{q^2-q_0^2}\, 
  \rho_T(\mathcal{Q})
  \; .
   \label{eq:propspecbos}
\end{align}
Plugging these expressions in Eq.~\eqref{eq:htlcontrdmid}, we obtain
\begin{align}
  \Gamma^{(5)\text{ HTL}}_\vartheta
  & \; = \; 
  \frac{ 
    g^2 g_\vartheta^2
  }{
    (2\pi)^4d_\vartheta\mD^2
  }
  \sum_{\tau_1=\pm 1} C_{\tau_1} I_{\tau_1}
  \int_{-\infty}^k \frac{{\rm d}q_0}{q_0}
  \int_{|q_0|}^{2k-q_0}{\rm d}q\,q\,
  (q^2-q_0^2)
  \nonumber \\
  & \; \times \; 
  \left[
    \rho_L(\mathcal{Q})
    +
    \left(1-\frac{q_0^2}{q^2}\right)
    \rho_T(\mathcal{Q})
  \right]
  \; .
  \label{eq:subtrstart}
\end{align}

This integral is analytically doable in 
the soft $k\gg q > \vert q^0 \vert$ limit: 
by changing variables from $q$ to $\qperp=\sqrt{q^2-q_0^2}$
and taking d$q_0$ to be the inner integration, 
its range can be stretched up to $+\infty$ and 
the $\qperp$ integration range up to $2k\,$,
committing  a subleading error in the soft limit. 
We can then use the light-cone sum rules derived 
in \cite{Aurenche:2002pd,CaronHuot:2008ni} 
and applied to $\opdim=5$ operators in 
\cite{Ghiglieri:2015nfa,Ghiglieri:2020mhm,Bouzoud:2024bom}.
This leads to
\begin{align}
  \Gamma^{(5)\text{ HTL}}_\vartheta
  & \; = \;
  \frac{ g^2 g_\vartheta^2 }{ (2\pi)^4d_\vartheta\mD^2 } 
  \sum_{\tau_1=\pm 1}C_{\tau_1} I_{\tau_1}
  \int_{0}^{2k}{\rm d}\qperp\,\qperp^3
  \nonumber \\[1mm]
  & \hspace{2cm} 
  \times
  \int_{-\infty}^{+\infty} 
  \frac{{\rm d}q_0}{q_0}
  \biggl[ 
    \rho_L(q_0,\qperp)
    +
    \frac{\qperp^2}{q^2}\rho_T(q_0,\qperp)
  \biggr]
  \nonumber \\[1mm]
  & \; = \; 
  \frac{ g^2 g_\vartheta^2 }{ (2\pi)^3d_\vartheta\mD^2 } 
  \sum_{\tau_1=\pm 1}C_{\tau_1} I_{\tau_1}
  \int_{0}^{2k}{\rm d}\qperp\,\qperp^3
  \left[
    \frac{1}{\qperp^2}-\frac{1}{\qperp^2+\mD^2}
  \right]
  \nonumber \\[1mm]
  & \; = \; 
  \frac{ g^2 g_\vartheta^2 }{ 16\pi^3d_\vartheta } 
  \sum_{\tau_1=\pm 1}C_{\tau_1} 
  I_{\tau_1}
  \ln\left(1+\frac{4k^2}{\mD^2}\right)
  \; .
  \label{eq:htlcontrd5}
\end{align}

%
\subsection{\texorpdfstring{$\opdim=4$}{d=4}  soft exchange}
\label{sec:dim4app}

Let us start from Eq.~\eqref{eq:htlcontrstart4}. 
By taking the $t$ parametrization it turns into
\begin{align}
  \Gamma^{(4)\text{ HTL}}_\vartheta
  \; = \; & 
  \frac{ C^{(4)} }{ 2(4\pi)^3d_\vartheta  k^2 }
  \int_{-\infty}^k {\rm d}q_0
  \int_{|q_0|}^{2k-q_0}{\rm d}q
  \int_{\qp}^\infty {\rm d}p
  \int_{-\pi}^{\pi}\frac{{\rm d}\varphi}{2\pi}\nonumber\\
  & \; \times \; 
  \frac{
    n_{-}(p)n_{-\tau}(k{-}q_0){\bar n}_{+}(p{-}q_0)
  }{
    n_\tau(k)
  }
  \left\vert\mathcal{M}^{(4)}_\mathrm{HTL}\right\vert^2
  \; ,
  \label{eq:htlcontrstart4t}
\end{align}
The fraction therein can be simplified as
\begin{equation}
  \label{statsimpl}
  \frac{
    n_{-}(p)n_{-\tau}(k{-}q_0){\bar n}_{+}(p{-}q_0)
  }{
    n_\tau(k)
  }
  \; = \;
  \frac{  f_{-\tau}(k)  }{  f_\tau(k)  }
  f_{-}(p)[1+ f_{+}(p)]
  \, + \,
  \mathcal{O}\biggl(\frac{q_0}{T}\biggr)
  \; .
\end{equation}
Using 
\begin{equation}
  \frac{f_{-\tau}(k)}{f_\tau(k)}
  \; = \; 
  1+2n_{-\tau}(k)
  \quad \text{ and } \quad
  f_{-}(p)[1+ f_{+}(p)]
  \; = \; 
  \frac12\bigl[f_{+}(p)+f_{-}(p)\bigr]
  \; ,
  \label{fermibose}
\end{equation}
we can rewrite Eq.~\eqref{statsimpl} as
\begin{equation}
  \label{statsimplfinal}
  \frac{
    n_{-}(p)n_{-\tau}(k{-}q_0){\bar n}_{+}(p{-}q_0)
  }{
    n_\tau(k)
  }
  \; = \;
  \biggl[
    \frac12+n_{-\tau}(k)
  \biggr]
  \bigl[
    f_{+}(p)+f_{-}(p)
  \bigr]
  \, + \, 
  \mathcal{O}\biggl(\frac{q_0}{T}\biggr)
  \; .
\end{equation}

Going back to Eq.~\eqref{eq:htlcontrstart4t}, 
taking the explicit expression of the matrix element squared 
in Eq.~\eqref{msqdim4},
performing the Dirac algebra therein 
and taking the soft $\mathcal{Q}\ll \mathcal{K},\mathcal{P}$ limit, 
we get
\begin{align}
  \Gamma^{(4)\text{ HTL}}_\vartheta
  \; = \; &
  \frac{
    g^2 g_\vartheta^2C^{(4)}
  }{
    64\pi^3d_\vartheta k
  } 
  \left[
    \frac{1}{2}+n_{-\tau}(k)
  \right]
  \int_{-\infty}^k {\rm d}q_0
  \int_{|q_0|}^{2k-q_0}{\rm d}q
  \underbrace{
    \int_{0}^\infty {\rm d}p \,p\,
    \bigl[f_{+}(p){+}f_{-}(p)\bigr]
  }_{= J_++J_-}
  \nonumber \\
  & \; \times \;
  \left\{
      \left(\frac{q_0-q}{q}\right)^2
      \left\vert S_R^{+}(\mathcal{Q})\right\vert^2
      +
      \left(\frac{q_0+q}{q}\right)^2
      \left\vert S_R^{-}(\mathcal{Q})\right\vert^2
  \right\}
  \; ,
  \label{eq:htlcontrd4mid}
\end{align}
with $J_\pm$ given in Eq.~\eqref{eq:IandJ}.
Here too we dropped the subleading terms in 
the $\mathcal{Q}\ll \mathcal{K},\mathcal{P}$ expansion
in going from Eq.~\eqref{eq:htlcontrstart4t} 
to \eqref{eq:htlcontrd4mid}.

Similarly to the $\opdim=5$ case, we can rewrite 
the squared moduli of retarded propagators in terms 
of the spectral function. 
Using Eq.~\eqref{htlS} one finds
\begin{equation}
  \left\vert S^{\pm}_R\right\vert^2
  \; = \; 
  \frac{
    2q^2\left(q\pm q_0\right)
  }{
    \pi\masym^2\left(q^2-q_0^2\right)
  } \,  \rho^{\pm}(\mathcal{Q})
  \; .
  \label{eq:propspecfer}
\end{equation}
Plugging this in Eq.~\eqref{eq:htlcontrd4mid} we then obtain
\begin{align}
  \Gamma^{(4)\text{ HTL}}_\vartheta
  & \; = \; 
  \frac{
    g^2 g_\vartheta^2C^{(4)}T^2
  }{
    128\pi^2d_\vartheta\masym^2k
  }  
  \left[\frac{1}{2}+n_{-\tau}(k)\right]
  \int_{-\infty}^k {\rm d}q_0
  \int_{|q_0|}^{2k-q_0}{\rm d}q
  \nonumber \\[2mm]
  &\hspace{3cm}
  \times
  \left[ 
    (q-q_0)\rho^{+}(\mathcal{Q})
    +
    (q+q_0)\rho^{-}(\mathcal{Q})
  \right]
  \; .
  \label{eq:htlcontrd4midd}
\end{align}

This expression can be evaluated following the techniques of 
\cite{Besak:2012qm,Ghiglieri:2013gia,Ghiglieri:2016xye}. 
This amounts to performing the same variable change and 
integration-range manipulation leading to 
Eq.~\eqref{eq:htlcontrd5}. 
The $q_0$ integral is then carried out analytically using 
the light-cone techniques of \cite{Besak:2012qm,Ghiglieri:2013gia}. 
This gives us 
\begin{align}
  \Gamma^{(4)\text{ HTL}}_\vartheta
  \; = \; &
  \frac{
    g^2 g_\vartheta^2C^{(4)}T^2
  }{
    128\pi^2d_\vartheta\masym^2k
  } 
  \left[
    \frac{1}{2}+n_{-\tau}(k)
  \right]
  \int_0^{2k}{\rm d}\qperp\, q_\perp
  \nonumber \\[2mm]
  & \times
  \int_{-\infty}^{+\infty}{\rm d}q_0
  \left[
    \left(1-\frac{q_0}{q}\right)\rho^{+}(\mathcal{Q})
    +
    \left(1+\frac{q_0}{q}\right)\rho^{-}(\mathcal{Q})
  \right]_{q=\sqrt{q_0^2+\qperp^2}}
  \nonumber \\[2mm]
  \; = \; &
  \frac{
    g^2 g_\vartheta^2 C^{(4)} T^2
  }{
    64\pi d_\vartheta k
  }
  \left[
    \frac{1}{2}+n_{-\tau}(k)
  \right]
  \int_0^{2k}{\rm d}\qperp\frac{\qperp}{\qperp^2+\masym^2}
  \nonumber \\[2mm]
  \; = \; &
  \frac{
    g^2 g_\vartheta^2 C^{(4)} T^2
  }{
    128\pi d_\vartheta k
  } 
  \left[
    \frac{1}{2}+n_{-\tau}(k)
  \right]
  \ln\left(1+\frac{4k^2}{\masym^2}\right)
  \; .
  \label{eq:htlcontrd4}
\end{align}

%
\section{The symmetric-phase SM model file}
\label{app_sm}

We provide a complete model file
for the SM at temperatures 
$T\gtrsim 160\;{\rm GeV}$ \cite{DOnofrio:2015gop}, 
where the  gauge group 
$
  G_{\rm SM}
  =
  {\rm SU}(3)_c \times {\rm SU}(2)_{\rm L}\times {\rm U}(1)_Y
$
is unbroken. 
This allows \myaut{} users to implement economical BSM extensions 
with little effort. 

We then take the opportunity to illustrate the  notation used 
for the 
particle content of the symmetric-phase SM and 
its \myaut{} implementation through \myfr{}. 
SM fields are denoted by
\begin{itemize}
  \item 
  Gauge bosons: $B$, $W$, $g$ are the
  ${\rm U}(1)_Y$, ${\rm SU}(2)_{\rm L}$, ${\rm SU}(3)_c$ 
  gauge bosons.
  \item 
  Leptons: $\ell_{\rm L}$ and $e_{\rm R}$, 
  where $\ell_{\rm L}$ is the doublet 
  $\big(\begin{smallmatrix} \nu \\ e\end{smallmatrix}\big)_{\rm L}$.
  They come in $n_{\rm G}=3$ generations. 
  There are no right-handed neutrinos. 
  \item 
  Quarks: $Q_{\rm L}$, $u_{\rm R}$ and $d_{\rm R}$  
  where $Q_{\rm L}$ is the doublet 
  $\big(\begin{smallmatrix} u \\ d\end{smallmatrix}\big)_{\rm L}$. 
  They come in $n_{\rm G}=3$ generations.
  \item 
  Scalars: the Higgs doublet $\phi$.
\end{itemize}

For what concerns interactions, 
in addition to the three gauge couplings we 
implement the Higgs quartic coupling 
$-\lambda(\phi^\dagger\phi)^2$ and 
the top Yukawa coupling
$
  - h_t\bar{Q}_{\mathrm{L}\,3} 
    u_{\mathrm{R}\,3}
    \tilde{\phi}
    +
    \mathrm{h.c.}
$, where the ${}_3$ index labels 
the third generation, e.g. $u_{\mathrm{R}\,3}=t_\mathrm{R}$.

%
\begin{table}[t]
\begin{center}
\begin{tabular}{|c|c|c|c|c|c|}
    \hline
    \rule{0pt}{3ex} Name & Representation & Self-conjugate? & Spin & \multicolumn{2}{c|}{Symbols}
    \\[4pt]
    \hline
    \rule{0pt}{3ex} $B$ & $\repr{1}{1}{0}$ & yes & $1$ & \texttt{V[1]} & \texttt{B}
    \\[4pt]
    \hline
    \rule{0pt}{3ex} $W$ & $\repr{1}{3}{0}$ & yes & $1$ & \texttt{V[2]} & \texttt{Wi}
    \\[4pt]
    \hline
    \rule{0pt}{3ex} $g$ & $\repr{8}{1}{0}$ & yes & $1$ & \texttt{V[3]} & \texttt{G}
    \\[4pt]
    \hline
    \rule{0pt}{3ex} $\ell_{\rm L}$ & $\repr{1}{2}{\frac{1}{2}}$ & no & $\frac{1}{2}$ & \texttt{F[1]} & \texttt{lL}
    \\[4pt]
    \hline
    \rule{0pt}{3ex} $e_{\rm R}$ & $\repr{1}{1}{1}$ & no & $\frac{1}{2}$ & \texttt{F[2]} & \texttt{eR}
    \\[4pt]
    \hline
    \rule{0pt}{3ex} $Q_{\rm L}$ & $\repr{3}{2}{-\frac{1}{6}}$ & no & $\frac{1}{2}$ & \texttt{F[3]} & \texttt{QL}
    \\[4pt]
    \hline
    \rule{0pt}{3ex} $u_{\rm R}$ & $\repr{3}{1}{-\frac{2}{3}}$ & no & $\frac{1}{2}$ & \texttt{F[4]} & \texttt{uR}
    \\[4pt]
    \hline
    \rule{0pt}{3ex} $d_{\rm R}$ & $\repr{3}{1}{\frac{1}{3}}$ & no & $\frac{1}{2}$ & \texttt{F[5]} & \texttt{dR}
    \\[4pt]
    \hline
    \rule{0pt}{3ex} $\phi$ & $\repr{1}{2}{-\frac{1}{2}}$ & no & $0$ & \texttt{S[1]} & \texttt{Phi}
    \\[4pt]
    \hline
\end{tabular}
\end{center}

\caption[a]{\small
  Particle content of the symmetric-phase Standard Model defined in 
  \texttt{analytical/models/symmetric.fr}, 
  see App.~\ref{app_sm} for an explanation of the notation used 
  in the ``Representation'' column and of the sub-columns in 
  the ``Symbols'' column.
}
\label{tab:smcontent}
\end{table}
%

We summarize this in Table~\ref{tab:smcontent} and give 
each particle's representation in $G_{\rm SM}$. 
We use the notation $\repr{\alpha}{\beta}{Y}$, 
where $\bm{\alpha}$ is the ${\rm SU}(3)_c$ 
representation, $\bm{\beta}$ is 
the ${\rm SU}(2)_{\rm L}$ representation and $Y$ is the hypercharge. 
In our convention the electromagnetic charge $Q$ reads $Q=T_3-Y$, 
where $T_3=\pm1/2$ for the two 
components of each ${\rm SU}(2)$ doublet.

The ``Symbols'' column in Table~\ref{tab:smcontent} gives 
the symbols used by \myaut{} to designate the particles. 
The left sub-column corresponds to the names that should 
be used to refer to the particles in the configuration file 
(see Sec.~\ref{sec:conf} and App.~\ref{app:cfg}) 
and in using the \myaut{} \mymath{} package 
(see Sec. \ref{sec:runmath}); 
they are inherited from the \myfr{} model file.
The right sub-column corresponds to the names used by \myaut{} in 
the generated PostScript files for the Feynman diagrams.

%
\section{The \myaut{} configuration file}
\label{app:cfg}

The \textit{configuration file} is documented extensively online. 
We summarize here its main points.
It contains 3 sections: \texttt{Tools}, \texttt{Model} and \texttt{Run}.
\begin{mdframed}[style=mylststyle]
\begin{lstlisting}[language=myYAML,style=YAMLstyle]
[Tools]
math = /path/to/math
feynrules = /path/to/feynrules-current
feynarts = /path/to/FeynArts-X.YY/FeynArts.m
formcalc = /path/to/FormCalc-X.YY/FormCalc.m
\end{lstlisting}
\end{mdframed}
The \texttt{Tools} section contains the paths to the tools used by 
the program, 
namely the location of the \texttt{WolframKernel} executable and 
the path of the
\myfr{},   \texttt{feynarts} and \texttt{formcalc} installations.

\begin{mdframed}[style=mylststyle]
\begin{lstlisting}[language=myYAML,style=YAMLstyle]
[Model]
modelpath = /path/to/the/model/file.fr
includeSM = yes/no
lagrangian = L
produced = X[n]
inbath = Y[i] Z[j] A[k]
assumptions = ...
replacements = ...
noneq = ...
flavorexpand = {X[i],Ind}, {-X[i],Ind}
\end{lstlisting}
\end{mdframed}

The \texttt{Model} section contains instructions 
about the computation. 
\begin{itemize}
  \item 
  The \texttt{modelpath} key is the path to the \textit{model file} 
  describing the model considered by the user; 
  \item 
  The \texttt{includeSM} key is boolean. 
  If set to \texttt{yes}, 
  \myaut{} will merge the user-provided model file 
  with the SM model file \texttt{symmetric.fr} 
  --- see Sec.~\ref{sec:validation} and App.~\ref{app_sm}. 
  This provides an easy way of studying a model where 
  the Standard Model is  extended with a few particles only; 
  \item 
  The \texttt{lagrangian} key is the variable containing 
  the Lagrangian in the model file. 
  If the key \texttt{includeSM} is set to \texttt{yes}, 
  then the names 
  \texttt{LGauge}, 
  \texttt{LFermion}, 
  \texttt{LHiggs}, 
  \texttt{LYukawa} and 
  \texttt{LSM} cannot be used as they would create a conflict 
  with already existing variables; 
  \item 
  The \texttt{produced} key is the model-file name for 
  the particle whose production rate must be computed. 
  It has the form \texttt{X[n]}. 
  While \myfr{} supports a large range of particle types, 
  \myaut{} is restricted to the following: 
  scalar fields \texttt{S}, 
  Majorana/Dirac fermions \texttt{F}, 
  vector fields \texttt{V} and 
  spin-two fields \texttt{T}. 
  In particular, spin-3/2 fields are currently unsupported;
  \footnote{%
    See our companion 
    paper~\cite{gravitinopaper} for our handling of gravitino production through its spin-1/2 Goldstino component.
  }
  \item 
  The \texttt{inbath} key describes the particles present in 
  the thermal bath.
  There are three possible ways to set this key, 
  as described in the documentation. 
  We emphasize that the code treats particles and 
  antiparticles separately. 
  Therefore, if the user wants to add or remove a particle 
  that is not self-conjugate, they should provide both 
  the name of the particle and the antiparticle. 
  \item
  The \texttt{assumptions} and \texttt{replacements} optional 
  keys contain a comma-separated list, 
  written the \textsc{Wolfram} language. 
  In the former case, one can provide information about 
  the real nature of some coupling constants: 
  \mymath{} assumes that all variables are complex by default 
  which could prevent further simplifications in the final result. 
  Similarly, replacement rules for parameters in 
  the model might be useful for writing intricate combinations 
  of parameters in a shorthand form. 
  \item 
  The \texttt{noneq} key contains the model-file symbol for 
  the bath-$\vartheta$ coupling constant. 
  \item 
  The \texttt{flavorexpand} optional key contains 
  a list of fermions that have 
  generation-depen\-dent Yukawa couplings and therefore thermal masses. 
  As explained at the end of Sec.~\ref{sec:masses}, 
  handling non-diagonal Yukawa couplings would require redefining 
  the fermion fields to find a diagonal basis. 
  This key should be formatted as a comma-separated list of 
  lists containing two elements: the name of the fermion and 
  the name of the generation index used in the Yukawa coupling. 
  Again, \myaut{} treats fermions and anti-fermions separately; 
  both particles need to be given. 
  Further details on the use cases of this key are given 
  in App.~\ref{app:yukawa}.    
\end{itemize}

\begin{mdframed}[style=mylststyle]
\begin{lstlisting}[language=myYAML,style=YAMLstyle]
[Run]
conf = yes/no
rules = yes/no
proc = yes/no
verbose = yes/no
autothermdir = /path/to/autotherm/
\end{lstlisting}
\end{mdframed}

The \texttt{Run} section contains information on how 
the code should be run. 
\begin{itemize}
  \item 
  The boolean \texttt{conf} key controls whether 
  the \texttt{.m} files used by the \anpart{} component 
  of \myaut{} should be generated. 
  This should be set to \texttt{yes} for the first run of 
  any model or if the configuration file has been changed. 
  \item 
  The boolean \texttt{rules} key tells \myaut{} to generate 
  the Feynman rules of the model. 
  This should be set to \texttt{yes} if the Feynman rules 
  have not been generated yet, or if the model file was changed. 
  \item 
  The boolean \texttt{proc} key tells \myaut{} to 
  compute the matrix elements squared 
  $\left\vert \mathcal{M}(\tau_1;\sigma_1,\sigma_2)\right\vert^2$, 
  as per Eq.~\eqref{eq:totalmat}, 
  as well as the thermal masses of the mediators 
  in IR-divergent processes. 
  \item 
  The boolean \texttt{verbose} key controls the level of 
  command-line output of \myaut{}. 
  In particular, setting it to \texttt{no} suppresses 
  all messages from \myfa{} and \myfc{}. 
  \item 
  The \texttt{autothermdir} key is the path to 
  the top-level directory of 
  the \myaut{} installation on the user's system.
\end{itemize}

%
\section{The \texttt{analytical\_pipeline} output files}
\label{app:json}

A successful execution of the \anpart{} component, either 
through \texttt{autotherm-run} or 
through \texttt{controller.analytical\_pipeline} 
will generate the following files and directories
\begin{itemize}
  \item 
  \texttt{config.m} and \texttt{config\_2.m}: 
  contain the same configuration information as \texttt{config.cfg}, 
  in a format that can be parsed by \mymath{}. 
  \item 
  \texttt{config\_totmat2.m}: 
  contains the result for 
  $\left\vert \mathcal{M}(\tau_1;\sigma_1,\sigma_2)\right\vert^2$. 
  This expression includes helper functions used by \myaut{} 
  to keep track of the processes.
  \item 
  \texttt{config\_mat2.m}: 
  contains the list of all computed processes 
  (of the form {\ttfamily \{X[i],X[j]\}->} {\ttfamily \{Y[k],Z[l]\}}) 
  with the associated $\left\vert \mathcal{M}\right\vert^2$. 
  Those results also contain the helper functions mentioned in 
  the previous item. 
  \item 
  \texttt{config\_result.json}: 
  contains the final result in a more human-readable format. 
  \item 
  \texttt{Diagrams/}: 
  contains the Feynman diagrams for all nonzero 
  processes in \textsc{PostScript} format. 
  Files are named \texttt{diagram\_p1\_p2\_p3\_p4.ps} 
  where \texttt{p1,p2,p3,p4} are the names for 
  the particles in the processes, 
  taken from the \texttt{PropagatorLabel} parameters in 
  the model file. 
  For antiparticles, the suffix \texttt{bar} is added to the name. 
  An example is shown in Fig.~\ref{fig:d4diags}. 
  The \TeX{} format can be obtained by setting \texttt{AUTdiagext} 
  to \texttt{".tex"} in a \textsc{Mathematica} 
  notebook when using \myaut{} as a standalone \textsc{Mathematica} 
  package, as outlined in Sec.~\ref{sec:runmath}. 
  See~\cite{Hahn:2000kx} for further instructions on 
  the handling of the \myfa{} \TeX{} output. 
\end{itemize}

We will now describe how the data inside 
the \texttt{config\_result.json} file is structured 
--- see Sec.~\ref{sec:runpython} for relevant context.

\begin{mdframed}[style=mylststyle]
\begin{lstlisting}[style=JSONstyle]
"mat2": {
    "(s1,s2,t1)": "...",
    ...
}
\end{lstlisting}
\end{mdframed}

The \texttt{mat2} section contains 
$\left\vert \mathcal{M}(\tau_1;\sigma_1,\sigma_2)\right\vert^2$. 
Note that they are classified by a $(\sigma_1,\sigma_2,\tau_1)$ tuple.
The helper functions have been removed from this result, except
for the cases discussed in App.~\ref{app:yukawa}.

\begin{mdframed}[style=mylststyle]
\begin{lstlisting}[style=JSONstyle]
"masses": {
    "X[i]": "...",
    "Y[i][1]": "...",
    "Y[i][2]": "...",
    ...
    "Y[i][n]": "...",
    ...
}
\end{lstlisting}
\end{mdframed}

The \texttt{masses} section contains a list of all mediators 
for processes that would cause a divergence with their thermal masses 
squared, normalized by the temperature, that is $m_T^2/T^2$. 
$m_T=\mD$ for gauge bosons, $m_T=\masym$ for fermions. 
In the latter case, thermal masses for mediators appearing in 
the \texttt{flavorexpand} 
list in the configuration file  are given generation by generation. 

\begin{mdframed}[style=mylststyle]
\begin{lstlisting}[style=JSONstyle]
"couplings": {
    "gauge": "g1,...",
    "noneq": "noneqnum",
    "others": "x1,x2,..."
}
\end{lstlisting}
\end{mdframed}

The \texttt{couplings} section contains comma-separated lists of all 
parameters appearing in 
$\left\vert \mathcal{M}(\tau_1;\sigma_1,\sigma_2)\right\vert^2$, 
split into the following categories:
\begin{itemize}
  \item 
  \texttt{gauge}: 
  those are all gauge groups couplings constants, 
  i.e. symbols that have been set as a \texttt{CouplingConstant} in 
  the definition of a gauge group in the model file.
  \item
  \texttt{noneq}: 
  this is the coupling parameter between $\vartheta$ and 
  the thermal bath, i.e. the value of the \texttt{noneq} key in 
  the configuration file.
  \item 
  \texttt{others}: 
  all other parameters appearing in the matrix elements squared.
\end{itemize}

\begin{mdframed}[style=mylststyle]
\begin{lstlisting}[language=myJSON,style=JSONstyle]
"flags": {
    "fermionflag": true/false
}
\end{lstlisting}
\end{mdframed}

The \texttt{flags} section contains additional information about 
the computation. 
For now, the only category is \texttt{fermionflag}. 
Its value is set to \texttt{true} when $t$- or $u$-channel 
fermion mediators with generation-dependent thermal mass contribute 
to $\Gamma_\vartheta$, and to \texttt{false} otherwise. 
See App.~\ref{app:yukawa} and 
the ``Majorana fermion DM'' model~\cite{Biondini:2020ric} 
example for details on the handling of this case. 

%
\section{Handling of generation-dependent Yukawa couplings}
\label{app:yukawa}

Let us assume for definiteness an interaction 
of the form $Y_{ij}\bar\psi_i \chi_j \phi+\text{h.c.}$,
with $\psi$ and $\chi$ two Weyl fermions of opposite chirality  
and $i$ and $j$ two indices in two (potentially different) 
generation spaces. $\phi$ is a scalar.

A first necessary step is to be carried out in the model file. 
The user then needs to define \emph{two} entries in its 
\texttt{M\$Parameters} section. 
The first entry, e.g. \texttt{ymat} for our $Y_{ij}$, 
must have its \texttt{ParameterType} set to \texttt{Internal}.
The second entry \emph{must} be named by appending the suffix 
\texttt{ext} to the name of the first. 
In our example, this would be \texttt{ymatext}. 
For this entry, \texttt{ParameterType} must be set 
to \texttt{External} and
\texttt{Value} to be a list of replacement rules for the values of 
the different elements of \texttt{ymat}.
This could be something along the lines of 
{\ttfamily ymatext[1,1]->y11,...}\,.

The contribution from this Yukawa coupling to the thermal mass of 
the $\psi$ field will be proportional to $(YY^\dagger)_{ii'}$. 
Conversely, the $\chi$ field receives 
a contribution proportional to $(Y^\dagger Y)_{jj'}$. 
Our \texttt{ThermalMass} routine 
described  in Sec.~\ref{sec:runmath} can only handle cases
where $(YY^\dagger)_{ii'}$ and $(Y^\dagger Y)_{jj'}$ are diagonal. 
If the 
both $i$ and $j$ live in the three-dimensional SM generation space, 
\texttt{ThermalMass} computes the thermal mass correctly and 
does not require the \texttt{flavorexpand} 
key to be set in the configuration file outlined 
in App.~\ref{app:cfg}. 
It can also handle cases where the scalar, 
rather than one of the two fermions, 
lives in that configuration space. 

Our implementation of the MSSM, as outlined in Sec.~\ref{sub:d5val}, 
provides a non-trivial check of our routine. 
\texttt{ThermalMass} indeed correctly determines all thermal masses 
for MSSM particles --- see e.g.~\cite{Rychkov:2007uq}. 
In particular, it finds the \emph{same} asymptotic masses 
for both superpartners within each chiral multiplet, 
including the $h_t$-dependent ones 
for third-generation $Q_\mathrm{L}$, $u_\mathrm{R}$ and 
their squark counterparts, 
in accordance with the arguments in~\cite{Caron-Huot:2008vbk}. 

%
\begin{figure}[t]

\begin{center}
  \includegraphics[width=0.6\linewidth,trim=0 7.2cm 0 0, clip]{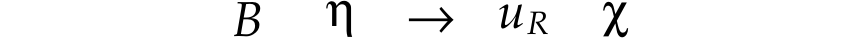}$\quad$
  \includegraphics[width=0.29\linewidth,trim=0 13.cm 0 0, clip]{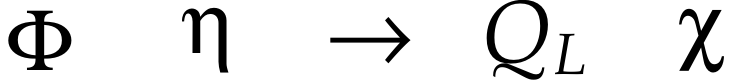}
  \includegraphics[width=0.6\linewidth,trim=0 0 0 3cm, clip]{figs/d4yukawa}$\quad$
  \includegraphics[width=0.29\linewidth,trim=0 0 0 4cm, clip]{figs/d4yukawa2}
\end{center}

\vspace{-2mm}

\caption[a]{\small
  Sample processes mediated by a $t$-channel $u_\mathrm{R}$ in 
  the model studied in~\cite{Biondini:2020ric}.
  Figure generated by \myaut{} through \myfa{}.
}
\label{fig:yukawa}
\end{figure}
%

The \texttt{flavorexpand} key becomes necessary whenever 
a fermion with a generation-dependent 
thermal mass mediates a $t$- or $u$-channel exchange. 
The ``Majorana fermion DM''
model described in Sec.~\ref{sub:d4val} is one such example. 
In Fig.~\ref{fig:yukawa} 
we show on the left the two tree-level diagrams for 
the process $B\bar\eta\to \bar u_\mathrm{R}\chi$, 
where 
$\chi$ is the Majorana DM candidate, 
$\eta$ is a colored scalar, 
$u_\mathrm{R}$ an up-type right-handed quark and 
$B$ the U(1) gauge boson --- such a process would 
also arise in supersymmetric scenarios where 
$\eta$ is a squark and $\chi$ a gaugino.
On the right we show the single diagram for 
the $\phi\bar\eta\to \bar Q_\mathrm{L}\chi$ process.

As Fig.~\ref{fig:yukawa}  shows, in these processes, 
along with many others not shown here, 
the $u_\mathrm{R}$ quark mediates a $t$-channel exchange; 
HTL resummation of the $u_\mathrm{R}$ propagator 
is therefore necessary for a finite $\Gamma_\chi$. 
As Eq.~\eqref{htlS} shows, the fermion HTL self-energy factorizes 
into the thermal mass times 
a universal function of $q_0/q$. 
In this case the thermal mass for $t_\mathrm{R}$ 
will be larger than that for $u_\mathrm{R}$ and $c_\mathrm{R}$, 
due to the top Yukawa $h_t$. 
Indeed, executing 
\texttt{ThermalMass} in a \textsc{Mathematica} notebook
as in Sec.~\ref{sec:runmath}, i.e. 

\begin{mdframed}[style=mylststyle]
\begin{lstlisting}[language=mymath,style=mathstyle]
ThermalMass[F[4]]
\end{lstlisting}
\end{mdframed} 
returns for the up-type right-handed quarks \texttt{F[4]}
\begin{mdframed}[style=mylststyle]
\begin{displaymath}
\hspace{-3cm}\left\{\frac{1}{9} \left(\text{g1}^2+3 \text{g3}^2\right),\frac{1}{9} \left(\text{g1}^2+3 \text{g3}^2\right),\frac{1}{9} \left(\text{g1}^2+3
\text{g3}^2\right)+\frac{\text{ht}^2}{4}\right\}
\end{displaymath}
\end{mdframed}
This is a \textsc{Mathematica} list with three entries, 
corresponding to the three generations. 
It correctly shows the top Yukawa contribution for $t_\mathrm{R}$.
We remark that we have implemented $h_t$ as a real valued coupling 
in our \texttt{symmetric.fr} model file. 

When evaluating the matrix elements squared for processes such as 
those in Fig.~\ref{fig:yukawa}, 
we then need to replace the generation-space $\delta_{ij}$ structure 
in  the $u_\mathrm{R}$ 
propagator with a $m_i\delta_{ij}$ one, 
with $m_i$ a bookkeeping coefficient. 
The matrix element squared 
will then be proportional to a $m_i m_{i'}^*$ matrix; 
when the 
fermion thermal mass matrix is diagonal, this matrix is diagonal too.
Hence 
$
  \vert\mathcal{M}_{ab\to c\chi}\vert^2
  =
  \sum_i \vert m_i\vert^2 \vert\mathcal{M}_{ab\to c\chi}\vert_{i}^2 
$.
By differentiating with respect to the squared moduli of 
the bookkeeping coefficients $m_i$ we then obtain 
the $\vert\mathcal{M}_{ab\to c\chi}\vert_{i}^2$ 
contribution of each  $u_{\mathrm{R}\,i}$ mediator to 
the matrix elements squared. 

In its current form, \myaut{}, namely its 
\texttt{ComputeMatrixElement} \textsc{Wolfram}-language routine, 
needs to be instructed to perform such a decomposition. 
This is done by setting 
\begin{mdframed}[style=mylststyle]
\begin{lstlisting}[language=myYAML,style=YAMLstyle]
flavorexpand = {F[4],Generation}, {-F[4],Generation}
\end{lstlisting}
\end{mdframed}
in the configuration file. 
Let us test this for the  processes in Fig.~\ref{fig:yukawa}. 
By executing these instructions in a \textsc{Mathematica} notebook: 
\begin{mdframed}[style=mylststyle]
\begin{lstlisting}[language=mymath,style=mathstyle]
clean = {AUThast[_] -> 1, AUTstatspart[___] -> 1};
pr1=ComputeMatrixElement[{V[1],-S[2]}->{-F[4],F[6]}]/.clean;
pr2=ComputeMatrixElement[{S[1],-S[2]}->{-F[3],F[6]}]/.clean;
Print[pr1]; 
Print[pr2];
\end{lstlisting}
\end{mdframed} 
we obtain as output
\begin{mdframed}[style=mylststyle]
\begin{displaymath}
\hspace{-5.5cm}
\left\{\frac{8 \text{c1}^2 \text{g1}^2 U y^2}{3 T},
\frac{8 \text{c2}^2 \text{g1}^2 U y^2}{3 T},\frac{8 \text{c3}^2 \text{g1}^2 U y^2}{3 T}\right\}
\end{displaymath}
\begin{displaymath}
\hspace{-9.4cm}
\left\{0,0,\frac{6 \text{c3}^2 \text{ht}^2 U y^2}{T}\right\}
\end{displaymath}
\end{mdframed}
In \myfc{} $U$ and $T$ are the Mandelstam invariants. 
As per \texttt{chiDM.fr}, we have  
$
  \mathcal{L}_\mathrm{int}
  =
  -y c_i\bar\chi u_\mathrm{R\,i}\bar\eta
$ + h.c.,
so that we can set $y$ as a \texttt{noneq} real number and 
the $c_i$ as \texttt{others} user-provided couplings.

These results show how the $B\bar\eta\to \bar u_\mathrm{R}\chi$ 
process receives an equal contribution from 
each intermediate $u_\mathrm{R}$ (trivially, this is 
also true for intermediate scalars), due to the universality 
of the gauge coupling to the $B$ boson. 
Conversely, the 
$\bar\phi\eta\to Q_\mathrm{L}\chi$ process is nonvanishing 
only when mediated by $t_\mathrm{R}$, 
due to the presence of the SM up-type Yukawa coupling. 

Whenever the \texttt{flavorexpand} key is nonempty and (some of)
the matrix elements squared become lists in generation space, 
the \anpart{} component of \myaut{} will introduce 
some helper functions in the \texttt{mat2} 
entries and set \texttt{fermionflag} to \texttt{true}
in the .json output file. When the \texttt{analytical\_pipeline}
function of the \texttt{controller} module parses this file,
it will handle this case differently, i.e.
\begin{mdframed}[style=mylststyle]
\begin{lstlisting}[language=Python,style=pystyle]
from analytical.controller import *
from numerical.manipulate import *

# If we already ran the field-theoretical pipeline
# set "conf", "rules" and "proc" in chiDM.cfg to "no"
(msqdictlist, couplingsdict,masslist) = analytical_pipeline("chiDM.cfg")

rateobjlist = [NumRate(msqdict,couplingsdict,masslist[i],1) for i, msqdict in enumerate(msqdictlist)]

\end{lstlisting}
\end{mdframed}
In this case the first and third elements returned by 
\texttt{analytical\_pipeline} are a list of dicts and 
a list of strings, rather than a dict and a string. 
These are lists in generation space and are thus fed by 
the final instruction of this block into three 
instances of the \texttt{NumRate} class. 
Note that \texttt{deg} has been set to one when 
creating the \texttt{NumRate} instances, 
in accordance with footnote~\ref{foot:weyl}. 

Finally, the rate can be obtained from \texttt{rateobjlist} 
by summing over generations  as follows
\begin{mdframed}[style=mylststyle]
\begin{lstlisting}[language=Python,style=pystyle]
NPTS=100
k=NumPy.geomspace(0.01,15,NPTS)
numrates=[NumPy.zeros(NPTS),NumPy.zeros(NPTS),NumPy.zeros(NPTS)]
for i in range(3):
    rateeval=rateobjlist[i].rate(k,yval,numcouplings,0)
    for j in range(3):
        numrates[j] += rateeval[j+1]
\end{lstlisting}
\end{mdframed}
where \texttt{yval} and \texttt{numcouplings} encode 
the values of $y$ and of the other couplings. 
The three entries of \texttt{numrates} are then three arrays 
containing the values of $\Gamma_\chi$ in the three schemes.

%
\section{Ad-hoc extensions to \myfa{} and \myfc{}}
\label{app:fafc}

This Section  describes ad-hoc 
changes made to the usual \myfa{}-\myfc{} pipeline for 
the purposes of \myaut{}.

%
\subsection{Polarization sum for massless spin-two particles}
\label{sub:graviton}

Computing the production rate of the graviton
requires summing the polarizations for this particle.
We will describe here the procedure used by \myaut{}.
Note that the current version of \myaut{} can only handle 
a single massless tensor among the external states.

Each external spin-2 particle is associated with 
a polarization tensor $\polvec{\mu\nu}{k}$,
where $\lambda=1,2$ indicates which polarization state 
the particle is in;
a massless spin-2 particle has two possible polarization states.

One can then show~\cite{Ghiglieri:2020mhm} that 
the polarization sum becomes
\begin{equation}
  \sum_\lambda
  \polvec{\mu\nu}{k} 
  \left[ \polvec{\alpha\beta}{k} \right]^{\star}
  \; = \; 
  L_{\mu\nu;\alpha\beta}
  \; ,
  \label{eq:polsumtens}
\end{equation}
where
\begin{align}
  L_{\mu\nu;\alpha\beta}
  & \; = \;
  \frac{
    \projt{\mu\alpha}\projt{\nu\beta}
    +
    \projt{\mu\beta}\projt{\nu\alpha}
    -
    \projt{\mu\nu}\projt{\alpha\beta}
  }{2}
  \; ,
  \label{eq:defl}
  \\[1mm]
  \projt{\mu\nu}
  & \; = \; 
  \delta_{\mu i}\delta_{\nu j}
  \left(\delta_{ij}-\frac{k_i k_j}{k^2}\right)
  \; = \; 
  - \eta_{\mu\nu}
  + \frac{
      \mathcal{U}_\mu\mathcal{K}_\nu+\mathcal{U}_\nu\mathcal{K}_\mu
    }{
      \mathcal{K}\cdot\mathcal{U}
    } 
  - \frac{
      \mathcal{K}_\mu\mathcal{K}_\nu
    }{
      \left(\mathcal{K}\cdot\mathcal{U}\right)^2
    }
  \; .
  \label{eq:defpt}
\end{align}
In Eq.~\eqref{eq:defpt} we expressed the projector $\projt{\mu\nu}$ 
in a more covariant fashion by introducing 
the four-velocity defining the medium rest frame 
$\mathcal{U}^\mu=(1,\mathbf{0})$. 
$\mathcal{U}$ will disappear from the final result because 
of gravitational gauge invariance. 

\myfc{} is currently unable to perform the polarization sum; 
it stops by rewriting 
the polarization tensor as the product of polarization vectors 
for massless spin-one particles 
--- see e.g.~\cite{Bjerrum-Bohr:2014lea,Holstein:2006bh}. 
That is
\begin{equation}
  \polvec{\mu\nu}{k}
  \; \equiv \; 
  \polvec{\mu}{k}\polvec{\nu}{k}
  \; .
  \label{eq:polstensor}
\end{equation}

Our function \texttt{TensorSum}, written in 
the \textsc{Wolfram} language, 
implements Eqs.~\eqref{eq:polsumtens}--\eqref{eq:defpt} and 
checks that the final expression is $\mathcal{U}$-independent.

%
\subsection{SU(2) and generation algebra}

The \myfa{} and \myfc{} tools are designed to 
compute matrix elements squared in the broken phase. 
While the SM model file included with \myfr{} does contain 
the ${\rm SU}(2)_\mathrm{L}$ gauge group, 
it represents the Standard model in the broken phase, 
where  ${\rm SU}(2)$ indices turn into flavor indices 
in the fermion sector. 
In the boson sector they label gauge, 
Goldstone and Higgs modes. 
In most use cases, 
fermions are also treated generation by generation.

In contrast, one of the primary use cases of \myaut{} is to study 
early-universe particles dynamics 
before EW symmetry breaking.
We therefore need to implement a way to properly deal with 
${\rm SU}(2)$ and generation algebra. 
The \texttt{symmetric.fr} model file provided 
with \myaut{} properly defines the ${\rm SU}(2)$ 
gauge group, 
including the relevant representations and structure constants. 
Within \texttt{autotherm.wl}, 
tracing over ${\rm SU}(2)$ and generation indices is 
handled by the \texttt{SUAlgebra} function 
for matrix elements squared and by 
the \texttt{AUTSUNSimplify} and \texttt{DoTheYukawa} 
functions for amplitudes appearing during 
the computation of thermal masses. 

During the development of \myaut{}, 
we noticed that \myfc{} handles polynomials 
of Kronecker deltas incorrectly. 
This is relevant for processes involving e.g. 
the quartic $W$ vertex.
We worked around this by replacing all Kronecker deltas 
for ${\rm SU}(2)$ indices in amplitudes by 
a dummy function \texttt{AUTIndexDelta}. 


{\small
%
\bibliographystyle{utphys}
\bibliography{ref}

\cleardoublepage
}

\end{document}